\documentclass[11pt,a4paper]{article}
\usepackage{jheppub}
\usepackage{amsmath,amssymb,amsfonts,bbold,mathdots, verbatim}
\usepackage{tikz}
\usetikzlibrary{decorations.markings}
\usepackage{mathrsfs}

\renewcommand{\O}{\ensuremath{\mathcal{O}}}

\newcommand{\e}{\epsilon}

\renewcommand{\L}{\mathcal{L}}
\newcommand{\bcL}{\bar{\mathcal{L}}}

\newcommand{\h}{{\bar h}}

\newcommand{\p}{\partial}

\renewcommand{\t}{\tau}
\newcommand{\s}{\sigma}
\newcommand{\D}{\Delta}
\newcommand{\refb}[1]{(\ref{#1})}
\newcommand{\w}{\omega}
\newcommand{\bw}{\bar{\omega}}
\newcommand{\<}{\langle}
\renewcommand{\>}{\rangle}

\newcommand{\non}{\nonumber}

\newcommand{\txb}{\textcolor{cyan}}

\renewcommand{\O}{\ensuremath{\mathcal{O}}}

\def\beaa{\begin{eqnarray*}}
\def\eeaa{\end{eqnarray*}}
\def\bea{\begin{eqnarray}}
\def\eea{\end{eqnarray}}
\def\be{\begin{equation}}
\def\ee{\end{equation}}
\newcommand{\bes}{\begin{subequations}}
\newcommand{\ees}{\end{subequations}}
\def\ba{\begin{align}}
\def\ea{\end{align}}

\def\vec{\overrightarrow}

\def\non{\nonumber}

\title{BMS Modular Diaries: Torus one-point function}

\author[a]{Arjun Bagchi,} \author[a, b]{Poulami Nandi,} \author[a]{Amartya Saha,} \author[a, c]{and Zodinmawia.} \author{\\}

\affiliation[a]{Indian Institute of Technology Kanpur, Kalyanpur, Kanpur 208016. INDIA. \\}

\affiliation[b]{Center for Quantum Mathematics and Physics, University of California, Davis, CA 95616 USA.  \\}

\affiliation[c]{Department of Physics, Mizoram University, Aizawl 796004, India.  \\}

\emailAdd{abagchi@iitk.ac.in, pnandi@ucdavis.edu, amartyas@iitk.ac.in, zdma84@gmail.com}

\abstract{Two dimensional field theories invariant under the Bondi-Metzner-Sachs (BMS) group are conjectured to be dual to asymptotically flat spacetimes in three dimensions. In this paper, we continue our investigations of the modular properties of these field theories. In particular, we focus on the BMS torus one-point function. We use two different methods to arrive at expressions for asymptotic structure constants for general states in the theory utilising modular properties of the torus one-point function. We then concentrate on the BMS highest weight representation, and derive a host of new results, the most important of which is the BMS torus block. In a particular limit of large weights, we derive the leading and sub-leading pieces of the BMS torus block, which we then use to rederive an expression for the asymptotic structure constants for BMS primaries. Finally, we perform a bulk computation of a probe scalar in the background of a flatspace cosmological solution based on the geodesic approximation to reproduce our field theoretic results.}

\begin{document}

\maketitle

\newpage

\section{Introduction}

\subsection{Preliminaries: Conformal symmetry} 

The techniques of conformal field theory (CFT) \cite{DiFrancesco:1997nk, Blumenhagen:2009zz} have been central to our understanding of various diverse subjects including the theory of phase transitions in statistical physics, the understanding of string theory from worldsheet symmetries, the study of quantum gravity in AdS (and dS) spacetimes through the celebrated gauge-gravity correspondence, as well as explorations in cosmology. These techniques hinge on the extra symmetries CFTs possess vis-a-vis usual relativistic quantum field theories (QFTs). Of particular importance is the procedure of conformal bootstrap \cite{Ferrara:1973yt, Polyakov:1974gs, Poland:2016chs, Simmons-Duffin:2016gjk}, which aims to solve CFTs by imposing an infinite set of consistency conditions on the theory that stem from crossing symmetry of the four-point function. This programme has seen a resurgence over the last decade, following the seminal work \cite{Rattazzi:2008pe}.

\medskip

\noindent 
In general $d$ dimensions, the symmetry algebra that underlies a CFT is $so(d,2)$ as opposed to $iso(d-1,1)$ of a Poincare-invariant QFT in the same dimension. In two dimensional CFTs, the underlying symmetry enhances to two copies of the infinite dimensional Virasoro algebra. Following the lead of Belavin, Polyakov and Zamolodchikov \cite{Belavin:1984vu}, over the years, 2d CFTs have been the most fertile of all theories for performing analytic calculations. The power of infinite symmetry coupled with the might of the machinery of complex analysis has helped with an enormous amount of analytic control on calculations in 2d CFTs in general, without having to refer to any particular Lagrangian description. Before the current resurgence mentioned above, conformal bootstrap techniques were used in 2d to exactly solve a class of CFTs, called the 2d Minimal models.  

\medskip

\noindent 
In addition to consistency from crossing symmetry of 4-point function, in 2d CFTs there are further constraints that arise from modular invariance on the torus. These help in constraining the possible parameter space in 2d CFTs even further. Modular invariance has the additional very useful feature of relating the low energy spectrum of the theory to the high energy spectrum. This leads to the famous Cardy formula for entropy counting in 2d CFTs \cite{Cardy:1986ie}, which spectacularly matches up with the Bekenstein-Hawking entropy of Banados-Teitelboim-Zanelli black holes in AdS$_3$, and serves as one of the very early successes of the holographic principle \cite{Strominger:1997eq,Carlip:1998qw}.

\subsection{Flat spacetimes, BMS and Holography} 

\noindent 
Holography for AdS$_3$ uses the power of the infinite dimensional Virasoro symmetries, which are also the asymptotic symmetries of the bulk spacetime \cite{Brown:1986nw}. However, infinite asymptotic symmetries are not exclusive to AdS$_3$. Bondi, van der Burg, Mezner and, independently, Sachs in the 1960's found rather surprisingly that the symmetries of 4d asymptotically flat spacetimes at null infinity ($\mathscr{I}^\pm$) enhances from the expected Poincare group to to an infinite dimensional group, named BMS after the discoverers \cite{Bondi:1962px, Sachs}. There has been a recent explosion in activity in research connecting the infrared physics of gauge theories and gravity to the BMS group \cite{Strominger:2017zoo}. The existence of an infinite dimensional asymptotic symmetry algebra is also true in 3d asymptotically flat spacetimes and the symmetry algebra in question is the BMS$_3$ \cite{Barnich:2006av}:
\bes{}\label{bms3}
\bea{}
&& [L_n, L_m] = (n-m) L_{n+m} + c_L (n^3-n) \delta_{n+m, 0} \\
&& [L_n, M_m] = (n-m) M_{n+m} + c_M (n^3-n) \delta_{n+m, 0} \\
&& [M_n, M_m] = 0.
\eea
\ees
Here $L_n$'s are the so-called super-rotations, which are the diffeomorphisms of the circle at infinity. $M_n$'s are called super-translations and these are angle dependent translations along the null directions of $\mathscr{I}^+$. $c_L, c_M$ are two allowed central extensions of the theory. For Einstein gravity, $c_L = 0, c_M= 1/4G$, where $G$ is the Newton's constant. One can generate other values of the central term by considering more general theories of gravity e.g. by adding gravitational Chern-Simons term to the Einstein-Hilbert action \cite{Bagchi:2012yk}. 
\begin{figure}[t]
\begin{center}
\includegraphics[scale=0.5]{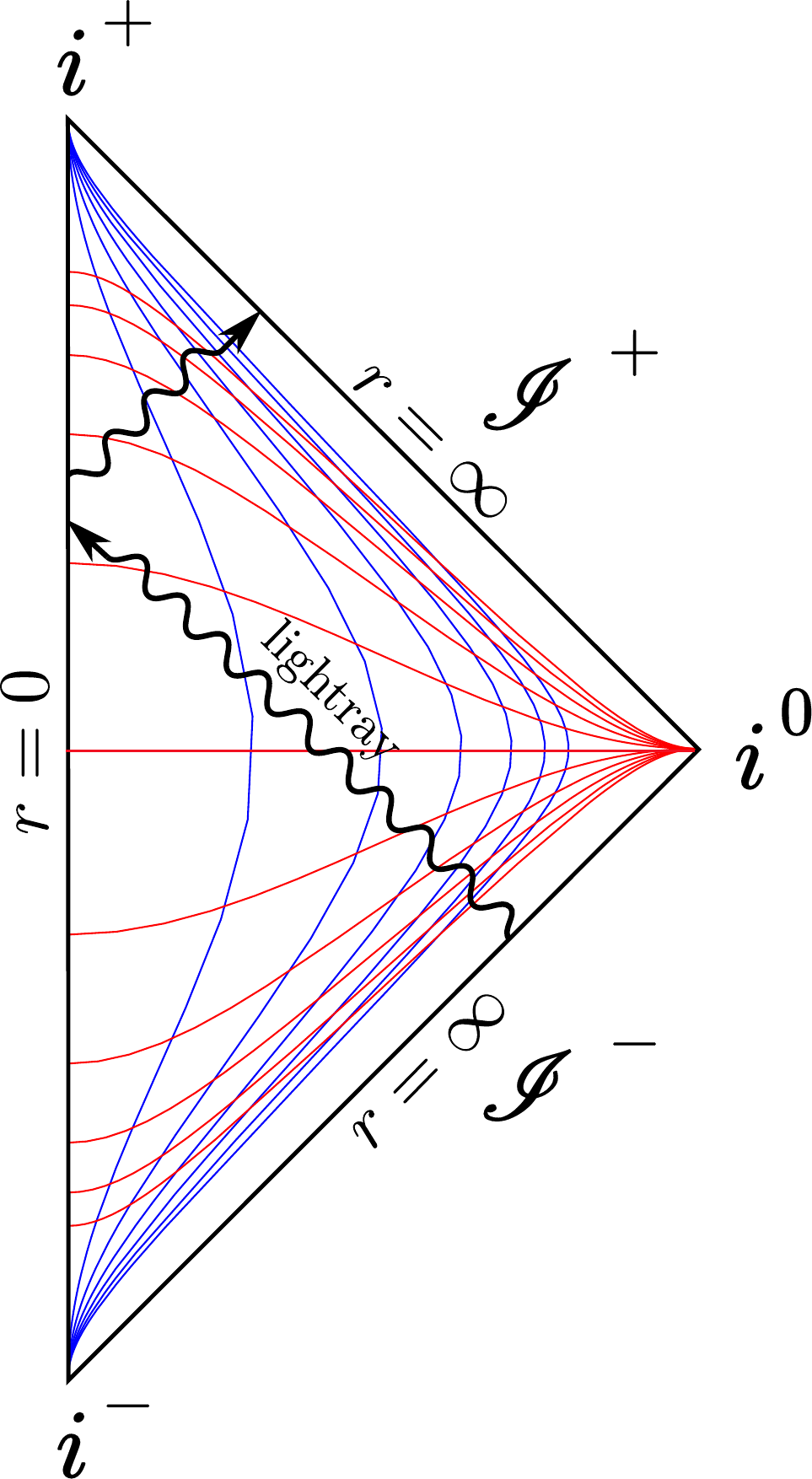}
\end{center}
\caption{Penrose diagram for three-dimensional Minkowski spacetime. }
\end{figure}

\medskip

\noindent It is natural, following lessons of AdS$_3$, to attempt constructions of holography in 3d asymptotically flat spacetimes, using the algebra \refb{bms3}. Specifically, the putative dual 2d field theory should be one which lives on the null boundary, say $\mathscr{I}^+$, and be invariant under the symmetry algebra \refb{bms3} \cite{Bagchi:2010zz, Bagchi:2012cy}. As we will go on to describe in the next section, the BMS$_3$ algebra can be obtained by an In{\"o}n{\"u}-Wigner contraction of two copies of the Virasoro algebra. This contraction which is the infinite radius limit of AdS, manifests itself on the boundary theory as an ultra-relativistic or a Carrollian limit on the parent 2d CFT, where the speed of light goes to zero. This rather strange limit results in the degeneration of the background (flat) metric that the field theory lives on and the Riemannian structures are replaced by Carrollian structures. These Carrollian manifolds are generically fibre-bundles. In the case of $\mathscr{I}^+$, the structure is of course a direct product ${\rm I\!R}_u \times S^1$, where $u$ is the null direction. This is a flat Carrollian manifold. We would be interested in 2d conformal field theories defined on these manifolds, the so called conformal Carrollian theories. Conformal Carrollian symmetries have been shown to be isomorphic to BMS symmetries \cite{Duval:2014uva}. 

\bigskip

\noindent In this paper, continuing with our earlier endeavours, we take forward our construction of the modular properties of BMS-invariant field theories (henceforth abbreviated as BMSFTs) \cite{Bagchi:2012xr, Bagchi:2013qva, Bagchi:2019unf}. As mentioned earlier, the BMS algebra can be arrived at from a limit of (two copies of) the Virasoro algebra. Hence it is conceivable that in this limit, some of the modular properties of 2d CFTs are also inherited by 2d BMSFTs. Following this line of argument, a BMS-Cardy formula \cite{Bagchi:2012xr} was proposed using these contracted modular properties{\footnote{See \cite{Barnich:2012xq} for an alternative derivation.}}. BTZ black holes are orbifolds of AdS$_3$. Similar orbifolds exist for 3d Minkowski spacetimes. The so-called shifted boost orbifold turns out to be the analogue of the non-extremal BTZ and is a cosmological solution with a cosmological horizon \cite{Cornalba:2003kd}. The thermodynamics of these cosmologies, also called Flat Space Cosmologies (FSCs) have been studied \cite{Bagchi:2012xr, Barnich:2012xq, Bagchi:2013lma, Detournay:2014fva} and in particular, the Bekenstein-Hawking entropy of these FSCs have been reproduced by the BMS-Cardy formula adding the significant feather to the cap of the programme of construction of 3d flat holography \cite{Bagchi:2012xr}. We will review modular properties and the BMS-Cardy formula in Sec.~2, and the FSC story briefly in Sec.~\refb{bulk}. 

Some other interesting advances in the study of holography for three dimensional asymptotic flat spacetimes include the construction of the flat limit of Liouville theory as an explicit putative boundary theory \cite{Barnich:2012rz}, matching of correlation functions between bulk and boundary for stress-energy tensors \cite{Bagchi:2015wna} and generic fields \cite{Hijano:2017eii}, aspects of entanglement entropy \cite{Bagchi:2014iea, Basu:2015evh, Hijano:2017eii, Jiang:2017ecm, Grumiller:2019xna, Fareghbal:2019czx, Godet:2019wje, Apolo:2020bld, Apolo:2020qjm}, understanding holographic reconstruction \cite{Hartong:2015usd}, and the flat version of the fluid-gravity correspondence \cite{Ciambelli:2018wre}. 

\subsection{Goal and summary of results of current work} 
The conformal bootstrap programme has been instrumental in constraining the landscape of known CFTs, and thereby relativistic QFT as renormalisation group flows away from CFTs, and ultimately aims to chart out all know and consistent relativistic theories. The main idea of the bootstrap is simple: any CFT is completely fixed by the dimension of primary operators and the structure constants characterising the three-point functions of primary operators. In 2d CFTs, an arbitrary correlation function or the partition function on an arbitrary Riemann surface can be completely fixed by this data. In two dimensions, imposing crossing symmetry of four-point functions on the sphere and modular covariance of one-point function on the torus implies the imposition of crossing and modular invariance to higher points and higher genus \cite{Moore:1988uz,Sonoda:1988fq}. Hence investigation of properties of the torus one-point function is an important programme, which was initiated recently by Kraus and Maloney in \cite{Kraus:2016nwo}. This has interesting holographic ramifications as well. Like the Cardy formula, the asymptotic behaviour of the structure constants or three-point coefficients of the theory is again fixed by the low lying spectrum. From the dual bulk, the field theory structure constants match with the corresponding expectation values in the dual BTZ geometry \cite{Kraus:2016nwo}. See \cite{Alkalaev:2016ptm} for a different slightly different bulk interpretation.  

\medskip

\noindent 
The BMS bootstrap programme has been initiated in \cite{Bagchi:2016geg} and elaborated in \cite{Bagchi:2017cpu}, together with other relevant developments \cite{Hijano:2017eii, Hijano:2018nhq, Lodato:2018gyp, Merbis:2019wgk}, in an effort to constrain possible 2d BMSFTs. We wish to take this programme of constraining BMSFTs further by focusing on modular invariance. Following \cite{Kraus:2016nwo}, the analysis of the torus one-point function has been generalised in many contexts, viz. to charged 2d CFTs \cite{Das:2017vej}, warped CFTs \cite{Song:2019txa}. In this paper, we work out the generalisation of the Kraus-Maloney analysis to the case of 2d BMS invariant field theories. In the process, we develop a lot of new machinery for BMSFTs, in particular derive expressions for the torus blocks of BMSFTs. Finally, we also show that the our expressions can be reproduced by looking at a one-point function in the putative dual FSC solution. 

\subsection{Outline of the paper and a quick look at the main results} 

Our paper is organised as follows. In Sec.~\refb{bmsft}, we review properties of 2d BMSFTs including in particular modular invariance of BMSFTs. In Sec~\refb{1pt}, we provide a definition of the BMS torus one-point function and derive its modular properties. 

\medskip

\noindent
In Sec~\refb{sconst}, we discuss the derivation of the asymptotic form of the BMS structure constants by exploiting the modular properties of the torus one-point function. Here we employ two methods, one a saddle-point analysis and the second based on inverting integral transforms that improves on the saddle point analysis. These methods don't assume much about the representations the states fall into in the sums used in the definition of the torus one-point function. Our principal result in this section is 
\bea\label{cipi}
C_{i p i} \approx D(\Delta_{\chi},\xi_{\chi}) C_{\chi p \chi}  \exp \left(-\frac{\Delta_i \xi _p }{2\xi_i }\right) \exp\left(-2\pi \frac{\D_i \xi_{\chi}}{\sqrt{2c_M \xi_i}}\right). \\
\non \eea
This above result is the outcome of the saddle point approximation and here presented for the case $c_L=0$. Here $C_{ipi}\equiv\<\D_i,\xi_i|\phi_p|\D_i,\xi_i\>$ is the three point-coefficient, where $\phi_p$ is a BMS primary with weights $\D_p$ and $\xi_p$.  $\phi_\chi$ is the lightest BMS primary in the spectrum above the vacuum and has weights $\D_\chi$ and $\xi_\chi$, and $D(\Delta_\chi,\xi_\chi)$ is its degeneracy. 

\smallskip

\noindent
There are further improvements on this result in Sec~\refb{sconst} to include a non-zero $c_L$, to increase the range of validity of the result and to include generalisations when the vacuum has a degeneracy as well as where the vacuum is not the lightest primary. We will give details of this in the main body of the paper. 

\medskip

\noindent 
In the latter half of the paper, we refine our analysis to look at the torus one-point function of BMS primaries in the highest weight representation. This entails the development of a lot of new results in the context of BMS highest weight representations in Sec.~\refb{hw}. Our primary achievement in this part of the paper outlined in Sec~\refb{tblock} which details the construction of the BMS torus blocks. These are defined by 
\be
\mathcal{F}_{\D_A,\xi_A,c_L,c_M}^{\D_p,\xi_p}(\sigma,\rho) = \frac{q^{-\D_A+\frac{c_L}{2}}y^{-\xi_A+\frac{c_M}{2}}}{C_{ApA}} {\rm Tr}_{\D_A,\xi_A}\left(\phi_p(0,1)q^{L_0-\frac{c_L}{2}}y^{M_0-\frac{c_M}{2}}\right).
\ee
Here the trace is over $\{A\}$, the collection of all the the primary states with dimension $(\D_A,\xi_A)$. Also $q=e^{2\pi i \s}, \ y=e^{2\pi i \rho}$ are the exponentiation of the BMS modular parameters $\s, \rho$. We present expressions of the BMS torus blocks in a certain limit of large BMS weights (large $\xi_A$) for these states. In this approximation, as detailed in the main body of the paper, the result for the BMS torus block reads: 
\bea
&& \mathcal{F}_{\D_A,\xi_A,c_L,c_M}^{\D_p,\xi_p}(\sigma,\rho)\equiv \sum_N q^N \mathcal{F}_{N}(\Delta_p,\xi_p;\Delta_A,\xi_A|c_L,c_M|\rho), \quad \text{where}\\
&& \mathcal{F}_{N}=\left(1+\frac{\xi_p(\Delta_p-1)}{2\xi_A}N\right)\widetilde{\text{dim}}_N +\pi i\rho\frac{\xi_p^2}{\xi_A} \sum_{k=0}^Np(N-k)p(k)(N-k)(N-2k) + \mathcal{O}(\xi_A^{-2}). \non
\eea
In the above, $p(N)$ is the partition of the integer $N$ and $\widetilde{\text{dim}}_N$ is the partition of the same integer $N$ using two colors. As an aside, using a differential equation arising from the quadratic Casimirs of the global BMS algebra (or the Poincare algebra), we also derive a closed form expression for the BMS global torus block. 
\medskip

\noindent 
Using all of these tools, in Sec~\refb{onepointsec}, we arrive at the asymptotic structure constants for the BMS highest weight primaries. The form of this primary structure constant is almost identical to the one obtained in the analysis for general states in the earlier section with the identification 
\be
c_L \to c_L - \frac{1}{6}
\ee
This indicates a quantum shift in the central term $c_L$ and no such correction in $c_M$, in keeping with recent literature \cite{Merbis:2019wgk}. We have some more remarks at the end of Sec~\refb{onepointsec}. 

\medskip

\noindent 
Sec.~\refb{bulk} contains a bulk computation that matches our field theory results, specifically \refb{cipi}. Here we first review the Flat Space Cosmology (FSC) solutions and then use a geodesic approximation to compute a probe one-point function in this FSC background. We end in Sec.~\refb{conc} with some discussions and future directions. Complementing the analysis in the main text, there are twelve (no, really twelve!) appendices which outline various detailed computations omitted in the main body of the paper. 

\bigskip \bigskip

\newpage

\section{BMS field theories and modular invariance}\label{bmsft}
In this section, we review aspects of BMS invariant field theories in 2d, or in short 2d BMSFTs. We start with algebraic preliminaries and then look at aspects of representation theory. We then discuss modular invariance in 2d BMSFTs. Since this is basically all review material, readers familiar with earlier work on BMSFT can skip over to the next section. 

\subsection{Algebra and contraction} 
As we have stated in the introduction, we will be interested in 2d quantum field theories invariant under the BMS$_3$ algebra, rewritten here for convenience:
\bes{}\label{bms}
\bea{}
&& [L_n, L_m] = (n-m) L_{n+m} + c_L (n^3-n) \delta_{n+m, 0} \\
&& [L_n, M_m] = (n-m) M_{n+m} + c_M (n^3-n) \delta_{n+m, 0} \\
&& [M_n, M_m] = 0.
\eea
\ees
This algebra \refb{bms} can be obtained from two copies of the Virasoro algebra 
\bes{}\label{vir}
\bea{}
&& [\L_n, \L_m] = (n-m) \L_{n+m} + \frac{c}{12}(n^3-n) \delta_{n+m, 0} \\
&& [\bcL_n, \bcL_m] = (n-m) \bcL_{n+m} + \frac{\bar{c}}{12}(n^3-n) \delta_{n+m, 0} \\
&& [\L_n, \bcL_m] = 0.
\eea
\ees
by an In{\"o}n{\"u}-Wigner contraction defined by 
\be \label{v2b}
L_n = \L_n - \bcL_{-n}, \quad M_n = \e\left( \L_n + \bcL_{-n} \right), \quad \e \to 0. 
\ee
We shall refer to this limit as the ultra-relativistic (UR) or Carrollian limit. In terms of coordinates, this has the interpretation of sending the speed of light to infinity in the dual field theory. We shall see this shortly. From the point of the bulk, the limit \refb{v2b} can be identified as taking the radius of AdS ($\ell$) to infinity. This can be easily seen by considering the identification $\e = 1/\ell$. Remembering that the Brown-Henneaux central terms of Einstein gravity in AdS$_3$ is $c= \bar{c}= 3 \ell/2G$, it is interesting to note here that the limit \refb{v2b} along with the identification of the contraction parameter to the inverse of the AdS radius, leads to 
\be
c_L=\frac{1}{12}(c - \bar{c})=0, \quad c_M= \lim_{\ell \to \infty} \frac{1}{12 \ell}(c + \bar{c}) = \frac{1}{4G}
\ee
for Einstein gravity, as mentioned in the introduction. This can also be obtained from an independent asymptotic symmetry analysis in 3d asymptotically flat spacetimes \cite{Barnich:2006av}. 

In order to see the ultrarelativistic nature of the limit \refb{v2b}, one can look at the Virasoro generators on the cylinder (which also corresponds to the conformal boundary of global AdS$_3$ spacetime). 
\be
\L_n = e^{in \w} \p_\w, \quad \bcL_n = e^{in \bw} \p_{\bw}, \quad \mbox{where} \quad \w, \bw = u \pm \phi \non
\ee
The linear combinations required for taking the limit \refb{v2b} means that on the coordinates, the contraction manifests itself as \cite{Bagchi:2012cy} 
\be{}
\phi \to \phi, u \to \e u \quad \e \to 0. 
\ee
In terms of velocities, this means $v/c \sim \s/\t \to \infty$, where $c$ is the speed of light, which is the same as a Carrollian or UR limit of $c\to 0$. In terms of the coordinates on the null cylinder $(u, \phi)$, the generators of the BMS algebra take the following form:
\be\label{genc}
L_n = ie^{in \phi} (\p_\phi + in u \p_u), \quad M_n = ie^{in \phi} \p_u.
\ee
It is of importance to note here that the coordinate $\phi$ representing the circle at infinity is compact, whereas the null direction $u$ which serves as the time direction in the dual field theory is non-compact. We will be interested in another representation of the BMS algebra where both directions are non-compact. This we will call the ``plane" representation. The generator on the plane take the form
\be\label{genp}
L_n =- x^{n+1}\p_x - (n+1) x^n t \p_t, \quad M_n = x^{n+1} \p_t.
\ee
It can be checked that these again close to form the (centreless) BMS algebra \refb{bms}. We can map the cylinder to the plane by the coordinate transformation
\be\label{c2p}
t= -iu e^{i\phi}, \quad x = e^{i \phi}
\ee
Again it is straight-forward to check that the generators \refb{genc} and \refb{genp} go into each other under this transformation \refb{c2p}. 

\subsection{A brief look at representation theory}
In 2d CFTs, states are labelled by the eigenvalues of the $\L_0$ and $\bcL_0$ operators: 
\be
\L_0|h, \h\> = h |h, \h\>, \quad \bcL_0|h, \h\> = \h |h, \h\>.
\ee
Demanding the spectrum to be bounded from below, and noting that $\L_n, \bcL_n$ for $n>0$ lowers the $(h, \h)$ values, we are led to the definition of a primary state 
\be
\L_n|h, \h\>_p = \bcL_n|h, \h\>_p = 0, \quad \forall n>0.
\ee
A Virasoro highest weight module is built out of these primary states by acting with raising operators. An arbitrary descendant state is constructed by hitting a primary state with a bunch of raising operators $\L_{-n}, \bcL_{-n}$ for $n>0$. The primary states form the basis of these highest weight representations and as we mentioned in the introduction, a particular CFT is completely fixed by the spectrum of its primary states and the coefficients associated with the three point functions of these primary states. 

Looking at the structure of the BMS algebra \refb{bms}, we could attempt a similar construction of highest weight representations \cite{Bagchi:2009pe}. Now the states would be labelled under $L_0$ and $M_0$. 
\be
L_0|\D, \xi\>= \D |\D, \xi\>, \quad M_0|\D, \xi\> = \xi |\D, \xi\>. 
\ee
Demanding that the spectrum of $\D$ be bounded from below, we can define a BMS primary $|\D, \xi\>_p$ as 
\be
L_n |\D, \xi\>_p = M_n |\D, \xi\>_p = 0 \quad \forall n>0.
\ee
In a manner analogous to the 2d CFT case, we can build BMS highest weight modules from these primary states by acting with raising operators $L_{-n}, M_{-n}$ for $n>0$. 

\medskip

\noindent It is important to notice that the Virasoro highest weight representation does not contract to the BMS highest weight representation in the UR limit \refb{v2b}. To see this directly, we will look at the primary state conditions. We first notice that there is no problem with the labelling of the states in the limit. 
\bes
\bea
&& (\L_0 - \bcL_0) |h, \h\> = (h - \h) |h, \h\> \Rightarrow L_0  |\D, \xi\>= \D |\D, \xi\>, \, \mbox{where} \ \D= h - \h,  \\
&& (\L_0 + \bcL_0) |h, \h\> = (h + \h) |h, \h\> \Rightarrow M_0  |\D, \xi\>= \xi |\D, \xi\>, \, \mbox{where} \ \xi= \e(h + \h).
\eea
\ees
In the above, the state $|h, \h\>$ goes over to the state $|\D, \xi\>$ in the limit. Now let us look at the Virasoro primary state conditions:
\bes
\bea
\L_n |h, \h\> = (L_n + \frac{1}{\e} M_n) |h, \h\> &=& 0, \, \, \forall \, n>0 \Rightarrow M_n  |\D, \xi\>= 0, \, \, \forall \, n>0 \\
\bcL_n |h, \h\> = (- L_{-n} + \frac{1}{\e} M_{-n}) |h, \h\> &=& 0, \, \, \forall \, n>0 \Rightarrow M_{-n}  |\D, \xi\>= 0, \, \, \forall \, n>0.
\eea
\ees
So, Virasoro highest weight representations under the UR contraction become BMS representations with the following characteristics
\be
L_0  |\D, \xi\>= \D |\D, \xi\>, \, M_0  |\D, \xi\>= \xi |\D, \xi\>, \,  M_n  |\D, \xi\>= 0 \ \forall n \neq 0. 
\ee
These representations are the so-called induced representations, studied in detail in \cite{Barnich:2014kra, Barnich:2015uva, Campoleoni:2016vsh}. It should be pretty obvious that these representations are very different from the above described BMS highest weight representations. 

\medskip

\noindent The BMS highest weights are reproduced from another limit of the Virasoro which of course again yield the same algebra. This is a non-relativistic (NR) contraction, as opposed to the UR one described earlier. This different contraction is given by \cite{Bagchi:2009pe}
\be
L_n = \L_n + \bcL_n, \quad M_n = -\e (\L_n - \bcL_n)
\ee
The isomorphism between the algebras obtained by these two contractions is a surprising fact in two dimensions \cite{Bagchi:2010zz}. Following this limit on the cylinder Virasoro generators, we can see that the speed of light in this contraction goes to infinity thus justifying the non-relativistic nomenclature. Showing that the highest weights go to highest weights is also a simple exercise. 

\subsection{BMS modular invariance}\label{bmsmod}
In a 2d CFT, the partition function of the theory is defined as
\be
Z_{\mbox{\tiny{CFT}}}(\t, \bar{\t}) = \mbox{Tr} \left(q^{\L_0 - c/24} \ \bar{q}^{\bcL_0 - \bar{c}/24}\right).
\ee
Here $\t, \bar{\t}$ are the modular parameters and $q= e^{2\pi i \t}, \, \bar{q} = e^{-2\pi i \bar{\t}}$. In a similar vein, we will define the partition function of a 2d BMSFT as  
\be
Z_{\mbox{\tiny{BMS}}}(\s, \rho) = \mbox{Tr} \left(e^{2\pi i \s(L_0 - c_L/24)} \ e^{2\pi i \rho(M_0 - c_M/24)}\right).
\ee
Here the BMS modular parameters are $\s, \rho$. We demand that the BMS partition function arises as a smooth limit from the CFT partition function. This necessitates the following identification: 
\be\label{v2bmp}
\t = \s + \e \rho, \, \bar{\tau} = \s - \e \rho.
\ee
The modular transformation on a 2d CFT reads 
\be
\t \to \frac{a \t + b}{c \t + d}, \quad \bar{\tau} \to \frac{a \bar{\tau} + b}{c \bar{\tau} + d}
\ee
Following the identification \refb{v2bmp}, we find the BMS version of modular transformations \cite{Bagchi:2013qva}:
\be
\s \to \frac{a \s + b}{c \s + d}, \quad \rho \to \frac{\rho}{(c \s + d)^2}.
\ee
Demanding that the CFT partition function be invariant under modular transformations, particularly the S-transformation 
$$ \t \to - \frac{1}{\t}, $$
one relates the very high energy spectrum of the theory to the low energy sector and in particular derives an expression for the entropy of states of the theory, given by Cardy's famous formula
\be\label{car}
S_{\mbox{\tiny{Cardy}}} = 2 \pi \left( \sqrt{\frac{c h}{6}} + \sqrt{\frac{\bar{c} \h}{6}} \right),
\ee
where $h, \h$ are conformal weights of the states and $c, \bar{c}$ are the central charges of the 2d CFT. Using similar arguments and the BMS S-transformation
$$ \s \to - \frac{1}{\s}, \, \rho \to \frac{\rho}{\s^2}, $$ 
we arrive at the BMS-Cardy formula \cite{Bagchi:2013qva}
\be\label{bcar}
S_{\mbox{\tiny{BMS-Cardy}}} = 2 \pi \left( c_L \sqrt{\frac{2 \xi}{c_M}} + \D \sqrt{\frac{2 c_M}{\xi}} \right).
\ee
One can also obtain the above formula \refb{bcar} directly from \refb{car} by looking carefully at the limit \cite{Riegler:2014bia,Fareghbal:2014qga}\footnote{This crucially involves a minus sign between the two factors in parenthesis in \refb{car}, a fact which can be attributed to the inner horizon of the original BTZ solution from where the bulk dual FSC is obtained.} or by an exchange of spatial and temporal circles on the BMS torus following methods outlined in \cite{Detournay:2012pc} (see e.g. \cite{Jiang:2017ecm}). The validity of \refb{bcar} can be verified by the fact that this formula correctly reproduces the Bekenstein-Hawking entropy of the putative bulk dual Flatspace Cosmological solutions. We will have more to say about this when we do a bulk calculation in Sec~\refb{bulk}. 

\bigskip 

\section{BMS one-point function on the torus} \label{1pt}
Having set the stage, we now move on to our object of interest in the current paper, the torus one-point function of a 2d BMSFT. We expect that in keeping with properties of 2d CFTs, the modular covariance of this quantity, together with crossing symmetry on the sphere, would crucial to ensuring higher point crossing symmetry and higher genus modular invariance in a 2d BMSFT. 

\subsection{Torus 1-pt function: Definition} 
For a local primary field $\phi_{\rm{cyl}}(u, \phi)$ (where cyl in subscript denotes `cylinder') on the BMS torus with modular parameters $\sigma,\rho$, we define a {\em{BMS torus one-point function}} as:
\be
 \langle \phi_{\rm cyl}(u, \phi)\rangle_{(\sigma, \rho)} = {\rm Tr}\left[ \phi_{\rm{cyl}}(u, \phi)  e^{-\beta H} e^{i \theta P }\right],
\ee
where $H$ is the Hamiltonian which generates translation along the longitude of the torus and $P$ generates translation along the meridian (circumference) of the torus. We think of the torus as a cylinder with length $\beta$ in which the two ends of the cylinder are glued together. Before gluing, we twist one of the ends by an angle $\theta$. 

\medskip

\noindent
We will start out by showing that $\langle \phi_{\rm cyl}(u, \phi)\rangle_{(\sigma, \rho)} $ is translation invariant. To see this, let us take the basis states to be orthonormal states $|E_i,P_i\rangle$ which are simultaneous eigenstates of $H$ and $P$
\be
H|E_i,P_i\rangle =  E_i |E_i,P_i\rangle ,\,\,\,\, P|E_i,P_i\rangle =  P_i |E_i,P_i\rangle.
\ee
Then 
\be
 \langle \phi_{\rm cyl}(u, \phi)\rangle_{(\sigma, \rho)} = \sum_i \langle E_i,P_i| \phi_{\rm{cyl}}(u, \phi)  e^{-\beta H} e^{i \theta P }|E_i,P_i\rangle.
\ee
Using the fact that $P$ and $H$ are operators for spatial and temporal translations we have
\bea
 \langle \phi_{\rm cyl}(u+u^{\prime}, \phi+\phi^{\prime})\rangle_{(\sigma, \rho)} &=& \sum_i \langle E_i,P_i| e^{iu^{\prime} H} e^{i \phi^{\prime} P }\phi_{\rm{cyl}}(u, \phi) e^{-iu^{\prime} H} e^{-i \phi^{\prime} P } e^{-\beta H} e^{i \theta P }|E_i,P_i\rangle \cr
 &=& \sum_i e^{iu^{\prime} E_i} e^{i \phi^{\prime} P_i } e^{-iu^{\prime} E_i} e^{-i \phi^{\prime} P_i } \langle E_i,P_i| \phi_{\rm{cyl}}(u, \phi)  e^{-\beta H} e^{i \theta P }|E_i,P_i\rangle \cr
 &=& \langle \phi_{\rm cyl}(u, \phi)\rangle_{(\sigma, \rho)}. \non
\eea
Thus, $\langle \phi_{\rm cyl}(u, \phi)\rangle_{(\sigma, \rho)}$ is invariant under translation. Using this, we may write   
\be
\langle \phi_{\rm cyl}(u, \phi)\rangle_{(\sigma, \rho)} = \langle \phi_{\rm cyl}(0,0)\rangle_{(\sigma, \rho)}. 
\ee
As we have seen earlier, we can map the cylinder to a plane with coordinates $(t,x) $ by
\be\label{ptc}
t=-iue^{i\phi},\quad\quad x=e^{i\phi}.
\ee
Since we have built the torus by identifying the ends of a cylinder, this map stays the same.

The operator $H$ and $P$ can be written in terms of operators $L_0$ and $M_0$ on the plane. Then, the one point functions can be written as  
\begin{align}
\langle \phi_{\rm cyl}(u, \phi)\rangle_{(\sigma, \rho)}={\rm{Tr}}\left[\phi_{\rm{cyl}}(u, \phi)e^{2\pi i\sigma(L_0-\frac{c_L}{2})}e^{2\pi i \rho(M_0-\frac{c_M}{2})}\right].
\end{align}
The primary field in the cylinder $\phi_{\rm cyl}(u,\phi)$ are related to the primary field on the plane $\phi(x,t)$ as
\be\label{ptceq}
 \phi_{\rm cyl}(u,\phi) = e^{i\phi\D}e^{-i u\xi}\phi(t,x).
\ee
The detailed derivation along with the BMS$_3$ transformation for primary fields are done in appendix \ref{Finite ${BMS}_3$ transformations}.
Thus the one-point function is given by
\be
 \boxed{\langle \phi_{\rm cyl}(u, \phi)\rangle_{(\sigma, \rho)} = {\rm{Tr}}\left[\phi(0,1)\ e^{2\pi i\sigma(L_0-\frac{c_L}{2})}e^{2\pi i \rho(M_0-\frac{c_M}{2})}\right].}
 \label{one_point_function}
\ee
This will be our working definition of the BMS torus one-point function. 

\subsection{Torus 1-pt function: Modular transformation property}
We will now derive the modular transformation properties of a BMS torus one-point function. We will take inspiration from 2d CFTs, like reviewed in Sec.~{\refb{bmsmod}}. Modular transformation in relativistic CFT is given by
\be 
\tau \rightarrow \gamma. \tau \equiv \frac{a\tau+b}{c\tau+d},\quad \bar{\tau} \rightarrow \gamma. \bar{\tau} \equiv \frac{a\bar{\tau}+b}{c\bar{\tau}+d},  
\ee
where $a, b, c, d$ are integers satisfying $ad-bc=1$. Under this transformation, the elliptic co-ordinates $(w,\bar{w})$ of the torus transforms as
\be
w\rightarrow \gamma. w\equiv \frac{w}{c\tau+d},\quad \bar{w}\rightarrow \gamma. \bar{w}\equiv \frac{\bar{w}}{c\bar{\tau}+d}. 
\label{elliptic}
\ee
Thus primary fields transforms as 
\bea
\phi_{h,\bar{h}}(\gamma.w,\bar{\gamma.w})|_{\gamma.\tau} &=&  \left(\frac{\partial (\gamma.w)}{\partial w}\right)^{-h}\left(\frac{\partial (\gamma.\bar{w})}{\partial\bar{w}}\right)^{-\bar{h}}\phi_{h,\bar{h}}(w,\bar{w})
\eea
Using this and the modular invariance of torus partition function, we can deduce the following transformation property for one point torus function
\bea
\<\phi_{h,\bar{h}}(\gamma.w,\gamma.\bar{w})\>_{(\gamma.\tau,\gamma.\bar{\tau})} = (c\tau+d)^{-h}(c\bar{\tau}+d)^{-\bar{h}}\<\phi_{h,\bar{h}}(w,\bar{w})\>_{(\tau,\bar{\tau})}
\label{cft_1pt_tran}
\eea

\medskip

\noindent To find the transformation property of BMS torus one-point function under BMS modular transformation, we adopt a similar prescription. As stated before, the BMS version of modular transformation reads
\be
\sigma \rightarrow \gamma.\sigma = \frac{a\sigma +b }{c\sigma + d},\quad\quad \rho \rightarrow \gamma.\rho=\frac{\rho}{(c\sigma+d)^2}. 
\ee
For the coordinate $u$ and $\phi$, we can find the analog of \eqref{elliptic} for by taking its UR limit. 
For this, we have to use the contraction
\be
(w,\bar{w})\rightarrow \epsilon u \pm \phi, \quad\quad(\tau,\bar{\tau})\rightarrow \sigma \pm  \epsilon\rho.
\ee
We know that 
\be
u =\frac{w+\bar{w}}{2\epsilon},\quad\quad \phi = \frac{w-\bar{w}}{2}. 
\ee
Thus 
\be
\gamma.u =  \frac{\gamma.w+\gamma.\bar{w}}{2\epsilon},\quad\quad \tilde{\phi}\equiv\gamma.\phi = \frac{\gamma.w - \gamma.\bar{w}}{2}.
\ee
We expand $ \gamma.w$ and $\gamma.\bar{w}$ in powers of $\e$ with the above definitions to obtain: 
\bea
&& \gamma.w = \frac{w}{c\tau + d} = \frac{\epsilon u + \phi}{c(\sigma + \epsilon \rho) + d} 
 =\epsilon \left(\frac{u}{c\sigma + d}-\frac{\phi\rho}{(c\sigma +d)^2}\right) + \frac{\phi}{c\sigma + d} +\mathcal{O}(\epsilon^2). \\
&& \gamma.\bar{w} = \epsilon \left(\frac{u}{c\sigma + d}-\frac{\phi\rho}{(c\sigma +d)^2}\right) - \frac{\phi}{c\sigma + d} +\mathcal{O}(\epsilon^2).
\eea
Thus we have
\be
\gamma.u =  \frac{u}{c\sigma + d}-\frac{\phi\rho}{(c\sigma +d)^2},\quad \quad \gamma.\phi= \frac{\phi}{c\sigma + d}.
\label{bms_ellip_trans}
\ee
Note that this is a finite BMS transformation of the form: $u \rightarrow u \partial_{\phi} f(\phi) + g(\phi),\,\phi \rightarrow f(\phi)$.

\bigskip \bigskip

\noindent The BMS primary field with weight $\Delta$ and $\xi$ transform under
$u \rightarrow \gamma.u,\,\phi \rightarrow \gamma.\phi$ 
as
\be
\phi_{\D,\xi}(\gamma.u,\gamma.\phi)|_{\gamma.\sigma,\gamma.\rho}= (\partial_{\phi}(\gamma.\phi))^{-\D}e^{\xi \frac{\partial_{\phi}(\gamma.u)}{\partial_{\phi}(\gamma.\phi)}}\phi_{\D,\xi}(u,\phi)_{(\sigma,\rho)}.
\ee
Thus the one point function for BMS primary field would transform as
\bea 
\<\phi_{\D,\xi}(\gamma.u,\gamma.\phi)\>_{(\gamma.\sigma,\gamma.\rho)} &=& (\partial_{\phi}(\gamma.\phi))^{-\D}e^{\xi \frac{\partial_{\phi}(\gamma.u)}{\partial_{\phi}(\gamma.\phi)}}\<\phi_{\D,\xi}(u,\phi)\>_{(\sigma,\rho)}.
\label{bms_1pt_tran}
\eea
Substituting \eqref{bms_ellip_trans} in the above equation we get
\bea 
\<\phi_{\D,\xi}(\gamma.u,\gamma.\phi)\>_{(\gamma.\sigma,\gamma.\rho)}
&=& (c\sigma + d)^{\D}e^{- \frac{\xi c\rho}{(c\sigma+d)}} \<\phi_{\D,\xi}(u,\phi)\>_{(\sigma,\rho)}.
\eea
In particular for modular S-transformation $(a=0,b=-1,c=1,d=0)$, we have
\be
\left<\phi_{\D,\xi}\right>_{(-1/\sigma, \ \rho/\sigma^2)} = \sigma^{\D} e^{-\frac{\xi\rho}{\sigma}} \<\phi_{\D,\xi}\>_{(\sigma,\rho)}.
\label{s_mod_trans}
\ee
We will use the above to derive an asymptotic form of the BMS structure constants in the next section{\footnote{We note that analogues of the above formulae could have of course been derived completely in the NR limit, with a flip in the time and space directions.}}.

\bigskip

\newpage

\section{Asymptotic form of BMS structure constants} \label{sconst}
In this section, we present two different ways of obtaining the asymptotic form of the BMS structure constants. The first is a saddle-point method, which generalises the BMS-Cardy analysis of e.g. \cite{Bagchi:2012xr} from zero-point functions, i.e. partition functions, to one-point functions on the torus. The second relies on inverting a combination of a Laplace and Fourier transformations and is an improvement on the saddle point analysis. 

\medskip

\noindent It is important to state here that the analyses in both methods are independent of basis. We will just assume that the states of the BMSFT are eigenstates of $L_0$ and $M_0$, and the forms of the BMS modular transformations discussed in the previous section. So, our analysis should be independent of which of the two representations discussed earlier, the highest weight or the induced, the states of the BMSFT fall into. This is a statement we don't prove in our paper, as we will not work with induced representations. We will later see that an explicit computation with BMS primaries and highest weight representations, based on the construction of BMS torus blocks yields identical results when the appropriate limits are considered. 

\medskip

\noindent
The statement about the equivalence of the two representations may seem like an over-reach at this point, but we point the reader to our earlier work on BMS characters in \cite{Bagchi:2019unf}, which remarkably showed that the expressions for characters in these two very different looking representations are actually identical. We expect a similar story to emerge here for the torus one-point function as well. In later sections, we will use the BMS torus blocks to rewrite our torus 1-point function. The leading piece of the expression of the BMS torus blocks in the highest weight representation is again the character. It is not unreasonable to expect that a similar (but admittedly more difficult) calculation of the induced BMS torus block in the same limit may also yield the character as the leading piece. These are one and the same as shown in \cite{Bagchi:2019unf}. We will not have anything to say explicitly about calculations in the induced representation in this work. But the enigmatic equivalence between the two very different representations is something we wish to return to in the near future. 

\medskip

\noindent
Now, we concentrate on finding the asymptotic form of the BMS structure constants from the modular properties of the one-point function that we derived in the previous section. We first use a saddle-point analysis. 

\subsection{The saddle-point method} 
\label{sec:simultaneous_L_M}
Let us emphasise again that the trace on the left hand side of \eqref{one_point_function} does not depend on the representation which the states of the 2d BMSFT fall into. If we are using a bases $|\D_i,\xi_i\>$ which are both eigenstates of $L_0$ and $M_0$
\be
L_0|\D_i,\xi_i\>=\D_i|\D_i,\xi_i\>,\quad\quad  M_0|\D_i,\xi_i\>=\xi_i|\D_i,\xi_i\>,
\ee
which is true for both the highest weight and the induced representations, we see that the one-point function for a primary field $\phi_p$ is given by{\footnote{The notion of a primary state would of course depend on the representation and definition we are working with stems from the highest weight representation. So this may be a point of dispute. We can choose to define the 1-point function of a torus by \refb{t1pt}, which will make this a representation independent statement.}}
\be\label{t1pt}
\langle \phi_p\rangle_{(\sigma, \rho)} = \sum_{i}\<\D_i,\xi_i|\phi_p(0,1)|\D_i,\xi_i\> D(\D_i,\xi_i)e^{2\pi i\sigma(\D_i-\frac{c_L}{2})}e^{2\pi i \rho(\xi_i-\frac{c_M}{2})},
\ee
where $D(\D_i,\xi_i)$ is the density of states for $|\D_i,\xi_i\>$. We called $\<\D_i,\xi_i|\phi_p(0,1)|\D_i,\xi_i\>$ a one-point coefficient and we will use $C_{ipi}$ to denote it. Thus
\be
\langle \phi_p\rangle_{(\sigma, \rho)} = e^{-2\pi i \left(\sigma\frac{c_L}{2}+\rho\frac{c_M}{2}\right)}\sum_{i} D(\Delta_i,\xi_i)C_{ipi}e^{2\pi i(\sigma\Delta_i+\rho \xi_i)}.
\label{one_point_function_rep1}
\ee 
In this section we will deduce the form of three-point coefficient 
\be
C_{ipi}\equiv\<\D_i,\xi_i|\phi_p(0,1)|\D_i,\xi_i\>,
\ee 
for large $\D_i$ and $\xi_i$.  Sometime we will drop the index $i$ and use the notation $C_{\D p \D }\equiv\<\D_,\xi|\phi_p(0,1)|\D,\xi\>$. 
We define
\be\label{tilp}
 \widetilde{\langle \phi_p\rangle}_{(\sigma, \rho)}=e^{2\pi i \left(\sigma\frac{c_L}{2}+\rho\frac{c_M}{2}\right)}\,\langle \phi_p\rangle_{(\sigma, \rho)}.
\ee
Thus from \eqref{one_point_function_rep1} and \refb{tilp} we have
\be
\widetilde{\langle \phi_p\rangle}_{(\sigma, \rho)}= \sum_{i} D(\Delta_i,\xi_i)C_{ipi}e^{2\pi i(\sigma\Delta_i+\rho \xi_i)},
\label{phitilde}
\ee
The S-modular transformation property of $\widetilde{\langle \phi_p\rangle}_{(\sigma, \rho)}$ can be worked out from that of $\langle \phi_p\rangle_{(\sigma, \rho)}$ given in \eqref{s_mod_trans}
\bea
\widetilde{\langle \phi_p\rangle}_{{(-\frac{1}{\sigma},\frac{\rho}{\sigma^2})}}&=& e^{2\pi i \left(-\frac{1}{\sigma}\frac{c_L}{2}+\frac{\rho}{\sigma^2}\frac{c_M}{2}\right)}\left<\phi_p\right>_{(-\frac{1}{\sigma},\frac{\rho}{\sigma^2})} \cr
&=& e^{2\pi i \left(-\frac{1}{\sigma}\frac{c_L}{2}+\frac{\rho}{\sigma^2}\frac{c_M}{2}\right)}\sigma^{\D_p} e^{-\xi_p\frac{\rho}{\sigma}} \<\phi_p\>_{(\sigma,\rho)}\cr
&=& e^{2\pi i \left(-\frac{1}{\sigma}\frac{c_L}{2}+\frac{\rho}{\sigma^2}\frac{c_M}{2}\right)}\sigma^{\D_p} e^{-\xi_p\frac{\rho}{\sigma}} e^{-2\pi i \left(\sigma\frac{c_L}{2}+\rho\frac{c_M}{2}\right)}\,\widetilde{\langle \phi_p\rangle}_{(\sigma, \rho)}\cr
&=& e^{-2\pi i \left(\frac{1}{\sigma}\frac{c_L}{2}-\frac{\rho}{\sigma^2}\frac{c_M}{2}+\sigma\frac{c_L}{2}+\rho\frac{c_M}{2}\right)}e^{-\xi_p\frac{\rho}{\sigma}}\sigma^{\D_p}\widetilde{\langle \phi_p\rangle}_{(\sigma, \rho)}.
\eea
Thus the S-modular transformation of $\widetilde{\langle \phi_p\rangle}_{(\sigma, \rho)}$ is given by
\be
\widetilde{\langle \phi_p\rangle}_{(\sigma, \rho)} =  e^{2\pi i \left(\frac{1}{\sigma}\frac{c_L}{2}-\frac{\rho}{\sigma^2}\frac{c_M}{2}+\sigma\frac{c_L}{2}+\rho\frac{c_M}{2}+\frac{\xi_p}{2\pi i}\frac{\rho}{\sigma}-\frac{1}{2\pi i}\D_p\log \sigma\right)}\widetilde{\langle \phi_p\rangle}_{{(-\frac{1}{\sigma},\frac{\rho}{\sigma^2})}}.
\ee
\newpage \noindent 
We invert \eqref{phitilde} to obtain
\bea
 D(\Delta,\xi)C_{\Delta p \Delta} &=& \int d\sigma d\rho \,e^{-2\pi i(\sigma\Delta+\rho \xi)} \widetilde{\langle \phi_p\rangle}_{(\sigma, \rho)}\cr
  &=& \int d\sigma d\rho \, e^{-2\pi i(\sigma\Delta+\rho \xi)} \, e^{2\pi i \left(\frac{1}{\sigma}\frac{c_L}{2}-\frac{\rho}{\sigma^2}\frac{c_M}{2}+\sigma\frac{c_L}{2}+\rho\frac{c_M}{2}+\frac{\xi_p}{2\pi i}\frac{\rho}{\sigma}-\frac{1}{2\pi i}\D_p\log \sigma\right)}\widetilde{\langle \phi_p\rangle}_{{(-\frac{1}{\sigma},\frac{\rho}{\sigma^2})}}\cr
  &=& \int d\sigma d\rho \,\widetilde{\langle \phi_p\rangle}_{{(-\frac{1}{\sigma},\frac{\rho}{\sigma^2})}} \,\,e^{f(\sigma,\rho)},
  \label{M_integral}
\eea
where 
\be
f(\sigma,\rho)=2\pi i \left(\frac{1}{\sigma}\frac{c_L}{2}-\frac{\rho}{\sigma^2}\frac{c_M}{2}+\sigma\frac{c_L}{2}+\rho\frac{c_M}{2}+\frac{\xi_p}{2\pi i}\frac{\rho}{\sigma}-\frac{1}{2\pi i}\D_p\log \sigma-\sigma\Delta-\rho \xi\right). 
\ee
For large $(\D, \xi)$, we use the saddle point approximation to evaluate the integral \eqref{M_integral}
\be
 D(\Delta,\xi)C_{\Delta p \Delta} \approx  \widetilde{\langle \phi_p\rangle}_{{(-\frac{1}{\sigma_c},\frac{\rho_c}{\sigma^2_c})}} \,e^{f(\sigma_c,\rho_c)},
 \label{saddle_approx}
\ee
where $\sigma_c$ and $\rho_c$ are the critical points of the function $f(\sigma,\rho)$: $$\partial_{\sigma}f(\sigma_c,\rho_c)=0,\,\partial_{\rho}f(\sigma_c,\rho_c)=0.$$ These are given by:
 \bes\bea
&& -\frac{c_L}{2 \sigma ^2}+\frac{c_L}{2}+\frac{\rho  c_M}{\sigma ^3}-\Delta +\frac{i \Delta _p}{2 \pi  \sigma }+\frac{i \rho  \xi _p}{2 \pi  \sigma ^2}=0, \\
&& -\frac{c_M}{2 \sigma ^2}+\frac{c_M}{2}-\xi -\frac{i \xi _p}{2 \pi  \sigma }=0.
\eea
\ees
We have two solutions for the above equations. For $(\Delta, \xi)\gg(c_L,c_M,\D_p,\xi_p)$, these are given by
\be
(\sigma_{c}, \rho_c) \approx \left(i\sqrt{\frac{c_M}{2\xi}}, i\frac{ \left( \xi  c_L - \Delta  c_M  \right)}{ (2 \xi )^{3/2} \sqrt{c_M}}\right); 
\quad 
(\tilde{\sigma}_c, \tilde{\rho}_c)  \approx \left(-i\sqrt{\frac{c_M}{2\xi}}, -i\frac{ \left(\xi  c_L - \Delta  c_M \right)}{(2 \xi )^{3/2} \sqrt{c_M}}\right). 
\ee
Among this, $(\sigma_{c},\rho_c)$ is the solution we are interested in. It can be checked that the saddle-point approximation breaks down for the other solution and we will disregard it. For more details of the validity of the saddle-point analysis, we point the interested reader to \cite{Bagchi:2019unf}. Substituting $(\sigma_{_c},\rho_c)$ in \eqref{saddle_approx}, we have
\be
 D(\Delta,\xi)C_{\Delta p \Delta} \approx \widetilde{\langle \phi_p\rangle}_{{(-\frac{1}{\sigma_c},\frac{\rho_c}{\sigma^2_c})}} i^{-\D_p}\left(\frac{2\xi}{c_M}\right)^{\frac{\D_p}{2}} e^{2\pi\left(\sqrt{\frac{\xi}{2c_M}}c_L+\Delta\sqrt{\frac{c_M}{2\xi}}\right)}e^{\frac{\xi _p}{2} \left(-\frac{\Delta }{\xi }+\frac{c_L}{c_M}\right)}.
 \label{MC}
\ee
Now
\be
-\frac{1}{\sigma_c} = i\sqrt{\frac{2\xi}{c_M}},\quad\quad \frac{\rho_c}{\sigma_c^2} = i\sqrt{\frac{\xi}{2c_M}} \left(\frac{\D}{\xi}-\frac{c_L}{c_M}\right) 
\ee
So, we have
\bea
 \widetilde{\langle \phi_p\rangle}_{{(-\frac{1}{\sigma_c},\frac{\rho_c}{\sigma^2_c})}} &=& \sum_{i} D(\Delta_i,\xi_i)C_{i p i}e^{2\pi i(-\frac{1}{\sigma_c}\Delta_i+\frac{\rho_c}{\sigma_c^2} \xi_i)}\cr 
 &=& \sum_{i} D(\Delta_i,\xi_i)C_{i p i}\,e^{-2\pi\sqrt{\frac{2\xi}{c_M}}\Delta_i-2\pi\sqrt{\frac{\xi}{2c_M}} \left(\frac{\D}{\xi} - \frac{c_L}{c_M}\right) \xi_i}. \label{tp}
\eea
In the above equation, $C_{i p i}=\<\D_i,\xi_i|\phi_p(0,1)|\D_i,\xi_i\>$ is zero when $|\D_i,\xi_i\>$ is the vacuum $|0,0\>$. Hence the vacuum would not contribute to the sum on the right hand side of \refb{tp}. Since the exponent is decreasing very fast due to large value of $\Delta$ and $\xi$, the leading contribution comes from lightest state in the spectrum, excluding vacuum i.e, state with smallest values of the conformal weights. We denote this state by $|\chi\>$ and its weights by $(\D_{\chi},\xi_{\chi})$. (Equivalently, $|\chi\> = |\D_{\chi},\xi_{\chi}\>$.) Thus 
\be
 \widetilde{\langle \phi_p\rangle}_{{(-\frac{1}{\sigma_c},\frac{\rho_c}{\sigma^2_c})}} \approx D(\Delta_{\chi},\xi_{\chi})C_{\chi p \chi}\,e^{-2\pi\left(\sqrt{\frac{2\xi}{c_M}}\Delta_{\chi}+\sqrt{\frac{\xi}{2c_M}} \left(-\frac{c_L}{c_M} + \frac{\D}{\xi}\right) \xi_{\chi}\right)}.
\ee
Substituting the above equation in \eqref{MC}, we have
\bea
 D(\Delta,\xi)C_{\Delta p \Delta} &\approx &  D(\Delta_{\chi},\xi_{\chi})C_{\chi p \chi} i^{-\D_p}\left(\frac{2\xi}{c_M}\right)^{\frac{\D_p}{2}} e^{2\pi\left(\sqrt{\frac{\xi}{2c_M}}c_L+\Delta\sqrt{\frac{c_M}{2\xi}}\right)}e^{\frac{\xi _p}{2} \left(-\frac{\Delta }{\xi }+\frac{c_L}{c_M}\right)}\cr
 &&\times \,e^{-2\pi\left(\sqrt{\frac{2\xi}{c_M}}\Delta_{\chi}+\sqrt{\frac{\xi}{2c_M}} \left(\frac{\D}{\xi}-\frac{c_L}{c_M}\right) \xi_{\chi}\right)}.
\eea
Then the three-point coefficient is given by
\bea
C_{\Delta p \Delta} &\approx & \frac {D(\Delta_{\chi},\xi_{\chi})}{ D(\Delta,\xi)}C_{\chi p \chi} i^{-\D_p}\left(\frac{2\xi}{c_M}\right)^{\frac{\D_p}{2}} e^{2\pi\left(c_L \sqrt{\frac{\xi}{2c_M}}+\Delta\sqrt{\frac{c_M}{2\xi}}\right)}e^{\frac{\xi _p}{2} \left(-\frac{\Delta }{\xi }+\frac{c_L}{c_M}\right)}\non\\
 &&\times \,e^{-2\pi\left(\sqrt{\frac{2\xi}{c_M}}\Delta_{\chi}+\sqrt{\frac{\xi}{2c_M}} \left(\frac{\D}{\xi} -\frac{c_L}{c_M} \right) \xi_{\chi}\right)}.
\eea
The density of states $D(\Delta,\xi)$ for large $\D$ and $\xi$ is known (this is basically just the BMS-Cardy formula \cite{Bagchi:2012xr, Bagchi:2013qva,Bagchi:2019unf}) and given by
\be
D(\Delta,\xi) \approx D(0,0)e^{2\pi \left(c_L\sqrt{\frac{\xi}{2c_M}}+\Delta\sqrt{\frac{c_M}{2\xi}}\right)}.  
\ee
Substituting this in the previous equation we get

\bigskip

\noindent {\fbox{
\addtolength{\linewidth}{+2\fboxsep}%
 \addtolength{\linewidth}{+2\fboxrule}%
 \begin{minipage}{\linewidth}
\bea\label{asymptotic_carlip}
C_{i p i} \approx \frac {D(\Delta_{\chi},\xi_{\chi})}{ D(0,0)}C_{\chi p \chi} i^{-\D_p}\left(\frac{2\xi_i}{c_M}\right)^{\frac{\D_p}{2}} e^{\frac{\xi _p}{2} \left(-\frac{\Delta_i }{\xi_i }+\frac{c_L}{c_M}\right)} e^{-2\pi\left(\sqrt{\frac{2\xi_i}{c_M}}\Delta_{\chi}+\sqrt{\frac{\xi_i}{2c_M}} \left(\frac{\D_i}{\xi_i}-\frac{c_L}{c_M}\right) \xi_{\chi}\right)}. \cr
\eea
\end{minipage}
}}

\bigskip

\noindent This formula is the first of our main results of this paper. In the next subsection, we will improve on this result with a more refined analysis.

\medskip

\noindent
For the case when $|\phi_p\>$ and the lightest state $|\chi\>$ are both scalars, we have $\D_p=0$ and $\D_{\chi}=0$. Additionally, for Einstein gravity in 3d asymptotically flat spacetimes, we also have $c_L=0$ in the BMS algebra. For this special case we get
\be
C_{i p i} \equiv \<\D_i,\xi_i|\phi_p(0,1)|\D_i,\xi_i\>\approx \frac {D(\Delta_{\chi},\xi_{\chi})}{ D(0,0)}C_{\chi p \chi}  e^{-\frac{\xi _p}{2}\frac{\Delta_i }{\xi_i }}e^{-2\pi \frac{\D_i}{\sqrt{2c_M \xi_i}}\xi_{\chi}}.
\label{three_point_ceoff_limit}
\ee
We would be comparing the dual bulk analysis in Sec.~\refb{bulk} to the above formula. 

\subsection{An improved method}\label{onepointsec1}
In the previous subsection, we derived an asymptotic form of the one-point coefficient $C_{ipi}\equiv \<\D_i,\xi_i|\phi_p(0,1)|\D_i,\xi_i\>$ where we used bases that are simultaneous eigenstates of $L_0$ and $M_0$. Here we will obtain the asymptotic formula using a different method that is based on inverting integral transformations. As before, the BMS one-point function is given by 
\be{}
\langle \phi_p\rangle_{(\sigma, \rho)}=\text{Tr}\left(\phi_p(0,1)e^{2\pi i\sigma(L_0-\frac{c_L}{2})}e^{2\pi i\rho(M_0-\frac{c_M}{2})}\right)=\sum_{i} D(\D, \xi) q^{\Delta_i-\frac{c_L}{2}}y^{\xi_i-\frac{c_M}{2}}C_{ipi}, \non
\ee
where we have $q= e^{2\pi i \s}$ and $y= e^{2\pi i \rho}$. 
From \eqref{s_mod_trans}, $S$-transformation of $\langle \phi_p\rangle$ is given as:
\bes
\bea
\langle \phi_p\rangle_{(\sigma, \rho)} &=& \sum_{i} D(\Delta_i,\xi_i)C_{ipi}e^{2\pi i\sigma(\Delta_i-\frac{c_L}{2})}e^{2\pi i\rho(\xi_i-\frac{c_M}{2})} \label{a1} \\
&=&\sigma^{-\Delta_p}e^{\xi_p\frac{\rho}{\sigma}}\sum_{i} D(\D_i,\xi_i)C_{ipi} e^{-\frac{2\pi i}{\sigma}(\Delta_i-\frac{c_L}{2})}e^{2\pi i\frac{\rho}{\sigma^2}(\xi_i-\frac{c_M}{2})}. \label{a2} 
\eea
\ees
We choose $\sigma=i\beta$ and consider the limit $\beta\rightarrow0^+$ and $|\rho|\rightarrow0^+$. In this limit, the summation in \refb{a1} over the discrete\footnote{Let's denote the set of all $\Delta_A$ as $\Delta$ and the set of all $\xi_A$ as $\xi$. We assume the spectra to be such that any mapping from (domain) $\Delta$  to (codomain) $\xi$ is a surjection; otherwise a little reformulation is needed.} spectra can be approximated as a double-integral:
\be
\langle \phi_p\rangle_{(\sigma, \rho)}  = \int_{\xi_\chi}^\infty d\xi_i\int_{\Delta_\chi}^\infty d\Delta_i \ D(\Delta_i,\xi_i)C_{ipi} e^{-2\pi \beta(\Delta_i-\frac{c_L}{2})}e^{2\pi i\rho(\xi_i-\frac{c_M}{2})} \label{q1}
\ee
while \refb{a2} can be approximated by keeping only the most-dominating summand, which is the contribution due to the lightest-primary (having the lowest $\Delta$) $\chi$ of the theory such that $C_{\chi p\chi}\neq0$:
\be
\langle \phi_p\rangle_{(\sigma, \rho)}  = (i\beta)^{-\Delta_p} e^{-i\xi_p\frac{\rho}{\beta}} D(\Delta_\chi,\xi_\chi)C_{\chi p\chi} e^{-\frac{2\pi }{\beta}(\Delta_\chi-\frac{c_L}{2})}e^{-2\pi i\frac{\rho}{\beta^2}(\xi_\chi-\frac{c_M}{2})} \label{q2}
\ee
We now denote 
$$T_p(\Delta_i,\xi_i)\equiv D(\Delta_i,\xi_i)C_{ipi}.$$
The above equations \refb{q1}, \refb{q2} then lead to the following equality
\bea
\int_{0}^\infty d{\xi_i^\prime} \ \int_{0}^\infty d{\Delta_i^\prime} && T_p^\prime({\Delta_i^\prime},{\xi_i^\prime}) e^{-2\pi \beta{\Delta_i^\prime}}e^{2\pi i\rho{\xi_i^\prime}} \cr 
&& \hspace{-1cm} = (i\beta)^{-\Delta_p}T_p(\Delta_\chi,\xi_\chi) e^{2\pi (\beta-\frac{1}{\beta})(\Delta_\chi-\frac{c_L}{2})}e^{-2\pi i\rho\{(1+\frac{1}{\beta^2})(\xi_\chi-\frac{c_M}{2})+\frac{\xi_p}{2\pi \beta} \}},
\label{double_integral_fin}
\eea
The equation \refb{double_integral_fin} can be thought of as a Laplace transformation of ${T}_p^\prime({\Delta_i^\prime},{\xi_i^\prime})$ with respect to ${\Delta_i^\prime}$ and one-sided Fourier transformation of the same of ${T}_p^\prime({\Delta_i^\prime},{\xi_i^\prime})$ with respect to ${\xi_i^\prime} \ (\xi_i^\prime \geqslant0)$. We will invert the integral transformations \footnote{In hindsight, we know that $T_p^\prime({\Delta_i^\prime},{\xi_i^\prime})$ is of exponential order, so the inversion as done is justified.} one-by-one. We will first concentrate on the Fourier transform. This leads us to
\bea 
\int_{0}^\infty d\Delta_i^\prime {T}_p^\prime({\Delta_i^\prime},{\xi_i^\prime}) e^{-2\pi \beta\Delta_i^\prime}&=& \frac{T_p(\Delta_\chi,\xi_\chi)}{(i\beta)^{\Delta_p}} e^{2\pi (\beta-\frac{1}{\beta})(\Delta_\chi-\frac{c_L}{2})} \int_{-\infty}^{\infty}d(\text{Re}\,\rho)e^{-2\pi i\rho[(1+\frac{1}{\beta^2})(\xi_\chi-\frac{c_M}{2})+\frac{\xi_p}{2\pi \beta}]} \non\\
&=& \frac{T_p(\Delta_\chi,\xi_\chi)}{(i\beta)^{\Delta_p}}  e^{2\pi (\beta-\frac{1}{\beta})(\Delta_\chi-\frac{c_L}{2})}e^{2\pi\,(\text{Im}\,\rho) (\xi_i^\prime+(1+\frac{1}{\beta^2})(\xi_\chi-\frac{c_M}{2})+\frac{\xi_p}{2\pi\beta})} \non\\
&& \quad \times \delta\left[\xi_i^\prime+(1+\frac{1}{\beta^2})(\xi_\chi-\frac{c_M}{2})+\frac{\xi_p}{2\pi\beta}\right].
\eea
Let us denote the two roots of the argument of the $\delta$-function as $\beta_{\pm}$:
\bea 
\beta_\pm = \frac{-\frac{\xi_p}{2\pi}\pm \sqrt{\left(\frac{\xi_p}{2\pi}\right)^2+4(\frac{c_M}{2}-\xi_\chi)(\xi_i-\frac{c_M}{2})}}{2(\xi_i-\frac{c_M}{2})}.
\eea
where $\xi_i = \xi_i'+\xi_\chi$. Then, using the property of the $\delta$-function, we are led to  
\bea
\hspace{-0.5 cm} \int_{0}^\infty d\Delta_i^\prime {T}_p^\prime({\Delta_i^\prime},{\xi_i^\prime}) e^{-2\pi \beta\Delta_i^\prime}&=&\sum_{\star=+,-}\frac{T_p(\Delta_\chi,\xi_\chi)}{(i\beta_\star)^{\Delta_p}} e^{2\pi (\beta_\star-\frac{1}{\beta_\star})(\Delta_\chi-\frac{c_L}{2})}\frac{\delta(\beta-\beta_\star)}{|\frac{c_M-2\xi_\chi}{\beta_\star^3}-\frac{\xi_p}{2\pi\beta_\star^2}|}.
\label{to_invert_1}
\eea
We started out with the assumption that $\beta$ is real. Hence, for consistency, we require $\beta_\pm$ to be real. 

\bigskip

\begin{figure}[t]
\begin{center}
\begin{tikzpicture}[decoration={markings,
mark=at position 1.8cm with {\arrow[line width=1pt]{>}},
mark=at position 7.7cm with {\arrow[line width=1pt]{>}}}]
\draw[help lines,->] (-3,0) -- (3,0) coordinate (xaxis);
\draw[help lines,->] (0,-3.0) -- (0,3.0) coordinate (yaxis);
\path[draw,line width=0.8pt,postaction=decorate] (0.8,-2.4) -- (0.8,2.4) arc (90:270:2.4);
\node[below] at (xaxis) {$x$};
\node[left] at (yaxis) {$y$};
\draw (0.4,0) node{$\bullet$};
\node at (0.4,0.3) {$x_0$};
\node at (1.3,-0.3) {$C_{x=b}$};
\node at (1.9,2.4) {$z=b+i\infty$};
\node at (1.9,-2.4) {$z=b-i\infty$};
\node at (-1.8,1.7) {$C_{r\rightarrow\infty}$};
\end{tikzpicture}
\end{center}
\caption{Contour used for inverting the Laplace transformation}
\label{fig:contour}
\end{figure}
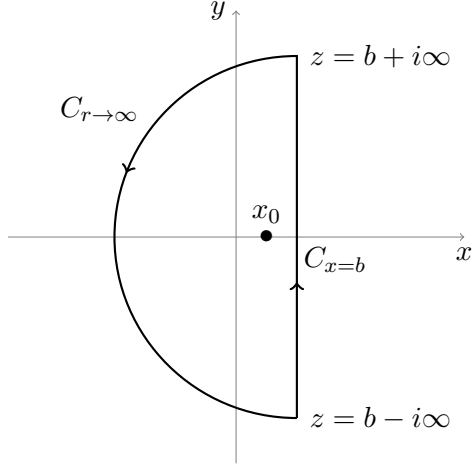

\noindent Now we concentrate on the Laplace transformation. To invert the Laplace-transform in \eqref{to_invert_1}, we need to know how to analytically continue the $\delta$-function which is defined for real arguments. Let's consider the following equivalence of the Cauchy integral theorem and the $\delta$-function (with $x_0$ being real and $F$ being an entire ($\equiv{C}^\infty$) function such that $\lim_{r\rightarrow\infty}F(re^{i\theta})=0$):
\begin{align}
\frac{1}{2\pi i}\oint_C dz\frac{F(z)}{z-x_0}=F(x_0)=\int_{-\infty}^b dxF(x)\delta(x-x_0).
\end{align}
where $C$ is the contour in Figure \ref{fig:contour} (relevant for inverse-Laplace transformation) enclosing $x_0$ where $-\infty<x_0<b$.
The semicircle part $C_{r\rightarrow\infty}$ of the closed contour has the equation $|z-b|=\lim_{r\rightarrow\infty}r$ and by the demanded property of $F(z)$, the contribution from $C_{r\rightarrow\infty}$ vanishes. Thus we are left with:
\begin{align}
\int_{-\infty}^b dxF(x)\delta(x-x_0)=\frac{1}{2\pi i }\int_{b-i\infty}^{b+i\infty} dz\frac{F(z)}{z-x_0}.
\end{align}
As $F(z)$ is any function belonging to the class of functions having the above specified properties, we conclude that for the problem (and contour) in hand, it is justified to write:
\begin{align}
\delta(x-x_0)\xrightarrow [\text{continuation}]{\text{Analytic}}\frac{1}{2\pi i }\frac{1}{z-x_0}.
\label{delta_ac}
\end{align}
Inverting the Laplace Laplace-transformation in \eqref{to_invert_1} using \eqref{delta_ac}, we have
\bea
\hspace{-0.5cm}\frac{{T}_p^\prime({\Delta_i^\prime},{\xi_i^\prime})}{T_p(\Delta_\chi,\xi_\chi)} &=&\sum_{\star=+,-}\frac{e^{2\pi (\beta_\star-\frac{1}{\beta_\star})(\Delta_\chi-\frac{c_L}{2})}}{(i\beta_\star)^{\Delta_p} } \frac{1}{|\frac{c_M-2\xi_\chi}{\beta_\star^3}-\frac{\xi_p}{2\pi\beta_\star^2}|}\int_{b-i\infty}^{b+i\infty} \frac{d\beta}{i}\frac{e^{2\pi \beta\Delta_i^\prime}}{2\pi i (\beta-\beta^{*})},
\eea
where, $b>\beta_+$, so that all of the singularities of the integrand lie at the left of the vertical integration-path. Also, when $|\beta-b|=\lim_{r\rightarrow\infty}r$,
\begin{align}
\lim_{r\rightarrow\infty}r\frac{e^{2\pi \Delta_A^\prime(b+r\cos{\theta}+ir\sin\theta)}}{b+re^{i\theta}-\beta_\star}=0
\end{align}
as along the semi-circle $C_{r\rightarrow\infty}$, $\cos\theta<0$ being on the left of the vertical path $C_{x=b}$ and $\Delta_A^\prime\geqslant0$. Thus, we can close the contour from the left to get the depicted contour $C$ whose semi-circle part doesn't contribute to the integral. Hence,
\bea
\frac{{T}_p^\prime({\Delta_i^\prime},{\xi_i^\prime})}{T_p(\Delta_\chi,\xi_\chi)}&=& \sum_{\star=+,-}(i\beta_\star)^{-\Delta_p} e^{2\pi (\beta_\star-\frac{1}{\beta_\star})(\Delta_\chi-\frac{c_L}{2})}\frac{1}{|\frac{c_M-2\xi_\chi}{\beta_\star^3}-\frac{\xi_p}{2\pi\beta_\star^2}|}\oint_{C} \frac{d\beta}{i}\frac{e^{2\pi \beta\Delta_i^\prime}}{2\pi i (\beta-\beta^{*})} \cr
&=&\sum_{\star=+,-}(i\beta_\star)^{-\Delta_p}e^{2\pi (\beta_\star-\frac{1}{\beta_\star})(\Delta_\chi-\frac{c_L}{2})}\frac{e^{2\pi \beta_\star\Delta_i^\prime}}{i|\frac{c_M-2\xi_\chi}{\beta_\star^3}-\frac{\xi_p}{2\pi\beta_\star^2}|}.
\label{laplace_fin_0}
\eea
Thus we note that upon inverting the double-integral transformation in \eqref{double_integral_fin}, the resultant inverse integrand has very sharp ($\delta$-function like) peaks at two different real values $\beta_\pm$ that are the sole contributors to the inverse transformation \eqref{laplace_fin_0}. 

To make the initial consideration of $\beta\rightarrow0^+$ consistent with the final result, $|\beta_\pm|\rightarrow0^+$ must be satisfied. This is same as the demand that  $|\beta_\pm|\ll1$. We also require $\beta$ to be real. Thus, we require $\xi_i\gg c_M-\xi_\chi\pm \frac{\xi_p}{2\pi}$ with the assumption of the existence of a $\xi_\chi<\frac{c_M}{2}$. This implies that $$\xi_i\gg \frac{c_M}{2}\pm \frac{\xi_p}{2\pi}.$$ In this limit, $$\beta_\pm\approx-\frac{\xi_p}{4\pi\xi_i}\pm\sqrt{\frac{{c_M}-2\xi_\chi}{2\xi_i}}.$$ 
As $|\beta_\pm|\rightarrow0^+$, in each summand in \eqref{laplace_fin_0} the most dominating contribution comes from the factor $e^{2\pi (-\frac{1}{\beta\star})(\Delta_\chi-\frac{c_L}{2})}$. Also, as $\sigma$ lies on the upper-half plane, to start with, we had $\beta>0$; hence, it is desirable to have $\beta_-$'s contribution to \eqref{laplace_fin_0} negligible in comparison to that of $\beta_+$. So, we also need to assume the existence of a $\Delta_\chi<\frac{c_L}{2}$. Hence, we finally reach
\bea
&&\frac{{T}_p({\Delta_i},{\xi_i})}{T_p(\Delta_\chi,\xi_\chi)}\approx\frac{(-i)^{\Delta_p+1}}{\beta_+^{\Delta_p-2}}e^{ \frac{2\pi}{\beta_+}(\frac{c_L}{2}-\Delta_\chi)}\frac{e^{2\pi \beta_+(\Delta_i-\frac{c_L}{2})}}{\frac{c_M-2\xi_\chi}{\beta_+}-\frac{\xi_p}{2\pi}}\nonumber\\
&\approx& \frac{(-i)^{\Delta_p+1}}{2\xi_i}{\left(\frac{\xi_i}{\frac{c_M}{2}-\xi_\chi}\right)}^{\frac{\Delta_p-1}{2}}
e^{\frac{\xi_p}{2}\left(-\frac{\Delta_i-\frac{c_L}{2}}{\xi_i}+\frac{\frac{c_L}{2}-\Delta_\chi}{\frac{c_M}{2}-\xi_\chi}\right)+2\pi[\sqrt{\frac{\frac{c_M}{2}-\xi_\chi}{\xi_i}}(\Delta_i-\frac{c_L}{2})+\sqrt{\frac{\xi_i}{\frac{c_M}{2}-\xi_\chi}}(\frac{c_L}{2}-\Delta_\chi)]}.\non\\
\label{eqn:tpbytp}
\eea
\textit{Density of states}: Let us call the lightest primary field in the theory $\O$. This may be the vacuum, but we are further generalising our results by considering cases where it may not be so and $\Delta_\chi\geqslant\Delta_\O\geqslant0$. Taking a cue out of Liouville theory, a BMS version of Liouville theory could be amenable to such a generalisation. We assume the existence of $\Delta_\O<\frac{c_L}{2}$ and $\xi_\O<\frac{c_M}{2}$. Clearly, putting $\Delta_p=0=\xi_p$ directly in \eqref{eqn:tpbytp} and replacing the primary field $\chi$ by $\O$, we get the asymptotic formula for density of states in the large $\xi_i$ limit:
\begin{align}
\frac{D(\Delta_i,\xi_i)}{D(\Delta_\O,\xi_\O)}=\frac{-i}{2\xi_i}\sqrt{\frac{\frac{c_M}{2}-\xi_\O}{\xi_i}} \exp 2\pi \left({\sqrt{\frac{\frac{c_M}{2}-\xi_\O}{\xi_i}}\left(\Delta_i-\frac{c_L}{2}\right)+\sqrt{\frac{\xi_i}{\frac{c_M}{2}-\xi_\O}}\left(\frac{c_L}{2}-\Delta_\O\right)}\right).
\label{den_states_large_xi}
\end{align}
Substituting \eqref{den_states_large_xi} in \eqref{eqn:tpbytp}, we find an asymptotic formula for the 3-point coefficient $C_{ipi}$ in the large $\xi_i$ limit:

\bigskip

\noindent {\fbox{
\addtolength{\linewidth}{+2\fboxsep}%
 \addtolength{\linewidth}{+2\fboxrule}%
 \begin{minipage}{\linewidth}
\bea\label{impr}
\hspace{-0.5cm} C_{ipi}&\approx &(-i)^{\Delta_p}C_{\chi p\chi}\frac{D(\Delta_\chi,\xi_\chi)}{D(\Delta_\O,\xi_\O)}\left(\frac{2\xi_i}{c_M-2\xi_\chi}\right)^{\frac{\Delta_p-1}{2}}\sqrt{\frac{2\xi_i}{c_M-2\xi_\O}}\,e^{\frac{\xi_p}{2}\left(-\frac{2\Delta_i-c_L}{2\xi_i}+\frac{c_L-2\Delta_\chi}{c_M-2\xi_\chi}\right)}\nonumber\\
&&\times e^{2\pi\left(\sqrt{\frac{2\xi_i}{c_M-2\xi_\chi}}\left(\frac{c_L}{2}-\Delta_\chi\right)-\sqrt{\frac{2\xi_i}{c_M-2\xi_\O}}\left(\frac{c_L}{2}-\Delta_\O\right)+\left(\sqrt{\frac{c_M- 2\xi_\chi}{2\xi_i}}-\sqrt{\frac{ c_M- 2\xi_\O}{2\xi_i}}\,\right)\left(\Delta_i-\frac{c_L}{2}\right)\right)}. \\
\non
\eea
\end{minipage}
}}

\bigskip

\noindent Note that the above asymptotic formula is valid irrespective of the order of magnitude of $\frac{\Delta_i}{\xi_i}$. Furthermore, if we take the field $(\D_\O,\xi_\O)$ to be the vacuum and consider the limit in which $c_M \gg \xi_{\chi}$, then we get back the asymptotic formula \eqref{asymptotic_carlip} obtained in the previous section using a saddle-point analysis. So, in this sense, the range of validity of the above formula is broader than that of \eqref{asymptotic_carlip}.

\newpage

\section{Machinery for BMS highest weights}\label{hw}
We have, in the preceding section, derived the asymptotic form of the BMS torus one-point function using general analyses, that did not rely on the explicit representations the states of the BMSFT fall into, except for the fact that the states were simultaneous eigenstates of $L_0$ and $M_0$. We now want to concentrate on BMS highest weight representations and use this to find the expression for the torus one-point function of primary fields. In this section and the next, we will build machinery for our ultimate calculation, which will be described in Sec~\refb{onepointsec}.  

\subsection{BMS highest weight modules}
As a basis for calculating the trace in the partition function or the one-point function on the torus \refb{one_point_function}, we can use the BMS primary states and their descendants. As we have briefly reviewed in Sec.~2, the primary states are eigenstates of $L_0$ and $M_0$
\be
L_0 |\D_A,\xi_A\>= \D_A |\D_A,\xi_A\>,\quad\quad  M_0 |\D_A,\xi_A\>= \xi_A |\D_A,\xi_A\>,
\ee
which are annihilated by $L_n$ and $M_n$  for $n>0$
\be
L_n |\D_A,\xi_A\> =0,\quad\quad  M_n |\D_A,\xi_A\> =0,\quad n>0.
\ee
By acting the operator $L_{-n}$ and $M_{-n}$ on the primary states we could raise the eigenvalue of $L_0$. Such states which are obtained by acting a series of $L_{-n}$ and $M_{-n}$ operators on the primary states are called descendant states. A BMS module of a primary state $|\D_A,\xi_A\>$ is the set which contain the primary states and all its descendant. States in the module have the form
\be
L_{-1}^{k_1}L_{-2}^{k_2}....L_{-l}^{k_l} M_{-1}^{q_1}M_{-2}^{q_2}....M_{-r}^{q_r}|\D_A,\xi_A \> \equiv L_{\vec{k}}M_{\vec{q}}|\D_A,\xi_A \> ,
\ee
where $\vec{k}=(k_1,k_2,....,k_l)$ and $\vec{q}=(q_1,q_2,....,q_r)$ and its $L_0$ eigenvalue is given by
\be
L_0 L_{\vec{k}}M_{\vec{q}}|\D,\xi \> = (N + \Delta_A) L_{\vec{k}}M_{\vec{q}}|\D_A,\xi_A \>, \quad \mbox{where} \ N=\sum_{i}ik_i + \sum_j jq_j.  
\ee 
$N$ is called the level of the state. To simplify or shorten equations, we will sometimes use the shorthand notation $|\D_A,\xi_A\> \equiv |A\>$.

We can group the states according to their level because inner product of sates having different level is zero. In general we can write the states at level $N$ as
\be
|\Psi\>= \left(\prod_{j=1}^{N} (L_{-j})^{N_j-k_j} \right)\left(\prod_{j=1}^{N} (M_{-j})^{k_j} \right)  |A\>,
\ee
with $\sum_j j N_j=N$. In the table below we shown the basis states for up to level 3. 
\begin{table}[t]
\begin{tabular}{|l|l|}
\hline 
Level & States\tabularnewline
\hline 
\hline 
N=0 & $|\Psi_{1}\rangle=|\Delta_A,\xi_A\rangle$\tabularnewline
\hline 
N=1 & $|\Psi_{1}\rangle=L_{-1}|\Delta_A,\xi_A\rangle,\,\,\,|\Psi_{2}\rangle=M_{-1}|\Delta_A,\xi_A\rangle$\tabularnewline
\hline 
N=2 & $\begin{array}{l}
|\Psi_{1}\rangle=L_{-1}^{2}|\Delta_A,\xi_A\rangle,\,\,\,|\Psi_{2}\rangle=L_{-2}|\Delta_A,\xi_A\rangle,\,\,\,|\Psi_{3}\rangle=L_{-1}M_{-1}|\Delta_A,\xi_A\rangle,\,\,\, \\
|\Psi_{4}\rangle = M_{-2}|\Delta_A,\xi_A\rangle,\,\,\,
|\Psi_{5}\rangle=M_{-1}^{2}|\Delta_A,\xi_A\rangle
\end{array}$\tabularnewline
\hline 
N=3 & $\begin{array}{l}
|\Psi_{1}\rangle=L_{-1}^3|\Delta_A,\xi_A\rangle,\,\,\,|\Psi_{2}\rangle=L_{-1}L_{-2}|\Delta_A,\xi_A\rangle,\,\,\,|\Psi_{3}\rangle=L_{-1}^2M_{-1}|\Delta_A,\xi_A\rangle,\\
|\Psi_{4}\rangle=L_{-3}|\Delta_A,\xi_A\rangle,\,\,\,|\Psi_{5}\rangle=L_{-2}M_{-1}|\Delta_A,\xi_A\rangle,\,\,\,|\Psi_{6}\rangle=L_{-1}M_{-2}|\Delta_A,\xi_A\rangle,\\
|\Psi_{7}\rangle=M_{-3}|\Delta_A,\xi_A\rangle,\,\,\,|\Psi_{8}\rangle=L_{-1}M_{-1}^2|\Delta_A,\xi_A\rangle,\,\,\,|\Psi_{9}\rangle=M_{-1}M_{-2}|\Delta_A,\xi_A\rangle,\\|\Psi_{10}\rangle=M_{-1}^3|\Delta_A,\xi_A\rangle
\end{array}$\tabularnewline
\hline 
\end{tabular}
\caption{States in the BMS module upto level 3}
\label{BMS_states}
\end{table}
We will use $|\D_A,\xi_A;N,i\>$ to denote the $i$-th basis state at level $N$. Sometimes, we will simply this further as $|A;N,i\>$. 

\subsection*{Gram matrix and their inverse}
For a given module we can form a matrix by taking the inner product of the basis states. This is known as the Gram matrix. Since states in different levels are orthogonal to each other, the Gram matrix will be block diagonal with each block for each level. We use the notation $K^{(N)}$ for the Gram matrix at level $N$. Since we have two basis states at level 2, the The Gram matrix for level 1 will be a $2\times 2$ matrix
\begin{align}
K^{(1)} = \left[ \begin{array}{cc}
\langle\Delta_A,\xi_A|L_{1}L_{-1}|\Delta_A,\xi_A\rangle & \langle\Delta_A,\xi_A|L_{1}M_{-1}|\Delta_A,\xi_A\rangle\\
\langle\Delta_A,\xi_A|M_{1}L_{-1}|\Delta_A,\xi_A\rangle & \langle\Delta_A,\xi_A|M_{1}M_{-1}|\Delta_A,\xi_A\rangle
\end{array}   \right] 
&= \left [ \begin{array}{cc}
2\Delta_A & 2\xi_A\\
2\xi_A & 0
\end{array}   \right].
\label{GM_level1}
\end{align} 
The Gram matrix $K^{(2)}$ for level 2 is a $5\times 5$ matrix given by
\bea
K^{(2)} = \left[  \begin{array}{ccccc}
4\Delta_A(2\Delta_A+1) & 6\Delta_A & 4\xi_A(2\Delta_A+1) & 6\xi_A & 8\xi^{2}_A\\
6\Delta_A & 4\Delta_A+6c_L & 6\xi_A & 4\xi_A+6c_M & 0\\
4\xi_A(2\Delta_A+1) & 6\xi_A & 4\xi^{2}_A & 0 & 0\\
6\xi_A & 4\xi_A+6c_M & 0 & 0 & 0\\
8\xi^{2}_A & 0 & 0 & 0 & 0
\end{array} \right].
\label{GM_level2} 
\eea 
Note that the Gram matrix has a peculiar triangular structure in which all the matrix elements below the anti-diagonal elements are zero. This is due to way we order our basis \cite{Bagchi:2019unf}. In fact for any level we can order our basis states such that the Gram matrix will have the following form: 
\be
K^{(N=\text{odd})} = \left(\begin{array}{ccccc}
K_{1,1} &  & \cdots &  & A_{\widetilde{\rm dim}_N}\\
 &  &  & \iddots & 0\\
\vdots &  & \iddots &  & \vdots\\
 & A_{2}\\
A_{1} & 0 & \cdots &  & 0
\end{array}\right) ,
\label{oddmatrix}
\ee
\be
K^{(N=\text{even})}=\left(\begin{array}{cccccccccc}
K_{1,1} & \cdots & \cdots &  & \cdots &  & \cdots &  & \cdots & A_{\widetilde{{\rm dim}}_{N}}\\
\vdots & \ddots &  &  &  &  &  &  & \iddots & 0\\
 &  & \ddots &  &  &  &  & A_{k+l} &  & \vdots\\
 &  &  & D_{1} & 0 & \cdots & 0\\
\vdots &  &  & 0 & D_{2} & \cdots & 0 &  &  & \vdots\\
\vdots &  &  & \vdots & \vdots & \ddots & \vdots &  &  & \vdots\\
 &  &  & 0 & 0 & \cdots & D_{l}\\
 &  & A_{k} &  &  &  &  & 0\\
\vdots & \iddots &  &  &  &  &  &  & \ddots & \vdots\\
A_{1} & 0 & \cdots &  & \cdots &  & \cdots &  & \cdots & 0
\end{array}\right).
\label{evenmatrix}
\ee
In the above, the non-zero Gram matrix in the anti-diagonal line, labelled by $A$, and the diagonal elements labelled $D$ which appear for even $N$ are of the form 
\be
\langle A|(\prod_{j=N}^1 M_j^{N_j-k_j})(\prod_{j=N}^1 L_j^{k_j})(\prod_{j=1}^N L_{-j}^{N_j-k_j})(\prod_{j=1}^N M_{-j}^{k_j})|A\rangle, 
\ee
where $\sum_{j=1}^{N}jN_j = N$. These are the matrix elements for which the number of $L$-operators and number of $M$-operators in the inner product are equal. In particular, the elements of $D$ are the norm of self conjugate states $(\prod_{j=1}^N L_{-j}^{N_j})(\prod_{j=1}^N M_{-j}^{N_j})|A\rangle$ (states which looks the same when $L$ and $M$ are exchanged, for example, state like $L_{-1}M_{-1}|A\>$). For more details about the above structure, we refer the interested reader to \cite{Bagchi:2019unf}. 

\medskip  

In a later section, we would like to take the large $\xi_A$ limit of the BMS torus blocks (which are of course to be defined later). So, let us point out some important properties of the matrix elements of the form given above. In explicitly calculating the matrix elements, we make use of the BMS algebra. The highest power in $\xi_A$ comes from the term is which each $L_{j}$ contracted with $M_{-j}$ and each $M_{j}$ contracted with $L_{-j}$ to form $M_0$ which will then act on the vacuum to give $\xi_A$. Thus the highest power in $\xi_A$ is equal to the number of $L$-operators or $M$-operators inside the inner product. This is given by $\sum_{i=1}^N N_j$.
We also note down the following factorization property of the inner product
\bea 
\langle A| \prod_{j=N}^1 M_j^{N_j-k_j} \prod_{j=N}^1 L_j^{k_j} \prod_{j=1}^N L_{-j}^{N_j-k_j} \prod_{j=1}^N M_{-j}^{k_j}|A\rangle =\prod_{j=1}^N \langle A| M_j^{N_j-k_j}L_{-j}^{N_j-k_j}|A\>\<A| L_j^{k_j}M_{-j}^{k_j}|A\rangle. \cr
\eea 
The terms in the above equation are given by
\bea
\<A| M_j^{N_j-k_j}L_{-j}^{N_j-k_j}|A\> &=& (N_j-k_j)!\<A|[M_j,L_{-j}]|A\> = (N_j-k_j)! (\xi_A -c_L (j^3-j)),\non \\
\<A| L_j^{k_j}M_{-j}^{k_j}|A\> &=& k_j!(\xi_A -c_L (j^3-j)),
\eea
where the factorial account for the number of ways to contracts the $L$ and $M$ operators.
From these properties we can deduce another factorisation formula which we will use later. Before we present a formula for generic matrix elements, we demonstrate it with the following example
\bea
\<A| M_2 M_{1}^3 L_{2}^2L_{-1}^3L_{-2}M_{-2}^2|A\> &=& 3 \<A|M_{1}L_{-1}|A\>\<A| M_2 M_{1}^2L_{2}^2L_{-1}^2L_{-2}M_{-2}^2|A\> \cr
\<A| M_2 M_{1}^3 L_{2}^2L_{-1}^3L_{-2}M_{-2}^2|A\> &=& \<A|M_{2}L_{-2}|A\>\<A|M_{1}^3 L_{2}^2L_{-1}^3M_{-2}^2|A\> \cr
\<A| M_2 M_{1}^3 L_{2}^2L_{-1}^3L_{-2}M_{-2}^2|A\> &=& 2\<A|L_{2}M_{-2}|A\>\<A|M_2 M_{1}^3 L_{2}L_{-1}^3L_{-2}M_{-2}|A\>.
\eea
In general we have
\bea
\hspace{-1cm}\frac{\langle A| \prod_{j=N}^1 M_j^{N_j-k_j-\delta_{j,r}} \prod_{j=N}^1 L_j^{k_j} \prod_{j=1}^N L_{-j}^{N_j-k_j-\delta_{j,r}} \prod_{j=1}^N M_{-j}^{k_j} |A\rangle}{\langle A| \prod_{j=N}^1 M_j^{N_j-k_j} \prod_{j=N}^1 L_j^{k_j} \prod_{j=1}^N L_{-j}^{N_j-k_j} \prod_{j=1}^N M_{-j}^{k_j}|A\rangle} =  \frac{(N_r-k_r)^{-1}}{\<A|M_r L_{-r}|A\>}.
\label{gram_matrix_property3}
\eea
\bea
\hspace{-1cm} \frac{\langle A| \prod_{j=N}^1 M_j^{N_j-k_j} \prod_{j=N}^1 L_j^{k_j-\delta_{j,r}} \prod_{j=1}^N L_{-j}^{N_j-k_j} \prod_{j=1}^N M_{-j}^{k_j-\delta_{j,r}}|A\rangle}{\langle A| \prod_{j=N}^1 M_j^{N_j-k_j} \prod_{j=N}^1 L_j^{k_j} \prod_{j=1}^N L_{-j}^{N_j-k_j} \prod_{j=1}^N M_{-j}^{k_j}|A\rangle} 
= \frac{1}{k_r\<A|M_r L_{-r}|A\>}.
\label{gram_matrix_property4}
\eea

\bigskip

\noindent We use $K_{(N)}$ to denote the inverse of the Gram matrix $K^{(N)}$ and specify the matrix elements by $K^{ab}_{(N)}$. Due to the structure of the matrix given above, the inverse Gram matrix $K_{(N)}$ will have the following structure:
\be
K_{(N=\text{odd})} = \left(\begin{array}{ccccc}
0 & \cdots & 0 & \cdots & \frac{1}{A_{\widetilde{\rm dim}_N}}\\
0 &  &  & \iddots & \vdots\\
\vdots &  & \iddots &  & \vdots\\
 & \frac{1}{A_{2}}\\
\frac{1}{A_{1}} & \iddots & \cdots &  & K^{\widetilde{\rm dim}_N,\widetilde{\rm dim}_N}
\end{array}\right) ,
\label{oddmatrix_in}
\ee
\be
K_{(N=\text{even})}=\left(\begin{array}{cccccccccc}
0 & \cdots & \cdots &  & \cdots &  & 0 &  & \cdots & \frac{1}{A_{\widetilde{{\rm dim}}_{N}}}\\
\vdots & \ddots &  &  &  &  &  &  & \iddots & \\
 &  & \ddots &  &  &  &  & \frac{1}{A_{k+l}} &  & \vdots\\
 &  &  & \frac{1}{D_{1}} & 0 & \cdots & 0\\
\vdots &  &  & 0 & \frac{1}{D_{2}} & \cdots & 0 &  &  & \vdots\\
0 &  &  & \vdots & \vdots & \ddots & \vdots &  &  & \vdots\\
 &  &  & 0 & 0 & \cdots & \frac{1}{D_{l}}\\
 &  & \frac{1}{A_{k}} &  &  &  &  & \ddots \\
\vdots & \iddots &  &  &  &  &  &  & \ddots & \vdots\\
\frac{1}{A_{1}} &  & \cdots &  & \cdots &  & \cdots &  & \cdots & K^{\widetilde{\rm dim}_N,\widetilde{\rm dim}_N}
\end{array}\right).
\label{evenmatrix_in}
\ee

\medskip 

\noindent Note that for matrix elements in the anti-diagonal line as well as the diagonal elements of the matrix at the centre for even level, $K_{(N)}^{ab}$ are simply given by $1/{K^{(N)}_{ab}}$. This fact will be important when we study large $\xi_A$ limit of the BMS torus blocks. Even for non-zero, non-anti-diagonal elements $K_{(N)}^{ab}$, the leading term in $\xi_A$ is given by $1/K^{(N)}_{ab}$.

\subsection{BMS mode expansion}
Inspired from 2d CFT, we propose that a local field $\phi_p(t,x)$ with dimensions $(\Delta,\xi)$ in a 2d BMSFT on the 2d Euclidean plane (endowed with a degenerate metric) has the following mode-expansion
\begin{align}
\phi_p(t,x)=\sum_{m\in \mathbb{Z}}\sum_{n\in \mathbb{Z}-\mathbb{N}} {x}^{-\Delta-m}e^{\frac{t}{x}\xi}(-1)^n{\left(\frac{t}{x}\right)}^{-n}\phi_{m,n},
\label{mode_exp}
\end{align}
where the quasi-primary field satisfies the following Hermitian conjugation relation
\begin{align}
\phi_p^{\dagger}(t,x)=x^{-2\Delta}e^{2\xi\frac{t}{x}}\phi_p\left(\frac{1}{x},\frac{t}{x^2}\right).
\end{align}
It is interesting to note here the $n$ index in \eqref{mode_exp} runs over negative integers only. This condition is imposed on $n$ to prevent blowing up of $\phi_p$ at $(t=0,x=1)$ which is mapped to the point $(0,0)$ on the cylinder.
Putting the mode-expansion ansatz into the above conjugation equation, we get:
\begin{align}
\sum_{m\in \mathbb{Z}}\sum_{n\in \mathbb{Z}-\mathbb{N}} x^{-\Delta-m}e^{\xi\frac{t}{x}}(-1)^n{\left(\frac{t}{x}\right)}^{-n}\phi_{m,n}^{\dagger}=\sum_{m\in \mathbb{Z}}\sum_{n\in \mathbb{Z}-\mathbb{N}} {x}^{-\Delta+m}e^{\xi\frac{t}{x}}(-1)^n{\left(\frac{t}{x}\right)}^{-n}\phi_{m,n}.
\end{align}
Comparing the coefficients on the both sides, we conclude that the BMS quasi-primary modes satisfy the following Hermitian-conjugation condition:
\begin{align}
\phi_{m,n}^{\dagger}=\phi_{-m,n}
\end{align}
The asymptotic `in'-state $|\Delta,\xi\rangle$ is created from the vacuum $|0\rangle$ of the theory as below:
\begin{align}
|\Delta,\xi\rangle\equiv \lim_{x\to 0}\lim_{\frac{t}{x}\to 1}\phi_p(t,x)|0\rangle.
\end{align}
Clearly, for the `in' states to be well-defined, the vacuum must satisfy
\begin{align}
\phi_{m,n}|0\rangle=0 \text{   , for $m>-\Delta$}\Longleftrightarrow\langle0|\phi_{m,n}=0 \text{   , for $m<\Delta$}.
\end{align}
From the plane-representation of $BMS_3$ algebra, we know that for a local primary field $\phi_p$ on 2d Euclidean plane (for $n\geqslant -1$)
\bea
[L_n,\phi_{\D,\xi}(t,x)]&=&[x^{n+1}\partial_x + (n+1)x^n t\partial_t +(n+1)(\D x^n -n \xi x^{n-1} t )]\phi_{\D,\xi}(t,x) \cr
[M_n,\phi_{\D,\xi}(t,x)] &=& [-x^{n+1}\partial_t + (n+1)\xi x^n]\phi_{\D,\xi}(t,x).
\eea
Substituting the mode-expansion ansatz into the above representation and comparing the coefficients on both sides, we obtain the following mode-commutation relations (for $n\geqslant -1$) only for primary fields: 
\begin{align}
&[L_n,\phi_{j,k}]=[n(\Delta-1-k)-j]\phi_{j+n,k}+n^2\xi\phi_{j+n,k+1},  \quad \forall k\leqslant -1,\nonumber\\
&[L_n,\phi_{j,0}]=[n(\Delta-1)-j]\phi_{j+n,0},\nonumber\\
&[M_n,\phi_{j,k}]=n\xi\phi_{j+n,k}+(k-1)\phi_{j+n,k-1},  \quad \forall k\leqslant 0.
\label{commutation_mode}
\end{align}
Although we shall not be highlighting this in the main text, most of the intrinsic analysis of 2d BMSFT done in this paper, and elaborated in the appendices, would be done with the help of the mode expansions defined in this section. 

\newpage

\section{BMS Torus blocks}\label{tblock}
In this section, using the tools we have developed for the BMS highest weight states, we construct the torus blocks for 2d BMSFTs. We start by defining these blocks. We then go onto calculating the contribution for low orders and then work in a particular limit, which we will call the large $\xi_A$ limit, which will be crucial for our analysis of the one-point functions in the following section. 

\subsection{Highest weight torus blocks}
For our basis we can use the primary states and their descendants. Then taking the trace with respect to this basis in \eqref{one_point_function}, the one-point function for primary field $\phi_p$ is given by
\bea
\<\phi_p\>_{(\sigma,\rho)} &=& {\rm{Tr}}\left(\phi_p(0,1) e^{2\pi i\sigma (L_{0}-c_{L}/2)}e^{2\pi i\rho (M_{0}-c_{M}/2)} \right) \cr
&=& e^{-2\pi i (\sigma c_L/2+\rho c_M/2)}\sum_{a,b} K^{ab} \<\Psi_a| \phi_p(0,1) \, e^{2\pi i\sigma L_{0}}e^{2\pi i\rho M_{0}}|\Psi_b\>,
\eea
where $|\Psi_a\>$ are the primary states and their descendant. $K^{ab}$ is the element of the inverse Gram Matrix formed by taking inner products of the basis states. Since inner products between states in different modules vanish, we can separate out the contribution of each module $|\D,\xi\>$ to the one-point function. So, with a bit of hindsight, we may write the one point function as  
\bea
 \<\phi_p\>_{A} &=& e^{-\pi i (\sigma c_L+\rho c_M)}\sum_{A} D(\D_A,\xi_A) e^{2\pi i \sigma \D_A}  e^{2\pi i \rho \xi_A} \<A|\phi_p(0,1)|A\> \mathcal{F}_{\D_A,\xi_A,c_L,c_M}^{\D_p,\xi_p}(\sigma,\rho),\cr
 & \equiv & e^{-\pi i (\sigma c_L+\rho c_M)}\sum_{A} D(\D_A,\xi_A) e^{2\pi i \sigma \D_A}  e^{2\pi i \rho \xi_A} C_{ApA}\,\mathcal{F}_{\D_A,\xi_A,c_L,c_M}^{\D_p,\xi_p}(\sigma,\rho), \label{onepointwblocks}
\eea
where $\{A\}$ is the collection of all the the primary states and $D(\D_A,\xi_A)$ is the number of primary states (density of states, or, multiplicity) with dimension $(\D_A,\xi_A)$. As before, $C_{ApA} \equiv \<A|\phi_p(0,1)|A\>$ is the three point coefficient. We called $\mathcal F$ the BMS torus block. This is completely determined by BMS symmetry alone.  

To simplify our notation, let us use $q=e^{2\pi i \sigma}$ and $y=e^{2\pi i \rho}$. Then, the BMS torus block  $\mathcal F$ is formally given by
\bea
{\boxed{\mathcal{F}_{\D_A,\xi_A,c_L,c_M}^{\D_p,\xi_p}(\sigma,\rho) = \frac{{\rm Tr}_{\D_A,\xi_A}\left(\phi_p(0,1)q^{L_0-\frac{c_L}{2}}y^{M_0-\frac{c_M}{2}}\right)}{q^{\D_A-\frac{c_L}{2}}y^{\xi_A-\frac{c_M}{2}}C_{ApA}}.}}
 \label{bms_torus_block}
\eea
The trace, ${\rm Tr}_{\D_A,\xi_A}$, is over the states in the module of $|\D_A,\xi_A\>$. Thus we have
\bea
{\rm Tr}_{\D_A,\xi_A}\left(\phi_p(0,1)q^{L_0-\frac{c_L}{2}}y^{M_0-\frac{c_M}{2}}\right) = \sum_{N; i,j} K^{ij}_{(N)} \< A,N,i|\phi_p \ q^{L_0-\frac{c_L}{2}}y^{M_0-\frac{c_M}{2}}|A,N,j\>. 
\label{trace_module}
\eea
For a state at level $N$
\be
 L_0|A;N,i\> = (N+\D_A)|A;N,i\> \Rightarrow q^{L_0} |A;N,i\> = q^{N+\D_A} |A;N,i\>. 
\ee
Using this in \eqref{trace_module}, we get
\bea
{\rm Tr} \left( \phi_p(0,1)q^{L_0-\frac{c_L}{2}}y^{M_0-\frac{c_M}{2}}\right) = \sum_{N}q^{N+\D_A -\frac{c_L}{2}}y^{-\frac{c_M}{2}}\sum_{i,j} K^{ij}_{(N)} \<A;N,i|\phi_p y^{M_0}|A;N,j\>. 
\eea
Further substituting this in \eqref{bms_torus_block}, we have
\bea
  \mathcal{F}_{\D_A,\xi_A,c_L,c_M}^{\D_p,\xi_p}(\sigma,\rho) &=&  \sum_N q^N\sum_{i,j} K^{ij}_{(N)} \frac{\<A;N,i|\phi_p(0,1)y^{M_0}|A;N,j\>}{y^{\xi_A}C_{ApA}}\cr
  &\equiv& \sum_N q^N \mathcal{F}_{N}(\Delta_p,\xi_p;\Delta_A,\xi_A|c_L,c_M|\rho),
  \label{bms_block_fin}
\eea
where 
\bea 
\mathcal{F}_{N}(\Delta_p,\xi_p;\Delta_A,\xi_A|c_L,c_M|\rho)&=&\frac{1}{y^{\xi_A}C_{ApA}}\sum_{i,j} K^{ij}_{(N)} \<A;N,i|\phi_p(0,1)e^{2\pi i M_0}|A;N,j\>.
\label{F_N}
\eea
From \eqref{bms_block_fin} can see that $q^N \mathcal{F}_N$ is the contribution of level $N$ to the torus BMS block. In what follows, we give an explicit expression for the BMS torus block for the first couple of levels. 
 
\subsection*{$\mathcal{F}_N$ for level 1}
As an example let us calculate $\mathcal{F}_N$ for level 1.
For this level, the states are
\be
|\D_A,\xi_A;1,1\> \equiv |\Psi_1\>= L_{-1}|\D_A,\xi_A\>,\quad\quad\quad|\D_A,\xi_A;1,2\> \equiv|\Psi_2\> = M_{-1}|\D_A,\xi_A\>,
\ee
and the Gram matrix is given in \eqref{GM_level1}.
The inverse of the Gram matrix is
\begin{align}
K_{(1)} = \left[
\begin{array}{cc}
 0 & \frac{1}{2 \xi_A } \\
 \frac{1}{2 \xi_A } & -\frac{\Delta_A }{2 \xi ^2_A} \\
\end{array}
\right].
\end{align}
Now,
\bea 
&& \mathcal{F}_{1}(\Delta_p,\xi_p;\Delta_A,\xi_A|c_L,c_M|\rho)\cr
=&&\frac{1}{y^{\xi_A}C_{ApA}}\left( K^{11}_{(1)} \<\Psi_1|\phi_p(0,1)e^{2\pi i M_0}|\Psi_1\> + K^{12}_{(1)} \<\Psi_1|\phi_p(0,1)e^{2\pi i M_0}|\Psi_2\>\right.\cr
&&\quad\quad\quad\quad \quad\quad \left.
 + K^{21}_{(1)} \<\Psi_2|\phi_p(0,1)e^{2\pi i M_0}|\Psi_1\> + K^{22}_{(1)} \<\Psi_2|\phi_p(0,1)e^{2\pi i M_0}|\Psi_2\>
\right).
\label{block_level1_exp}
\eea
Since $K^{11}_{(1)}$ is zero, we don't have to calculate $\<\Psi_1|\phi_p(0,1)e^{2\pi i M_0}|\Psi_1\>$. The other three matrix elements are listed below
\bea
\<\Psi_1|\phi_p(0,1)e^{2\pi i M_0}|\Psi_2\> &=&  C_{ApA}\,y^{\xi_A} (\xi_p\D_p - \xi_p + 2\xi_A),\cr
\<\Psi_2|\phi_p(0,1)e^{2\pi i M_0}|\Psi_1\>&=& y^{\xi_A} C_{ApA}(\xi_p\D_p-\xi_p + 2\xi_A + 2\pi i \rho \xi_p^2),\cr
\<\Psi_2|\phi_p(0,1)e^{2\pi i M_0}|\Psi_2\> &=&  y^{\xi_A} C_{ApA} \, \xi_p^2. \label{eq5}
\eea
These can be obtained using the method given in Appendix \ref{sec:useful_identities} or as done in Appendix \ref{sec:matrxi_ele_level1}.
Substituting the above matrix elements and $K_{(2)}^{ij}$ in \eqref{block_level1_exp}, we obtain
\be
\mathcal{F}_{1}(\Delta_p,\xi_p;\Delta_A,\xi_A|c_L,c_M|\rho) =  2+\frac{\xi_p(\D_p-1)}{\xi_A}+ \pi i \rho \frac{\xi_p^2}{\xi_A}-\frac{\D_A}{2\xi^2_A}\xi_p^2 .
\label{level1_bms}
\ee
Note that this is a polynomial in $\xi_A$, $\D_A$, $\xi_p$ and $\D_p$. In later section we would be interested in the large $\xi_A$ limit. In this limit we could see that the leading term is 2, which is equal to the number of states and the subleading term is $\frac{\xi_p(\D_p-1)}{\xi_A}+ \pi i \rho \frac{\xi_p^2}{\xi_A}$. Similarly, we have calculated some of the details of the torus block for level 2. They  are described in appendix \ref{appendix:identities}.

\subsection{Large $\xi_A$ limit of BMS torus blocks}
\label{sec:large_xi}
While deriving an asymptotic form for the BMS 3-point coefficients, we will need to consider the torus block of a heavy exchanged operator with dimension $$\xi_A\gg c_M,\xi_p,\Delta_p.$$ 
We shall call this the {\em{large $\xi_A$ limit}}. It is not required to know the relative values (ratios) of $\xi_A$, $\Delta_A$ and $c_L$ for the same, so, the asymptotic formula thus obtained will be valid for all values of $\Delta_A$, irrespective of any comparison to $\xi_A$ and $c_L$. We will find the leading and the sub-leading terms in the asymptotic expansion in $\frac{1}{\xi_A}$ of the torus block $\mathcal{F}_N$ in the large $\xi_A$ limit along with $\xi_A\gg \Delta_A$. 

\subsubsection{Summary of results}
Since this particular part of the paper may become difficult to read, for the sanity of the reader who is mainly interested in the results, we will present the main results of the section. 

\bigskip

\noindent{\em{Leading term in BMS torus block}}

\smallskip

\noindent The leading term of the BMS torus block in the large $\xi_A$ limit is given by
\bea
  \mathcal{F}_{\D_A,\xi_A,c_L,c_M}^{\D_p,\xi_p}(\sigma,\rho) =  \sum_N q^N \widetilde{\text{dim}}_N + \mathcal{O}\left(\frac{1}{\xi_A}\right) = \frac{q^{\frac{1}{12}}}{\eta(\sigma)^2} + \mathcal{O}\left(\frac{1}{\xi_A}\right),
  \label{f_leading}
\eea
where $$\eta(\sigma)=e^{\frac{2\pi i\sigma}{24}}\prod_{n=1}^{\infty}(1-e^{2\pi i \sigma n})=q^{\frac{1}{24}}\prod_{n=1}^{\infty}(1-q^n).$$ 

\bigskip

\noindent{\em{Subleading term in BMS torus block}}
\smallskip

\noindent 
We calculate the sub-leading terms in the BMS torus one-point function by intrinsic methods in a 2d BMSFT and the combined result of the leading and sub-leading orders is given by: 
\be
\mathcal{F}_{N}=\left(1+\frac{\xi_p(\Delta_p-1)}{2\xi_A}N\right)\widetilde{\text{dim}}_N +\pi i\rho\frac{\xi_p^2}{\xi_A} \sum_{k=0}^Np(N-k)p(k)(N-k)(N-2k) + \mathcal{O}(\xi_A^{-2}).
\ee
Here $\mathcal{F}_{N} = \mathcal{F}_{N}(\Delta_p,\xi_p;\Delta_A,\xi_A|c_L,c_M|\rho)$ which is defined by \refb{F_N}. $p(N)$ counts the number of partitions of a given number $N$. The formula above is the result of a long and tedious calculation and a number theoretic conjecture \refb{conjecture} which we provide evidence for in Appendix \ref{secconjecture}. We reproduce the same answer in Appendix \ref{gensub} as a limit of analogous answers in 2d CFT (relevant details of which are reviewed in Appendix \ref{ApA}). 

\subsubsection{Back to details} 
We now dive head-long into the details of the computation of the BMS torus blocks. A number of appendices would give further details of the computations carried out in the following analysis. We make heavy use of the BMS mode expansions developed earlier to prove several of these relations. Some of the formulae useful for what we will calculate and their proofs are presented in Appendix C. 

\medskip

\noindent For computing the BMS torus block to an arbitrary order, one of the principal tasks is to compute the inverse Gram-matrix to that order. Following our discussions of the generic form of the Gram matrix in Sec. 3, we will investigate this in detail now. 

\medskip

\noindent 
We are at first interested in isolating the elements of the inverse Gram matrix that would contribute to our computation of the torus block. This matrix has all its elements vanishing above the anti-diagonal barring those that are situated at the positions corresponding to the Gram-matrix elements that are norm-squared of the self-conjugate states. So, we don't have to know the value of $\<A;N,a|\phi_p(0,1)e^{2\pi i M_0}|A;N,b\>$ for which $K^{ab}_{(N)}=0$ since they will not contribute to $\mathcal{F}_N$ in \eqref{bms_block_fin}. In other words we only have to know the value of  $\<A;N,a|\phi_p(0,1)e^{2\pi i \rho M_0}|A;N,b\>$ for which the corresponding Gram matrix elements 
\be
K_{ab}^{(N)} =\<A;N,a|A;N,b\> =  \langle A|(\prod_{j=1}^N M_j^{k_j^\prime})(\prod_{j=1}^N L_j^{N_j^\prime-k_j^\prime})(\prod_{j=1}^N L_{-j}^{N_j-k_j})(\prod_{j=1}^N M_{-j}^{k_j})|A\rangle,
\ee
lie on the anti-diagonal Gram-matrix-elements or vanishing Gram-matrix-elements lying below the anti-diagonal or are norm-squared of the self-conjugate states. Such a Gram-matrix element must have: $$\text{len }M\geqslant\text{min}\{\sum_{j=1}^NN_j,\sum_{j=1}^NN_j^\prime\}\geqslant\text{len }L.$$ 
Here $\text{len }M$ and $\text{len }L$ are the number of $M$-operators and $L$-operators in the inner product $\<A;N,a|A;N,b\>$ i.e.,  
\be 
\text{len }M=\sum_{j=1}^N (k_j+k_j^\prime),\quad \text{len }L=\sum_{j=1}^N (N_j + N_{j}^{\prime}-k_j-k_j^\prime).
\ee
We observed that in the large $\xi_A$ limit with $\xi_A\gg \Delta_A$, non-zero $K^{ab}_{(N)}$ will be of the order of $\mathcal{O}(\frac{1}{{\xi_A}^{\text{len }M}})$.
Next, we note that in the one point matrix
\begin{align}
\langle A|(\prod_{j=N}^1 M_j^{k_j^\prime})(\prod_{j=N}^1 L_j^{N_j^\prime-k_j^\prime})\phi_p(0,1)(\prod_{j=1}^N L_{-j}^{N_j-k_j})\prod_{j=1}^N (M_{-j}^{k_j})|A\rangle
\end{align}
the `uncontracted' $M$-operators give rise to factors of $\xi_p$ through the commutator with $\phi_p(0,1)$. 

\newpage

\noindent As an example let us look at $\<A|L_1L_2\phi_p(0,1)M_{-1}M_{-1}M_{-1}|A\>$. After a bit of calculation, we have
\bea
\label{uncontracted_M}
\<A|L_2L_1\phi_p(0,1) M_{-1}M_{-1}M_{-1}|A\> = && 3\xi_A [M_1,[\phi_p(0,1),M_{-1}]] \cr 
 &+& 9\<A|[M_1,[L_1[[\phi_p(0,1),M_{-1}],M_{-1}]]]|A\> \cr
 &+&  3 \xi_A \<A|[L_2,[[\phi_p(0,1),M_{-1}],M_{-1}]]|A\> \cr
 &+&  3\<A|[M_2,[[\phi_p(0,1),M_{-1}],M_{-1}]]|A\>\\ 
 &+&  \<A|[L_2,[L_1,[[[\phi_p(0,1),M_{-1}],M_{-1}],M_{-1}]]]|A\>.\non
\eea
The factor of $\xi_A$  is due to $M_0$ which we get due to contraction of $L$'s with $M$'s. The commutator of the operators with $\phi_p$ are due to all the uncontracted operators. Each $[\phi_p,M]$ and $[\phi_p,L]$ will give a factor of $\xi_p$ and $\D_p$ respectively.
Hence, the generic element $\<A;N,a|\phi_p(0,1)|A;N,b\>$
is a polynomial in both $\xi_A$ and $\xi_p$ (and also in $\Delta_A^\prime$ and $\Delta_p$), with each monomial's $\xi$-dependence being of the form $~{\xi_A^\prime}^r{\xi_p}^s$, such that $r+s=\text{len }M$. Clearly, $r$ is the number of `contracted' $M$-operators while $s$ denotes the number of `uncontracted' $M$-operators for each monomial, with no two monomials having the same $r$, $s$. This is because these monomials have varying number of available $L$-operators for contracting $M$-operators as some $L$-operators are used in commutation with the $\phi_p(0,1)$-field but the total number of $M$-operators in each of them is same($=\text{len }M$).

\medskip

\noindent
Note that $$\<A|L_1L_2M_{-1}M_{-1}M_{-1}|A\>=0$$ 
and we can see from \eqref{uncontracted_M} that the number of uncontracted M-operators is equal to 2 or more. We gave a proof in Appendix \ref{sec:minimum_uncontracted} that this is true for any $\<A;N,a|\phi_p(0,1)|A;N,b\>$ for which the Gram matrix vanish, i.e. 
$$ \<A;N,a|\phi_p(0,1)|A;N,b\> = 0 \quad \text{if} \, \<A;N,a|A;N,b\>=0 \, \text{and $\#$ of uncontracted} \, M_n>2.$$ So, the one point matrix elements corresponding to the vanishing Gram matrix will be a polynomial in which the exponents of $\xi_p$ for each term is greater than or equal to 2.

\medskip

\noindent
To this end, we note that the generic matrix element at level $N$ can be rewritten using \eqref{bch} as below
\begin{align}
&\langle A|(\prod_{j=N}^1 M_j^{k_j^\prime})(\prod_{j=N}^1 L_j^{N_j^\prime-k_j^\prime})\phi_p(0,1)y^{M_0}(\prod_{j=1}^N L_{-j}^{N_j-k_j})(\prod_{j=1}^N M_{-j}^{k_j})|A\rangle y^{-\xi_A}\nonumber\\
=&\langle A|(\prod_{j=N}^1 M_j^{k_j^\prime})(\prod_{j=N}^1 L_j^{N_j^\prime-k_j^\prime})\phi_p(0,1)(\prod_{j=1}^N (L_{-j}+2\pi i\rho jM_{-j})^{N_j-k_j})\prod_{j=1}^N (M_{-j}^{k_j})|A\rangle\nonumber\\
=&\langle A|(\prod_{j=N}^1 M_j^{k_j^\prime})(\prod_{j=N}^1 L_j^{N_j^\prime-k_j^\prime})\phi_p(0,1)(\prod_{j=1}^N L_{-j}^{N_j-k_j})\prod_{j=1}^N (M_{-j}^{k_j})|A\rangle\nonumber\\
&+(\text{terms with less right-$L$-string length and more right-$M$-string length}).
\end{align}
Clearly, for the matrix elements contributing to $\mathcal{F}_{N}(\Delta_p,\xi_p;\Delta_A,\xi_A|c_L,c_M|\rho)$, the first term is of the highest order($\leqslant\text{len }L$) in $\xi_A$ among all the terms (due to maximum number of $L$-operators and minimum number of $M$-operators, under the restriction $\text{len }M\geqslant\text{len }L$), though other terms may be of this same order. Thus, the contribution of a generic matrix element that corresponds to a vanishing Gram-matrix element lying below the anti-diagonal is $\mathcal{O}(\frac{1}{{\xi_A}^s})$ in the said limit, where $s$ is the number of uncontracted $M$-operators in the first term (clearly, other terms have $s^\prime\geqslant s$). We already mentioned that $s\geqslant2$ for vanishing Gram-matrix elements. Thus, in this limit, $K^{ab}_{(N)}\<A;N,a|\phi_p(0,1)y^{M_0}|A;N,b\>$ corresponding to any vanishing Gram-matrix element, do not contribute to the leading and sub-leading pieces of interest in $\mathcal{F}_N$.
Thus, only the matrix elements corresponding to the non-vanishing anti-diagonal Gram-elements or squared-norms of the self-conjugate states contribute to the said pieces in this limit. 

The generic matrix element corresponding to a non-vanishing anti-diagonal Gram-element is given by:
\begin{align}
&\langle A|(\prod_{j=N}^1 M_j^{N_j-k_j})(\prod_{j=N}^1 L_j^{k_j})\phi_p(0,1)y^{M_0}(\prod_{j=1}^N L_{-j}^{N_j-k_j})(\prod_{j=1}^N M_{-j}^{k_j})|A\rangle y^{-\xi_A}\nonumber\\
=&\langle A|(\prod_{j=N}^1 M_j^{N_j-k_j})(\prod_{j=N}^1 L_j^{k_j})\phi_p(0,1)(\prod_{j=1}^N (L_{-j}+2\pi i\rho jM_{-j})^{N_j-k_j})(\prod_{j=1}^N M_{-j}^{k_j})|A\rangle\nonumber\\
=&\langle A|(\prod_{j=N}^1 M_j^{N_j-k_j})(\prod_{j=N}^1 L_j^{k_j})\phi_p(0,1)(\prod_{j=1}^N L_{-j}^{N_j-k_j})(\prod_{j=1}^N M_{-j}^{k_j})|A\rangle+\sum_{r=1}^N r(N_r-k_r)\nonumber\\
&\times (2\pi i\rho)\langle A|(\prod_{j=N}^1 M_j^{N_j-k_j})(\prod_{j=N}^1 L_j^{k_j})\phi_p(0,1)(\prod_{j=1}^{N} L_{-j}^{N_j-k_j-\delta_{j,r}}(\prod_{j=1}^N M_{-j}^{k_j+\delta_{j,r}})|A\rangle\nonumber\\
&+(\text{terms with less right-$L$-string length and more right-$M$-string length})
\label{one_point_matrix_limit}
\end{align}
The corresponding element in the inverse Gram-matrix is $\mathcal{O}({\xi_A^{-\sum_{j=1}^NN_j}})$. It is simply the inverse of the corresponding anti-diagonal Gram-element. As the terms not explicitly written above are $\mathcal{O}(\xi_A^{\text{len }L})$, with $\text{len }L=\sum_{j=1}^NN_j-r$, with $r\geqslant2$, their contribution to $\mathcal{F}_{N}(\Delta_p,\xi_p;\Delta_A,\xi_A|c_L,c_M|\rho)$ is  $\mathcal{O}(\frac{1}{\xi_A^r})$; thus they don't contribute to the leading or sub-leading pieces in this limit. On the other hand, the first term has $r=0$ and the second one has $r=1$; hence they contribute.

\subsubsection{Leading term}
Now we will construct the leading term in the BMS torus block in the large $\xi_A$ expansion. For this, let us look at the first term in \eqref{one_point_matrix_limit}
\be 
\langle A|(\prod_{j=N}^1 M_j^{N_j-k_j})(\prod_{j=N}^1 L_j^{k_j})\phi_p(0,1)(\prod_{j=1}^N L_{-j}^{N_j-k_j})(\prod_{j=1}^N M_{-j}^{k_j})|A\rangle.
\label{leading_first_term}
\ee
The leading term in $\xi_A$ will be the one in which all the $L$-operators and $M$-operators are contracted with each other
 \bea 
&& \langle A|(\prod_{j=N}^1 M_j^{N_j-k_j})(\prod_{j=N}^1 L_j^{k_j})(\prod_{j=1}^N L_{-j}^{N_j-k_j})(\prod_{j=1}^N M_{-j}^{k_j})\phi_p(1,0)|A\rangle\cr
=&&  \langle A|(\prod_{j=N}^1 M_j^{N_j-k_j})(\prod_{j=N}^1 L_j^{k_j})(\prod_{j=1}^N L_{-j}^{N_j-k_j})(\prod_{j=1}^N M_{-j}^{k_j})|A\rangle\langle A|\phi_p(1,0)|A\rangle\cr
= && C_{ApA}  \langle A|(\prod_{j=N}^1 M_j^{N_j-k_j})(\prod_{j=N}^1 L_j^{k_j})(\prod_{j=1}^N L_{-j}^{N_j-k_j})(\prod_{j=1}^N M_{-j}^{k_j})|A\rangle.
\label{matrix_leading}
\eea
The inverse Gram matrix elements $K_{(N)}^{ab}$ for 
\be 
K_{ab}^{(N)}=\langle A|(\prod_{j=N}^1 M_j^{N_j-k_j})(\prod_{j=N}^1 L_j^{k_j})(\prod_{j=1}^N L_{-j}^{N_j-k_j})(\prod_{j=1}^N M_{-j}^{k_j})|A\rangle
\ee
are simply given by
\be
K_{(N)}^{ab} = \left[{\langle A|(\prod_{j=N}^1 M_j^{N_j-k_j})(\prod_{j=N}^1 L_j^{k_j})(\prod_{j=1}^N L_{-j}^{N_j-k_j})(\prod_{j=1}^N M_{-j}^{k_j})|A\rangle}\right]^{-1}. 
\ee
Thus we have
\bea
&& \text{Leading order in }K^{ab}_{(N)}\<A;N,a|\phi_p(0,1)y^{M_0}|A;N,b\>  \non\\ \cr
&& = y^{\xi_A}\frac{\langle A|(\prod_{j=N}^1 M_j^{N_j-k_j})(\prod_{j=N}^1 L_j^{k_j})(\prod_{j=1}^N L_{-j}^{N_j-k_j})(\prod_{j=1}^N M_{-j}^{k_j})|A\rangle\langle A|\phi_p(1,0)|A\rangle}{\langle A|(\prod_{j=N}^1 M_j^{N_j-k_j})(\prod_{j=N}^1 L_j^{k_j})(\prod_{j=1}^N L_{-j}^{N_j-k_j})(\prod_{j=1}^N M_{-j}^{k_j})|A\rangle} \non\\ \cr
&&= y^{\xi_A}\langle A|\phi_p(0,1)|A\rangle = y^{\xi_A}C_{ApA}.
\eea
Each non-zero anti-diagonal terms in the Gram matrix and diagonal elements corresponding to self-conjugate states contributes the same factor as above. 
Since we have $\widetilde{\text{dim}}_N$ number of such elements, the leading term in $\mathcal{F}_{N}$ is given by
\be
\text{Leading term} = \frac{1}{y^{\xi_A}C_{ApA}} y^{\xi_A}C_{ApA} \widetilde{\text{dim}}_N = \widetilde{\text{dim}}_N.
\ee 
So, we have
\be
\mathcal{F}_N =  \widetilde{\text{dim}}_N + \mathcal{O}\left(\frac{1}{\xi_A}\right).
\ee
Using this, we can see from \eqref{bms_block_fin} that
\bea
  \mathcal{F}_{\D_A,\xi_A,c_L,c_M}^{\D_p,\xi_p}(\sigma,\rho) =  \sum_N q^N \widetilde{\text{dim}}_N + \mathcal{O}\left(\frac{1}{\xi_A}\right) = \frac{q^{\frac{1}{12}}}{\eta(\sigma)^2} + \mathcal{O}\left(\frac{1}{\xi_A}\right),
  \label{f_leading}
\eea
where $$\eta(\sigma)=e^{\frac{2\pi i\sigma}{24}}\prod_{n=1}^{\infty}(1-e^{2\pi i \sigma n})=q^{\frac{1}{24}}\prod_{n=1}^{\infty}(1-q^n).$$ 
It is of interest to note here that if we modify the definition of the BMS torus block \refb{bms_torus_block} to keep the exponential factors $q^{\D_A-c_L/2}$ and $y^{\xi_A-c_M/2}$, the expression of the leading term BMS torus block is exactly the expression for the BMS character found in \cite{Bagchi:2019unf, Oblak:2015sea, Barnich:2015mui, Garbarz:2015lua}. 

\subsubsection{Subleading term}
Now, we focus our attention on the sub-leading pieces in the BMS torus block in our large $\xi_A$ expansion. We shall see that there are actually two sources for the sub-leading correction. We being our descriptions with an example. 

\medskip

\noindent The sub-leading contribution to \eqref{leading_first_term} is due to the terms in which all the $L$-operators and $M$-operators contract to produce $M_0$ except for one pair of $L$ and $M$. This will have one less $\xi_A$ as compared to the leading term. The un-contracted pair of $L$ and $M$ appear in the commutator with $\phi_p$ in the form $[L_k,[\phi_p(0,1),M_{-k}]]$ or $[M_k,[\phi_p(0,1),L_{-k}]]$. Let us consider an example: 
\bea
&&\text{Sub-leading term in } \<A|M_1^3L_2^2\phi_p(0,1)L_{-1}^3M_{-2}^3|A\> \non\\
&=& 9\<A|M_1^2L_2^2L_{-1}^2M_{-2}^3[M_1,[\phi_p(0,1),L_{-1}]|A\> + 6\<A|M_1^3L_2^1L_{-1}^3M_{-2}^2[L_2,[\phi_p(0,1),M_{-2}]|A\> \cr
&=& 9\<A|M_1^2L_2^2L_{-1}^2M_{-2}^3|A\>\<A|[M_1,[\phi_p(0,1),L_{-1}]|A\> + 6\<A|M_1^3L_2^1L_{-1}^3M_{-2}^2|A\>\<A|[L_2,[\phi_p(0,1),M_{-2}]|A\>. \non
\eea
The factor of 9 is the number of choosing a pair of $M_1$ and $L_{-1}$ that appear in the commutator $\<A|[M_1,[\phi_p(0,1),L_{-1}]|A\>$. Since we have 3 $M_1^3$ and 3 $L_{-1}^3$, the total of pairs is given by $3\times 3 =9$. Similarly the factor of 6 is the number of choosing a pair of $L_2$ and $M_{-2}$ that appear in the commutator $\<A|[L_2,[\phi_p(0,1),M_{-2}]|A\>$. 

\medskip

\noindent As mentioned, we will have two types of terms contributing at the next-to-leading order to the BMS torus block. Following the example above, we can deduce the first type of term, which arises from the sub-leading term for a generic anti-diagonal element \eqref{leading_first_term}.  This first type consist of terms with 
$$\<A|[M_j,[\phi_p(0,1),L_{-j}]|A\> \quad \text{or} \quad \<A|[L_j,[\phi_p(0,1),M_{-j}]|A\>.$$ 
We can see that there are $(N_j-k_j)^2$ ways of choosing a pair of $M_j$ and $L_{-j}$ and $k_j^2$ number of ways of choosing a pair of $L_j$ and $M_{-j}$. Thus the subleading term is given by
\bea
&&\text{Type 1 Sub-leading term (SLT$_1$)} = \non\\
&&\quad \sum_{r=1}^{N}(N_r-k_r)^2 \langle A|\prod_{j=N}^1 M_j^{N_j-k_j-\delta_{j,r}}\prod_{j=N}^1 L_j^{k_j}\prod_{j=1}^N L_{-j}^{N_j-k_j-\delta_{j,r}}\prod_{j=1}^N M_{-j}^{k_j}|A\rangle \<A|[M_r,[\phi_p(0,1),L_{-r}]]|A\> \cr
&& \quad +\sum_{r=1}^{N}k_r^2 \langle A|(\prod_{j=N}^1 M_j^{N_j-k_j})(\prod_{j=N}^1 L_j^{k_j-\delta_{j,r}})(\prod_{j=1}^N L_{-j}^{N_j-k_j})(\prod_{j=1}^N M_{-j}^{k_j-\delta_{j,r}})|A\rangle \<A|[L_r,[\phi_p(0,1),M_{-r}]]|A\>. \non\\
\eea
Now, we have
\be
  \<A|[M_r,[\phi_p(0,1),L_{-r}]]|A\> =
  \<A|[L_r,[\phi_p(0,1),M_{-r}]]|A\> = C_{ApA}\,r^2 \xi_p(\D_p-1) .
\ee
Substituting this in the previous equation we get
\bea
\text{SLT$_1$} &=&C_{ApA}\sum_{r=1}^{N}(N_r-k_r)^2 r^2 \xi_p(\D_p-1) \langle A|\prod_{j=N}^1 M_j^{N_j-k_j-\delta_{j,r}} \prod_{j=N}^1 L_j^{k_j}\prod_{j=1}^N L_{-j}^{N_j-k_j-\delta_{j,r}}\prod_{j=1}^N M_{-j}^{k_j}|A\rangle \cr
&& +C_{ApA}\sum_{r=1}^{N}k_r^2 r^2 \xi_p(\D_p-1) \langle A|\prod_{j=N}^1 M_j^{N_j-k_j}\prod_{j=N}^1 L_j^{k_j-\delta_{j,r}}\prod_{j=1}^N L_{-j}^{N_j-k_j}\prod_{j=1}^N M_{-j}^{k_j-\delta_{j,r}}|A\rangle. 
\eea
We have to multiply this by the corresponding inverse Gram matrix $K^{ab}_{(N)}$. As we have already pointed out, for the one point matrices of interest, these are simply given by $K^{ab}_{(N)}=\frac{1}{K_{ab}^{(N)}}$. 
So, we get
\bea
\text{SLT$_1$} &=& C_{ApA}\sum_{r=1}^{N}\frac{(N_r-k_r)^2 r^2 \xi_p(\D_p-1) \prod_{j=1}^N \<A|M_j^{N_j-k_j-\delta_{j,r}}L_{-j}^{N_j-k_j-\delta_{j,r}}|A\>\<A|L_j^{k_j}M_{-j}^{k_j}|A\>}{\prod_{j=1}^N \<A|M_j^{N_j-k_j}L_{-j}^{N_j-k_j}|A\>\<A|L_j^{k_j}M_{-j}^{k_j}|A\>}\non\\
&& +C_{ApA}\sum_{r=1}^{N}\frac{k_r^2 r^2 \xi_p(\D_p-1) \prod_{j=1}^N \<A|M_j^{N_j-k_j}L_{-j}^{N_j-k_j}|A\>\<A|L_j^{k_j-\delta_{j,r}}M_{-j}^{k_j-\delta_{j,r}}|A\>}{{\prod_{j=1}^N \<A|M_j^{N_j-k_j}L_{-j}^{N_j-k_j}|A\>\<A|L_j^{k_j}M_{-j}^{k_j}|A\>}},
\eea
Using \eqref{gram_matrix_property3} and \eqref{gram_matrix_property4} the above expression is given by
\bea
\text{SLT$_1$}&=& C_{ApA}\sum_{r=1}^{N}\frac{(N_r-k_r) r^2 \xi_p(\D_p-1) }{\<A|M_rL{-r}|A\>} + C_{ApA}\sum_{r=1}^{N}\frac{k_r r^2 \xi_p(\D_p-1) }{\<A|L_rM{-r}|A\>}.
\eea
In the large $\xi_A$ limit we have
\be
 \<A|L_rM{-r}|A\> = 2r\xi_A -c_M(r^3-r) \approx  2r\xi_A. 
\ee
Substituting this in the previous equation, we get
\bea
\text{SLT$_1$}&=& C_{ApA}\sum_{r=1}^{N}\frac{(N_r-k_r) r \xi_p(\D_p-1) }{2\xi_A} + C_{ApA}\sum_{r=1}^{N}\frac{k_r r \xi_p(\D_p-1) }{2\xi_A}  \cr 
&& \quad = C_{ApA} \frac{\xi_p(\D_p-1)}{2\xi_A} \sum_{r=1}^{N}rN_r = C_{ApA} \frac{\xi_p(\D_p-1) }{2\xi_A} N
\eea
This is only for one term. Thus the total subleading contribution of these Type 1 subleading terms in $\mathcal{F}_{N}$ due to \eqref{leading_first_term}, after taking care of  the factors of $C_{ApA}$ and $y^{\xi_A}$ is
\be
 \text{Total contribution from all SLT$_1$} = N\frac{\xi_p(\D_p-1) }{2\xi_A} \widetilde{\text{dim}}_N. 
\ee
The other sub-leading piece in the BMS torus block comes from the second term in \eqref{one_point_matrix_limit}
\bea
&&\text{Type 2 Sub-leading term (SLT$_2$)} = \non\\
&& \quad 2\pi i\rho \sum_{r=1}^N r(N_r-k_r)\langle A| \prod_{j=N}^1 M_j^{N_j-k_j} \prod_{j=N}^1 L_j^{k_j} \ \phi_p(0,1) \ \prod_{j=1}^{N} L_{-j}^{N_j-k_j-\delta_{j,r}} \prod_{j=1}^N M_{-j}^{k_j+\delta_{j,r}}|A\rangle \cr
\label{sublead_2}
&&
\eea
The leading term in this is will be subleading in $\mathcal{F}_N$. These are the term in which one $M_{j}$ from $ M_j^{N_j-k_j}$ and an $M_{-j}$ from $M_{-j}^{k_j+\delta_{j,r}}$ are used in the commutator with $\phi_p$ i,e., $[M_j,[\phi_p,M_{-j}]]$. With this, the remaining $M$-operators is equal to the number of $L$-operators and thus can be contracted with no uncontracted $M$-operators. More explicitly, this is given by
\bea
\text{SLT$_2$} = && 2\pi i\rho \sum_{r=1}^N r(N_r-k_r) \langle A|\prod_{j=N}^1 M_j^{N_j-k_j-\delta_{j,r}} \prod_{j=N}^1 L_j^{k_j} \prod_{j=1}^{N} L_{-j}^{N_j-k_j-\delta_{j,r}}\prod_{j=1}^N M_{-j}^{k_j}|A\rangle \non\\
&& \quad \times \<A|\phi(0,1)|A\>(N_r-k_r)(k_r+1) \<A|[M_r,[\phi_p(0,1),M_{-r}]]|A\>,
\eea
where $(N_r-k_r)(k_r+1)$ is the number of ways of choosing a pair of $M_r$ and $M_{-r}$. 
We have to multiply the above by the inverse Gram matrix elements. From our previous calculation we have
\bea
 && \frac{\langle A|(\prod_{j=N}^1 M_j^{N_j-k_j-\delta_{j,r}})(\prod_{j=N}^1 L_j^{k_j})(\prod_{j=1}^{N} L_{-j}^{N_j-k_j-\delta_{j,r}}(\prod_{j=1}^N M_{-j}^{k_j})|A\rangle}{\langle A|(\prod_{j=N}^1 M_j^{N_j-k_j})(\prod_{j=N}^1 L_j^{k_j})(\prod_{j=1}^{N} L_{-j}^{N_j-k_j}(\prod_{j=1}^N M_{-j}^{k_j})|A\rangle}\non\\
 &&\qquad =  \frac{1}{(N_r-k_r)\<A|[M_r,L_{-r}]|A\>} \approx \frac{1}{(N_r-k_r)2r\xi_A}.
\eea
Using $\<A|[M_r,[\phi_p(0,1),M_{-r}]]|A\>=r^2\xi_p^2$, the type 2 sub-leading term due to \eqref{sublead_2} is given by 
\bea
\text{SLT$_2$}=2\pi i\rho \sum_{r=1}^N \frac{\,r^3(N_r-k_r)^2(k_r+1)}{(N_r-k_r)2r\xi_A}= (2\pi i\rho)\frac{\xi_p^2}{2\xi_A}\sum_{r=1}^N r^2(N_r-k_r)(k_r+1).
\eea
Let us change our notation a bit, we will replace $N_r-k_r$ by $l_r$ and $k_r$ by $m_r$. With this new notation, we have 
\bea
\text{SLT$_2$}= (2\pi i\rho)\frac{\xi_p^2}{2\xi_A}\sum_{r=1}^N r^2l_r(m_r+1).
\eea
This is the contribution due to only one matrix elements. We have to sum up for all the sets $\{\{l_r\},\{m_r\}\}$ for which 
\be
\sum_{r=1}^N r(l_r+m_r)=N,
\ee 
Let us denote this set by $\mathcal{A}$. Thus the subleading term due to \eqref{sublead_2} is
\be
\text{Total contribution from all SLT$_2$}= (2\pi i\rho)\frac{\xi_p^2}{2\xi_A}\sum_{\{\{l_r\},\{m_r\}\}\in \mathcal{A}}\sum_{r=1}^N r^2l_r(m_r+1). 
\ee
Combining all our results, we get
\begin{align}
\mathcal{F}_{N}(\Delta_p,\xi_p;\Delta_A,\xi_A|c_L,c_M|\rho)=&\left(1+\frac{\xi_p(\Delta_p-1)}{2\xi_A}N\right)\widetilde{\text{dim}}_N\nonumber\\
&+(2\pi i\rho)\frac{\xi_p^2}{2\xi_A}\sum_{\{\{l_r\},\{m_r\}\}\in A}\sum_{r=1}^N r^2l_r(m_r+1)+\mathcal{O}(\frac{1}{\xi_A^2}).
\label{blocks_large_limit}
\end{align}

\subsection*{A conjecture and a rewriting} 
We will now conjecture an identity involving the partition of integers using two colours which will enable us rewrite our earlier equation \refb{blocks_large_limit}. Consider the partition of an integer $N$ using two colours in the form: $$\sum_{j=1}^Nj(l_j+m_j)=N, \quad \forall \, l_j,m_j\geqslant0.$$ 
Let us denote the set of all such different partitions as $A$ and denote an element in $A$ as $\{\{l_j\},\{m_j\}\}$. Now, we conjecture the following relation:
\begin{align}
\sum_{\{\{l_j\},\{m_j\}\}\in A}\,\,\sum_{j=1}^N j^2l_j(m_j+1)&=\frac{1}{2}\sum_{k=0}^Np(N-k)p(k)(N-2k)^2\cr
&= \sum_{k=0}^Np(N-k)p(k)(N-k)(N-2k).
\label{conjecture}
\end{align}
Although we don't have a proof of this conjecture, we have checked for the validity of \eqref{conjecture} upto level 5. Some of the details are described in appendix \ref{secconjecture}. Substituting \eqref{conjecture} in \eqref{blocks_large_limit}, we have

\bigskip

\noindent {\fbox{
\addtolength{\linewidth}{+2\fboxsep}%
 \addtolength{\linewidth}{+2\fboxrule}%
 \begin{minipage}{\linewidth}
 \bea
\mathcal{F}_{N}(\Delta_p,\xi_p;\Delta_A,\xi_A|c_L,c_M|\rho)=&&\left(1+\frac{\xi_p(\Delta_p-1)}{2\xi_A}N\right)\widetilde{\text{dim}}_N \\
&&+\pi i\rho\frac{\xi_p^2}{\xi_A} \sum_{k=0}^Np(N-k)p(k)(N-k)(N-2k) + \mathcal{O}\left(\frac{1}{\xi_A^2}\right). \non
\label{onepoint5}
\eea
\end{minipage}
}}

\bigskip

\subsubsection{The special case of $\xi_p=0$}\label{tbxip0}
As noted previously, the uncontracted $M$-operators in the relevant matrix elements of \eqref{F_N} give rise to factors of $\xi_p$. So, when $\xi_p=0$, only the matrix elements corresponding to the non-vanishing anti-diagonal Gram-elements or squared-norms of the self-conjugate states give non-vanishing contribution to $\mathcal{F}_{N}(\Delta_p,\xi_p;\Delta_A,\xi_A|c_L,c_M|\rho)$. From \eqref{one_point_matrix_limit} and \eqref{matrix_leading}, it is evident that for $\xi_p=0$:
\begin{align}
&\langle A|(\prod_{j=N}^1 M_j^{N_j-k_j})(\prod_{j=N}^1 L_j^{k_j})\phi_p(0,1)y^{M_0}(\prod_{j=1}^N L_{-j}^{N_j-k_j})(\prod_{j=1}^N M_{-j}^{k_j})|A\rangle y^{-\xi_A}\nonumber\\
&=\langle A|(\prod_{j=N}^1 M_j^{N_j-k_j})(\prod_{j=N}^1 L_j^{k_j})\phi_p(0,1)(\prod_{j=1}^N L_{-j}^{N_j-k_j})(\prod_{j=1}^N M_{-j}^{k_j})|A\rangle\nonumber\\
&\quad +(\text{terms with atleast 1 uncontracted $M$-operators, hence giving 0})\nonumber\\
&=\langle A|(\prod_{j=N}^1 M_j^{N_j-k_j})(\prod_{j=N}^1 L_j^{k_j})(\prod_{j=1}^N L_{-j}^{N_j-k_j})(\prod_{j=1}^N M_{-j}^{k_j})|A\rangle y^{-\xi_A}\langle A|\phi_p(0,1)|A\rangle+0.
\end{align}
Thus the leading term in $\xi_A$ is the exact result in a closed-form for $\xi_p=0$. Consequently, we have
\begin{align}
\mathcal{F}_{N}(\Delta_p,\xi_p=0;\Delta_A,\xi_A|c_L,c_M|\rho)_{A\neq Id.}=\widetilde{\text{dim}}_N =\sum_{n=0}^Np(n)p(N-n).
\end{align}
Thus the `torus block' for a primary with $\xi_p=0$ is given by
\begin{align}
\mathcal{F}^{\Delta_p,\xi_p=0}_{\Delta_A,\xi_A:A\neq Id.,c_L,c_M}(\sigma,\rho)=\sum_{N=0}^{\infty}q^N\widetilde{\text{dim}}_N=\frac{q^\frac{1}{12}}{\eta(\sigma)^2}.
\end{align}
So, for $\xi_p=0$, the torus block doesn't depend on $\Delta_p$. Hence, the one-point function for a primary-field\footnote{We restrict the BMSFT to only contain primary-fields with $\Delta_p>0$, with the exception of the Identity being the only primary-field with $\Delta_p=0$.} with $\Delta_p>0$ and $\xi_p=0$ is as follows (the vacuum block doesn't contribute for $\Delta_p>0$, as $\langle 0| \phi_p(1,0)|0\rangle=\delta_{\Delta_p,0}$)
\begin{align}
\langle \phi_{\Delta_p>0,\xi_p=0}\rangle_{(\sigma, \rho)}=\frac{q^{\frac{1}{12}-\frac{c_L}{2}}y^{-\frac{c_M}{2}}}{\eta(\sigma)^2}\sum_{A\neq Id.} D(\Delta_A,\xi_A)\langle A| \phi_p(0,1)|A\rangle q^{\Delta_A}y^{\xi_A}.
\label{one_pt_zero_xi}
\end{align}

\bigskip

\subsection{Minkowski torus block from Casimir equation}
We will now digress a bit from the main theme of the paper to discuss an important point, which we will not use later in the paper. In this section we will find the global BMS torus block or the Minkowski torus block, using differential equation from the Casimir operators. Of course, the name Minkowski torus block is because the BMS$_3$ algebra, restricted to the ``global" part $n=0, \pm1$ is just $iso(2,1)$, the Poincare algebra in three bulk dimensions. We are interested, as always, in a 2d field theory with $iso(2,1)$ as its symmetries. 

\medskip

\noindent We will adopt the ideas and methods of \cite{Kraus:2017ezw} with suitable modifications to obtain the BMS versions of the analogous 2d CFT results. These methods are also similar to the ones in \cite{Bagchi:2016geg} used to obtain the BMS four-point blocks. 

\medskip

\noindent We begin by rewriting the block a bit as shown below
\bea
 \mathcal{F}_{\D_A,\xi_A,c_L,c_M}^{\D_p,\xi_p}(q,y) &=& \frac{{\rm Tr}_{\D_A,\xi_A}\left(\phi_p(u,\phi)q^{L_0-\frac{c_L}{2}}y^{M_0-\frac{c_M}{2}}\right)}{q^{\D_A-\frac{c_L}{2}}y^{\xi_A-\frac{c_M}{2}}C_{ApA}}\cr
 &=&\frac{q^{-\D_A}y^{-\xi_A}}{C_{ApA}}{\rm Tr}_{\D_A,\xi_A}\left(q^{L_0}y^{M_0}\phi_p(u,\phi)\right)\cr
 &\equiv& \frac{q^{-\D_A}y^{-\xi_A}}{C_{ApA}}\tilde{\mathcal{F}}_{\D_A,\xi_A,c_L,c_M}^{\D_p,\xi_p}(q,y).
\eea
Note that $\phi_p(u,\phi)$ is the primary field on the cylinder and instead of expressing the blocks as a function of $\sigma$ and $\rho$, we are using the variables $q=e^{2\pi i \sigma}$ and $y=e^{2\pi i \rho}$. 
We also use the cyclicity property of trace to move $q^{L_0}y^{M_0}$ to the front (also remember that $L_0$ and $M_0$ commute).

Let $P_A$ be the projector operator for the BMS module of the highest weight state $|\D_A,\xi_A\>$ i.e., it acts as an identity operator for any state in the module and annihilate any other states which does not belong to the module. It is given by
\be
P_A = \sum_{N=0}^{\infty} K_{(N)}^{ij}|A;N,i\>\<A;N,j|. 
\label{projector}
\ee 
We can use $P_A $ to project out the contribution of the BMS module of $|\D_A,\xi_A\>$ to the one-point function. So, the blocks $\tilde{\mathcal{F}}_{\D_A,\xi_A,c_L,c_M}^{\D_p,\xi_p}(q,y)$ is given by
\bea
\tilde{\mathcal{F}}_{\D_A,\xi_A,c_L,c_M}^{\D_p,\xi_p}(q,y)= {\rm{Tr}}_{\D_A,\xi_A}\left(q^{L_0} y^{M_0} \phi_p(u,\phi)  \right)={\rm{Tr}}\left(P_A\,q^{L_0} y^{M_0} \phi_p(u,\phi)  \right),
 \label{torus_block_proj}
\eea
where $\rm{Tr}$, without the subscript $\D_A$ and $\xi_A$, is trace over the basis of the whole Hilbert space of the theory.

\medskip

\noindent Just like what we have in the case of four-point conformal block, in the large central charge limit, the leading term of the torus block in \eqref{torus_block_proj} is given by tracing over states of the form $L_{-1}^k M_{-1}^q|\D_A,\xi_A\>$. So, at level $N$, we have to consider only states of the form $\sum_{k=0}^N (L_{-1})^{N-k}(M_{-1})^k|\D_A,\xi_A\>$. Likewise, the states $|A;N,i\>$ appearing in the projector in \eqref{projector} will only be of this form in the large central charge limit. Thus, the global part of $\tilde{\mathcal{F}}_{\D_A,\xi_A,c_L,c_M}^{\D_p,\xi_p}(q,y)$ is schematically given by
\bea
 \tilde{f}_{\D_A,\xi_A,c_L,c_M}^{\D_p,\xi_p}(q,y) = \text{Tr}_g\left(P_A^g\,q^{L_0} y^{M_0}\phi_p(u,\phi)  \right)=\text{Tr}_{g,A}\left(q^{L_0} y^{M_0} \phi_p(u,\phi) \right),
 \label{global_block_def}
\eea
where the index $g$ in $\text{Tr}_g$ and $P_A^g$ means that we are tracing only over states of the form $L_{-1}^k M_{-1}^q|\D,\xi\>$ and the projector also consists only of states of this form. 
Likewise, $\text{Tr}_{g,A}$ is over states of the form $L_{-1}^k M_{-1}^q|\D_A,\xi_A\>$. 

\medskip

\noindent
The global subgroup of BMS is generated by the set of generators $\{L_{0,\pm 1} 1,M_{0,\pm 1}\}$ which from a closed sub-algebra. The quadratic Casimir's of these generators are given by
\bea
\mathcal{C}_1 &=& M_0^2-M_{-1}M_1 \\
\mathcal{C}_2 &=& 2L_0M_0-\frac{1}{2}(L_{-1}M_1+L_1M_{-1}+M_1L_{-1}+M_{-1}L_1)\cr
&=& 2L_0M_0 - L_{-1}M_1 - M_{-1}L_1 -2M_0.
\eea
Since $L_{-1}$ and $M_{-1}$ commutes with $\mathcal{C}_1$ and $\mathcal{C}_2$, the states $L_{-1}^k M_{-1}^q|\D_A,\xi_A\>$ are eigenstates of these operators
\be
 \mathcal{C}_{1,2} L_{-1}^k M_{-1}^q|\D_A,\xi_A\> = \lambda_{1,2} L_{-1}^k M_{-1}^q|\D_A,\xi_A\>,
\ee
with eigenvalues given by
\be
\lambda_1 = \xi_A^2, \quad \lambda_2 = 2\xi_A(\D_A-1). 
\ee
Thus we have
\be
  \mathcal{C}_{1,2} P_A^g = \lambda_{1,2} P_A^g,\quad  P_A^g \mathcal{C}_{1,2}= \lambda_{1,2} P_A^g.
\ee
Using the above equation in \eqref{global_block_def}, we obtain
\bea
\text{Tr}_g\left( P_A^g \,\mathcal{C}_{1,2}\,q^{L_0} y^{M_0}\phi_p(u,\phi)  \right) &=& \lambda_{1,2} \tilde{f}(q,y)\cr
\implies \text{Tr}_{g,A}\left(\mathcal{C}_{1,2}\,q^{L_0} y^{M_0}\phi_p(u,\phi)  \right) &=& \lambda_{1,2} \tilde{f}(q,y),
\eea
where we denote $\tilde{f}_{\D_A,\xi_A,c_L,c_M}^{\D_p,\xi_p}(q,y)$ by $\tilde{f}(q,y)$ to reduce cluttering of indices.
In the LHS, we can move $\mathcal{C}_{1,2}$ toward the right by commuting with $q^{L_0}$ and $y^{M_0}$. For this, we will make use of the identities given below
\bea
&& L_n q^{L_0}=q^{n+L_0}L_n,\quad  L_n y^{M_0} = y^{M_0}L_n + n\log y\, y^{M_0} M_n,\quad M_n q^{L_0}=q^{L_0+n}M_n.\cr
&&
\label{identities_qy}
\eea
Using the fact that
\be
L_0 q^{L_0} = q \partial_q q^{L_0}, \quad   M_0 y^{M_0} = y \partial_y y^{M_0},
\ee
the insertion of $L_0$ and $M_0$ is same as acting with differential operator  $q \partial_q$ and $y \partial_y$
\bea
 \text{Tr}_{g,A}\left(L_0\,q^{L_0} y^{M_0}\phi_p(u,\phi)  \right) &=& q \partial_q  \text{Tr}_{g,A}\left(q^{L_0} y^{M_0}\phi_p(u,\phi)\right),\cr
  \text{Tr}_{g,A}\left(M_0\,q^{L_0} y^{M_0}\phi_p(u,\phi)  \right) &=& y \partial_y  \text{Tr}_{g,A}\left(q^{L_0} y^{M_0}\phi_p(u,\phi)\right).
\eea
For the insertion of $M_n$, we use \eqref{identities_qy} as well as the cyclicity property of trace to obtain
\bea
  \text{Tr}_{g,A}&& \hspace{-.5cm}\left(M_n\,q^{L_0} y^{M_0}\phi_p(u,\phi)  \right) =  q^n \text{Tr}_{g,A}\left(\,q^{L_0} y^{M_0}M_n\phi_p(u,\phi)  \right) \non\\
  &=& q^n \text{Tr}_{g,A}\left(\,q^{L_0} y^{M_0}\phi_p(u,\phi)M_n  \right) + q^n \text{Tr}_{g,A}\left(\,q^{L_0} y^{M_0}[M_n,\phi_p(u,\phi)]  \right) \non\\
  &=& q^n \text{Tr}_{g,A}\left(M_n\,q^{L_0} y^{M_0}\phi_p(u,\phi)M_n  \right) + q^n \text{Tr}_{g,A}\left(\,q^{L_0} y^{M_0}\mathcal{M}_n \phi_p(u,\phi) \right)\non\\
  &=& q^n \text{Tr}_{g,A}\left(M_n\,q^{L_0} y^{M_0}\phi_p(u,\phi)M_n  \right) + q^n \mathcal{M}_n \text{Tr}_{g,A}\left(\,q^{L_0} y^{M_0} \phi_p(u,\phi) \right),
\eea
where the differential operator $\mathcal{M}_n$ is defined as
\be
[M_n,\phi_p(u,\phi)] \equiv \mathcal{M}_n \phi_p(u,\phi).
\ee
Similarly, we define the differential operator $\mathscr{L}_n$ by 
\be
[L_n,\phi_p(u,\phi)] \equiv \mathscr{L}_n \phi_p(u,\phi).
\ee
Rearranging the above equation we get
\be
\text{Tr}_{g,A}\left(M_n\,q^{L_0} y^{M_0}\phi_p(u,\phi)  \right)  = \frac{q^n}{1-q^n} \mathcal{M}_n \tilde{f}(q,y).
\ee
We also have
\bea
\text{Tr}_{g,A}\left(L_n\,q^{L_0} y^{M_0}\phi_p(u,\phi)  \right) =\frac{q^n}{1-q^n}  \mathscr{L}_n \tilde{f}(q,y) + \frac{q^n n \log y}{(1-q^n)^2}\mathcal{M}_n \tilde{f}(q,y).
\eea
We could insert an additional $M$ or $L$ in the above two equations to obtain the following identities (note that $n=1,-1$)
\be
 \text{Tr}_{g,A}\left(M_{-n}M_n\,q^{L_0} y^{M_0}\phi_p(u,\phi)  \right) = \frac{1}{(1-q^n)(1-q^{-n})}\mathcal{M}_n \mathcal{M}_{-n}\tilde{f}(q,y),
\ee
\newpage
\bea
 \text{Tr}_{g,A}\left(M_{-n}L_n\,q^{L_0} y^{M_0}\phi_p(u,\phi)  \right) =&& \frac{1}{(1-q^n)(1-q^{-n})}\mathscr{L}_n\mathcal{M}_{-n}\tilde{f}(q,y) \\
 && + \frac{2n q^n y}{1-q^n}\partial_y \tilde{f}(q,y)+ \frac{n \log y}{(1-q^n)^2(1-q^{-n})}\mathcal{M}_n\mathcal{M}_{-n}\tilde{f}(q,y), \non \\
 \text{Tr}_{g,A}\left(L_{-n}M_n\,q^{L_0} y^{M_0}\phi_p(u,\phi)  \right)
 =&& \frac{1}{(1-q^n)(1-q^{-n})}\mathcal{M}_n\mathscr{L}_{-n}\tilde{f}(q,y) \\
 && + \frac{2n q^n y}{1-q^n}\partial_y \tilde{f}(q,y)- \frac{n \log y}{(1-q^n)(1-q^{-n})^2}\mathcal{M}_n\mathcal{M}_{-n}\tilde{f}(q,y).\non
\eea
Now that we have the required identities, let us find the differential equation for $\tilde{f}(q,y)$ using the quadratic Casimir $\mathcal{C}_1$. For this we have
\bea
 \text{Tr}_{g,A}\left(\mathcal{C}_1\,q^{L_0} y^{M_0}\phi_p(u,\phi) \right) &=& \text{Tr}_{g,A}\left(M_0^2\,q^{L_0} y^{M_0}\phi_p(u,\phi) \right)-\text{Tr}_{g,A}\left(M_{-1}M_1\,q^{L_0} y^{M_0}\phi_p(u,\phi) \right),\cr
 &&
 \eea
 which gives us
 \bea
 \xi_A^2 f(q,y)&=& y\partial y \left(y\partial y f(q,y)\right) - \frac{1}{(1-q)(1-q^{-1})}\mathcal{M}_{1}\mathcal{M}_{-1}f(q,y).
\eea
We know that $\tilde{f}(q,y)$ is invariant under translation. Thus 
\be
\mathcal{M}_0 \tilde{f}(q,t)=\partial_u \tilde{f}(q,t) = 0.
\ee
 So, we may add $\mathcal{M}_0^2 \tilde{f}(q,y)=0$ in the above equation
  \bea
 \xi_A^2  \tilde{f}(q,y)&=& y\partial y \left(y\partial y  \tilde{f}(q,y)\right) + \frac{1}{(1-q)(1-q^{-1})}(\mathcal{M}_0^2-\mathcal{M}_{1}\mathcal{M}_{-1})\tilde{f}(q,y).
\eea
Then using 
\be
[\mathcal{C}_1,\phi_p(u,\phi)]=(\mathcal{M}_0^2-\mathcal{M}_{1}\mathcal{M}_{-1})\phi_p(u,\phi)=\xi_p^2 \phi_p(u,\phi),
\ee
in the above differential equation, and substituting $$Y=\log y, \quad \tilde{f}(q,Y)=\frac{q}{(1-q)^2}f(q,Y)$$ we obtain{\footnote{Interestingly, the differential equation stays the same in $\tilde{f}$ as well. }}: 
\be
\partial_Y^2 f(q,Y) - \left(\xi_p^2\frac{q}{(1-q)^2}+\xi_A^2\right)f(q,Y)=0.
\ee
The general solution of this equation is given by
\be
  f(q,y) = A(q)y^{\xi_A\sqrt{1+\frac{\xi_p^2}{\xi_A^2}\frac{q}{(1-q)^2}}}+B(q)y^{-\xi_A\sqrt{1+\frac{\xi_p^2}{\xi_A^2}\frac{q}{(1-q)^2}}}
\ee
In the above, we have resubistuted $Y$ in terms of $y$. We know that $ f(q,y)$ varies like $y^{\xi_A}$. Thus, we have to drop the second term in the above equation which is not of this form. Thus, we have
\be
  f(q,Y) = A(q)e^{Y\sqrt{\xi_A^2+\xi_p^2\frac{q}{(1-q)^2}}}=A(q)y^{\sqrt{\xi_A^2+\xi_p^2\frac{q}{(1-q)^2}}}.
\ee

\bigskip
\noindent Next, let us look at the differential equation obtained from the Casimir $\mathcal{C}_2$
\bea
2\xi_A(\D_A-1) \tilde{f}(q,y) &=& \text{Tr}_{g,A}\left((2L_0M_0 - L_{-1}M_1 - M_{-1}L_1 -2M_0)\,q^{L_0} y^{M_0}\phi_p(u,\phi)  \right)\cr
&=& 2q\partial_q \partial_Y \tilde{f}(q,Y) - 2\frac{1+q}{1-q}\partial_Y\tilde{f}(q,Y)+\frac{q}{(1-q)^2}(\mathcal{M}_1\mathscr{L}_{-1}+\mathscr{L}_1\mathcal{M}_{-1})\tilde{f}(q,Y)\cr
&& +Y\frac{q(1+q)}{(1-q)^3}\mathcal{M}_1\mathcal{M}_{-1}\tilde{f}(q,Y).
\eea
Translational invariance of $\tilde{f}(q,Y)$ gives us $\mathcal{M}_0\tilde{f}(q,Y)=\mathcal{M}_0^2\tilde{f}(q,Y)=\mathcal{M}_0\mathscr{L}_0\tilde{f}(q,Y)=0$. Thus, the above equation is same as
\bea
 && 2q\partial_q \partial_Y \tilde{f}(q,Y) - 2\frac{1+q}{1-q}\partial_Y\tilde{f}(q,Y)+\frac{q}{(1-q)^2}(-2\mathcal{M}_0\mathscr{L}_0 + \mathcal{M}_1\mathscr{L}_{-1}+\mathscr{L}_1\mathcal{M}_{-1} + 2 \mathcal{M}_0)\tilde{f}(q,Y)\cr
&& +Y\frac{q(1+q)}{(1-q)^3}(-\mathcal{M}_0^2+\mathcal{M}_1\mathcal{M}_{-1})f(q,Y)=2\xi_A(\D_A-1) f(q,y).
\eea
Then we substitute
\bea
&&(\mathcal{M}_0^2-\mathcal{M}_1\mathcal{M}_{-1})\tilde{f}(q,Y)=\xi_p^2 \tilde{f}(q,Y)\cr
&&(2\mathcal{M}_0\mathscr{L}_0 - \mathcal{M}_1\mathscr{L}_{-1}-\mathscr{L}_1\mathcal{M}_{-1} -2 \mathcal{M}_0)\tilde{f}(q,Y) = 2\xi_p(\D_p-1) \tilde{f}(q,Y), 
\eea
in the above equation to obtain
\be
q\partial_q \partial_Y \tilde{f}(q,Y) - \frac{1+q}{1-q}\partial_Y\tilde{f}(q,Y)-\left(\frac{q}{(1-q)^2} \xi_p(\D_p-1) +Y\frac{q(1+q)}{(1-q)^3}\frac{\xi_p^2}{2}\right)\tilde{f}(q,y)= \xi_A(\D_A-1) \tilde{f}(q,Y).
\ee
Substituting $\tilde{f}(q,Y)=\frac{q}{(1-q)^2}f(q,Y)$, the differential equation simplifies to 
\be
q\partial_q \partial_Y f(q,Y) -\left(\frac{q}{(1-q)^2} \xi_p(\D_p-1) +Y\frac{q(1+q)}{(1-q)^3}\frac{\xi_p^2}{2}\right)f(q,y)= \xi_A(\D_A-1) f(q,Y).
\label{diff_eqn_sim_1}
\ee
Now
\be
\partial_Y f(q,Y)= \sqrt{\xi_A^2+\xi_p^2\frac{q}{(1-q)^2}} f(q,Y) \equiv \tilde{A}(q) e^{Y\sqrt{\xi_A^2+\xi_p^2\frac{q}{(1-q)^2}}},
\ee
where
\be
 \tilde{A}(q) = \sqrt{\xi_A^2+\xi_p^2\frac{q}{(1-q)^2}} A(q).
\ee
Then we have
\be
q\partial_q \partial_Y f(q,Y)=  \frac{Y q \left(1+q\right) \xi _p^2 e^{Y \sqrt{\xi _A^2+\frac{q \xi _p^2}{\left(1-q\right)^2}}}}{\left(1-q\right)^3 \left(2 \sqrt{\xi _A^2+\frac{q \xi _p^2}{\left(1-q\right)^2}}\right)} \tilde{A}(q) + e^{Y\sqrt{\xi_A^2+\xi_p^2\frac{q}{(1-q)^2}}}\, q \partial_q \tilde{A}(q) 
\ee
Using this, the differential equation \eqref{diff_eqn_sim_1} further simplifies to
\be
q\partial_q \tilde{A}(q)-\frac{1}{\sqrt{\xi_A^2+\xi_p^2\frac{q}{(1-q)^2}}}\left( \frac{q}{(1-q)^2}\xi_p(\D_p-1) +\xi_A(\D_A-1)\right) \tilde{A}(q)=0.
\label{diff_eqn_sim_2}
\ee
Let $\tilde{A}(q)=G(q)\tilde{B}(q)$ such that 
\be
q\partial_q G(q) = \frac{q}{(1-q)^2\sqrt{\xi_p^2+\xi_p^2\frac{q}{(1-q)^2}}}\xi_p(\D_p-1) G(q).   
\ee
We can solve this to find $G(q)$ which is given by
\bea
G(q)
&=& K_1\left(2 \sqrt{\xi _A^2+\frac{q \xi _p^2}{(1-q)^2}}+\xi_p\frac{1+q}{1-q}\right)^{\D_p-1},
\eea
where $K_1$ is a constant.
Substituting $\tilde{A}(q)=G(q)\tilde{B}(q) \tilde{B}(q)$ in \eqref{diff_eqn_sim_2}, we get the differential equation for $\tilde{B}(q)$
\be
q \partial_q \tilde{B}(q) -  \frac{1}{\sqrt{\xi_A^2+\xi_p^2\frac{q}{(1-q)^2}}}\xi_A(\D_A-1)\tilde{B}(q)=0,
\ee
whose solution is
\be
\tilde{B}(q)=K_2\xi _p^{2(-\D_A+1)}\left(\sqrt{(1-q)^2 \xi _A^2+q \xi _p^2}+(q+1) \xi _A\right)^{2(-\D_A+1)}q^{\D_A-1}
\ee
Combining all our results we have
\bea
\tilde{f}(q,y) &=& \frac{K q}{(1-q)^2\sqrt{\xi_A^2+\xi_p^2\frac{q}{(1-q)^2}}}y^{\sqrt{\xi_A^2+\xi_p^2\frac{q}{(1-q)^2}}} \left(2 \sqrt{\xi _A^2+\frac{q \xi _p^2}{(1-q)^2}}+\xi_p\frac{1+q}{1-q}\right)^{\D_p-1}\cr
&&\times \,\left(\sqrt{(1-q)^2 \xi _A^2+q \xi _p^2}+(q+1) \xi _A\right)^{-2(\D_A-1)}q^{\D_A-1}.\cr
&&
\label{fin_sol_tgb}
\eea

\bigskip

\noindent Finally we would like to find the integration constant $K$. We will do this by expanding the above solution around $q=0$ and compare the first few terms with the contribution from level $0$ and $1$ to $\tilde{f}(q,y)$. The expansion of \eqref{fin_sol_tgb} up to order in $q^1$ (mod $q^{\D_A}$) is given by
\be
K\,2^{2-2 \Delta _A} \xi _A^{1-2 \Delta _A} \left(2 \xi _A+\xi _p\right){}^{\Delta _p-1} q^{\D_A}y^{\xi_A}\left(1+q\left(2+\frac{\xi_p(\D_p-1)}{\xi_A}+\pi i \rho \frac{\xi_p^2}{\xi_A}-\frac{\D_A}{2\xi_A^2}\xi_p^2\right)+\mathcal{O}(q^2)\right).
\ee
Each level comes with a factor of $q^{N+\D_A}$ and the contribution of level 0 and level 1 to the global block $\tilde{f}(q,y)$ is
\be
C_{ApA}\,q^0 q^{\D_A} y^{\xi_A}  + C_{ApA}\,q^{\D_A +1 }y^{\xi_A}\left(2+\frac{\xi_p(\D_p-1)}{\xi_A}+\pi i \rho \frac{\xi_p^2}{\xi_A}-\frac{\D_A}{2\xi_A^2}\xi_p^2\right).  
\ee
Comparing this with the above equation we can see that
\be
K =  C_{ApA}\,4^{\Delta _A-1} \xi _A^{2 \Delta _A-1}\left(2 \xi _A+\xi _p\right){}^{1-\Delta _p}. 
\ee
Substituting this back in \eqref{fin_sol_tgb}, the form of the global torus block is given by
\bea
 \tilde{f}_{\D_A,\xi_A,c_L,c_M}^{\D_p,\xi_p}(q,y) &=& C_{ApA}\,4^{\Delta _A-1} \xi _A^{2 \Delta _A-1}\left(2 \xi _A+\xi _p\right){}^{1-\Delta _p} \left(\sqrt{(1-q)^2 \xi _A^2+q \xi _p^2}+(q+1) \xi _A\right)^{-2(\D_A-1)} \cr
 && \times \frac{q^{\D_A}}{(1-q)^2\sqrt{\xi_A^2+\xi_p^2\frac{q}{(1-q)^2}}}y^{\sqrt{\xi_A^2+\xi_p^2\frac{q}{(1-q)^2}}} \left(2 \sqrt{\xi _A^2+\frac{q \xi _p^2}{(1-q)^2}}+\xi_p\frac{1+q}{1-q}\right)^{\D_p-1}\cr
&&
\eea 
This is our final expression for the global BMS torus block, or the Minkowski torus block. 

\medskip
Before ending this subsection, we would like to point out that there is some potential subtlety between the BMS global blocks and the large $c_M$ limit of the BMS blocks. The expectation from 2d CFT on general grounds is that in the large central charge limit, since the norm of the states at level 2 and higher depend inversely on the central charge, the Virasoro blocks would reduce to the global blocks. However, as shown in \cite{Alkalaev:2016fok}, this is not true for the torus CFT in contrast to the CFT on the sphere. One may introduce the so-called ``light blocks" which are defined as the leading approximation in $1/c$ expansion. On the other hand, by truncating Virasoro algebra to its global $sl(2)$ sub-algebra one is led to global block. On the sphere, these two functions are equal, while on torus and higher genus surfaces these are different. It is likely that BMS torus blocks share the same property, i.e. one generally distinguishes between BMS global and BMS light blocks for field theories defined on higher genus surfaces. We thank Konstantin Alkalaev for pointing this out to us. We hope to come back to this issue in the future and resolve it.    

\bigskip \bigskip

\newpage

\section{Structure constants for BMS highest weight primaries}\label{onepointsec}
We derived expressions for the BMS torus blocks in the previous section. Armed with this knowledge, we now move on to calculating the asymptotic behaviour of three-point structure constants $C_{ApA}$ for primary fields in the highest weight representations. We will evaluate this in the large $\xi_A$ limit, which was also the limit we used to derive the expressions of the BMS torus blocks.  

\medskip

\noindent We note that $C_{ApA}$ is the coordinate independent constant in the three point function $\<\phi_{\D_A,\xi_A}(t_1,x_1)\phi_p(t_2,x_2)\phi_{\D_A,\xi_A}(t_3,x_3)\>$. The analysis in this section will closely follow to the one used in Section \ref{onepointsec1}. We will find the asymptotic formula for a generic primary field $\phi_p$. First of all let us rewrite the S-modular transformation \eqref{s_mod_trans} as
\bea \label{701}
 \<\phi_{\D,\xi}\>_{(\sigma,\rho)}&=&\sigma^{-\D} e^{\frac{\xi\rho}{\sigma}}\left<\phi_{\D,\xi}\right>_{(-\frac{1}{\sigma},\frac{\rho}{\sigma^2})}.
\eea
We expand the left hand side (LHS) of \refb{701} in terms of the torus blocks of the primaries, where as the right hand side (RHS) of \refb{701} is expanded using the eigenbasis of $L_0$ and $M_0$ as in Section \ref{onepointsec1}. This leads to
\bea
&&e^{-2\pi i (\sigma c_L/2+\rho c_M/2)}\sum_{A} D(\D_A,\xi_A) e^{2\pi i \sigma \D_A}  e^{2\pi i \rho \xi_A} C_{ApA}\,\mathcal{F}_{\D_A,\xi_A,c_L,c_M}^{\D_p,\xi_p}(\sigma,\rho)\non\\
&& \quad =\sigma^{-\Delta_p}e^{\xi_p\frac{\rho}{\sigma}}\sum_{i} D(\Delta_i,\xi_i)C_{ipi}
e^{-\frac{2\pi i}{\sigma}(\Delta_i-\frac{c_L}{2})}e^{2\pi i\frac{\rho}{\sigma^2}(\xi_i-\frac{c_M}{2})}. 
\eea
We will choose $\sigma=i\beta$ and consider the limit $\beta\rightarrow0^+$ and $|\rho|\rightarrow0^+$. In this limit, as before, we replace the LHS by a double integral while the RHS is dominated by contribution from the lightest field $\chi$ such that $C_{\chi p \chi}\neq 0$:
\bea
&& e^{2\pi \beta \frac{c_L}{2}}e^{-2\pi i \rho c_M/2}\int d\D_A \int d\xi_A T_p(\D_A,\xi_A) e^{-2\pi \beta \D_A}  e^{2\pi i \rho \xi_A} \,\mathcal{F}_{\D_A,\xi_A,c_L,c_M}^{\D_p,\xi_p}(i\beta,\rho)\non\\
 && \quad  \approx (i\beta)^{-\Delta_p} e^{-i\xi_p\frac{\rho}{\beta}} T_p(\Delta_\chi,\xi_\chi)e^{-\frac{2\pi }{\beta}(\Delta_\chi-\frac{c_L}{2})}e^{-2\pi i\frac{\rho}{\beta^2}(\xi_\chi-\frac{c_M}{2})}.
 \label{non_zero_xip_ti}
\eea 
Here $T_i(\D_A,\xi_A) \equiv D(\D_i,\xi_i)C_{ipi}$. For small $\beta$, the LHS is dominated by torus blocks with large $\xi_A$. So, to a good approximation, we may replace the integral by integral over only large $\xi_A$. From \eqref{f_leading},  for large $\xi_A$ we have
\be
\mathcal{F}_{\D_A,\xi_A,c_L,c_M}^{\D_p,\xi_p}(\sigma,\rho)
=\frac{q^\frac{1}{12}}{\eta(\sigma)^2} + \mathcal{O}\left(\frac{1}{\xi_A}\right)+ \ldots 
\Rightarrow \mathcal{F}_{\D_A,\xi_A,c_L,c_M}^{\D_p,\xi_p}(i\beta,\rho)= \frac{e^{-\frac{2\pi\beta}{12}}}{\eta(i\beta)^2}+ \mathcal{O}\left(\frac{1}{\xi_A}\right)+ \ldots 
\ee
For the special case of $\xi_p=0$, as described in Sec~\ref{tbxip0}, there are no subleading corrections to the block. For a generic $\xi_p \neq 0$, we will work also in the leading $\xi_A$ approximation. So we will keep only the first term in the above equation and neglect the $\mathcal{O}(\frac{1}{\xi_A})$ term and other higher order terms in $\xi_A$. For this to be possible i.e., for the subleading and other higher order terms to be suppressed in comparison to the leading term, we need the additional requirement that $\beta^2 \xi_A \gg1$. This is shown in Appendix \ref{justif}. 

\medskip 

\noindent Next, we use the small $\beta$ asymptotic of the Dedekind eta-function to obtain
\be
 \mathcal{F}_{\D_A,\xi_A,c_L,c_M}^{\D_p,\xi_p}(i\beta,\rho) \approx 2\pi \beta \exp \left\{-\frac{\pi}{6}\left(\beta-\frac{1}{\beta}\right)\right\}+... 
\ee 
Substituting the above expression in \eqref{non_zero_xip_ti}, we have
\bea
&& e^{2\pi \beta \frac{c_L}{2}}e^{-2\pi i \rho \frac{c_M}{2}}2\pi \beta e^{\frac{\pi}{6}\left(-\beta+\frac{1}{\beta}\right)}\int d\D_A \int d\xi_A T_p(\D_A,\xi_A) e^{-2\pi \beta \D_A}  e^{2\pi i \rho \xi_A}\non\\
&& \quad \approx (i\beta)^{-\Delta_p} e^{-i\xi_p\frac{\rho}{\beta}} \ T_p(\Delta_\chi,\xi_\chi)e^{-\frac{2\pi }{\beta}(\Delta_\chi-\frac{c_L}{2})}e^{-2\pi i\frac{\rho}{\beta^2}(\xi_\chi-\frac{c_M}{2})}\non
\eea
This leads to
\bea
\int d\D_A^{\prime} \int d\xi_A^{\prime} \ T_p(\D_A^{\prime},\xi_A^{\prime}) e^{-2\pi \beta \D_A^{\prime}}  e^{2\pi i \rho \xi_A^{\prime}}
 \approx && \frac{\beta^{-\Delta_p-1}}{2\pi i^{\Delta_p}} T_p(\Delta_\chi,\xi_\chi) e^{2\pi \beta (-\frac{c_L}{2}+\frac{1}{12})} \\
 && \times e^{-2\pi i \rho (\frac{1}{\beta^2}(\xi_\chi-\frac{c_M}{2})-\frac{c_M}{2}+\frac{\xi_p}{2\pi \beta})}e^{-\frac{2\pi }{\beta}(\Delta_\chi-\frac{c_L}{2}+\frac{1}{12})} .\non
 \eea 
We multiply both sides with $e^{-2\pi i \xi_A (\text{Re}\,\rho)}$ and integrate over real part of $\rho$. On the LHS we have a $\delta(\xi_A^{\prime}-\xi_A)$ which when integrated over $\xi_A^{\prime}$ will give us  
$${\text{LHS}} = \int d\D_A^{\prime} T_p(\D_A^{\prime},\xi_A) e^{-2\pi \beta \D_A^{\prime}} e^{-2\pi (\text{Im}\,\rho) \xi_A}.$$ 
On the RHS, the integral over $(\text{Re}\,\rho)$ will give us a delta-function. Specifically, we have
\bea
\int_0^{\infty} d\D_A^{\prime}  T_p(\D_A^{\prime},\xi_A) e^{-2\pi \beta \D_A^{\prime}} &\approx& \frac{\beta^{-\Delta_p-1}}{2\pi i^{\Delta_p}} e^{2\pi \beta (-\frac{c_L}{2}+\frac{1}{12})}e^{-\frac{2\pi }{\beta}(\Delta_\chi-\frac{c_L}{2}+\frac{1}{12})} \ T_p(\Delta_\chi,\xi_\chi)\\
&& {\hspace{-2cm}} \times e^{2\pi (\text{Im}\,\rho)(\frac{1}{\beta^2}(\xi_\chi-\frac{c_M}{2})-\frac{c_M}{2}+\frac{\xi_p}{2\pi \beta}+\xi_A)}\delta\left\{\frac{1}{\beta^2}(\xi_\chi-\frac{c_M}{2})-\frac{c_M}{2}+\frac{\xi_p}{2\pi \beta}+\xi_A\right\}. \non
\eea
The two roots of the argument of the $\delta$-function are 
\be
\beta_{\pm}=\frac{-\frac{\xi _p}{2 \pi }\pm\sqrt{\left(\frac{\xi _p}{2 \pi }\right)^2+4 \left(\frac{c_M}{2}-\xi _{\chi }\right)\left(\xi _A-\frac{c_M}{2}\right)}}{2 \left(\xi _A-\frac{c_M}{2}\right)}.
\label{roots_beta_3} 
\ee
Using the property of the $\delta$-function we have
\bea
\int_0^{\infty} d\D_A^{\prime}  T_p(\D_A^{\prime},\xi_A) e^{-2\pi \beta \D_A^{\prime}}  = && \sum_{*=+,-}\frac{\beta^{-\Delta_p-1}}{2\pi i^{\Delta_p}}e^{2\pi \beta_* (-\frac{c_L}{2}+\frac{1}{12}) -\frac{2\pi }{\beta_*}(\Delta_\chi-\frac{c_L}{2}+\frac{1}{12})} T_p(\Delta_\chi,\xi_\chi) \non\\ 
&& \qquad \times e^{2\pi (\text{Im}\,\rho)}\frac{\delta(\beta-\beta_\star)}{|\frac{c_M-2\xi_\chi}{\beta_\star^3}-\frac{\xi_p}{2\pi\beta_\star^2}|}.
\eea
We then invert the above Laplace-transformation using the analytic continuation of the $\delta$-function \eqref{delta_ac} to obtain
\bea
\frac{T_p(\Delta_A,\xi_A)}{T_p(\Delta_\chi,\xi_\chi)} &=&\sum_{*=+,-}\frac{\beta^{-\Delta_p-1}}{2\pi i^{\Delta_p}}{\frac{e^{2\pi \beta_* (-\frac{c_L}{2}+\frac{1}{12})}e^{-\frac{2\pi }{\beta_*}(\Delta_\chi-\frac{c_L}{2}+\frac{1}{12})}}{|\frac{c_M-2\xi_\chi}{\beta_\star^3}-\frac{\xi_p}{2\pi\beta_\star^2}|}} \int_{b-i\infty}^{b+i\infty} \frac{d\beta}{i}\frac{e^{2\pi \beta\Delta_A^\prime}}{2\pi i (\beta-\beta^{*})}\cr
&=& \sum_{*=+,-}\frac{1}{2\pi}(i\beta_*)^{-\Delta_p-1} e^{2\pi \beta_* (-\frac{c_L}{2}+\frac{1}{12})}e^{-\frac{2\pi }{\beta_*}(\Delta_\chi-\frac{c_L}{2}+\frac{1}{12})} \frac{e^{2\pi \beta_*\Delta_A}}{|\frac{c_M-2\xi_\chi}{\beta_\star^3}-\frac{\xi_p}{2\pi\beta_\star^2}|}.\cr
&&
\eea
Note that we start out with the condition that $\beta$ is small and $\xi_A$ is large. Indeed, from \eqref{roots_beta_3} we see that for $\xi_A \gg \frac{\xi_p}{2\pi} \pm \frac{c_M}{2}$, $\beta_{\pm}$ is small. We also assume $\beta$ to be real for which we require  $\xi_{\chi} < \frac{c_M}{2}$. In these limit $$\beta_\pm\approx-\frac{\xi_p}{4\pi\xi_A}\pm\sqrt{\frac{{c_M}-2\xi_\chi}{2\xi_A}}.$$ 
We also require the contribution from $\beta_+$ to dominate and the contribution from $\beta_-$ to be negligible. This is achieved by the condition that $$\Delta_{\chi}<\frac{c_L}{2}-\frac{1}{12}.$$ So, we have 
\bea
 \frac{T_p(\Delta_A,\xi_A)}{T_p(\Delta_\chi,\xi_\chi)}  
 &\approx &  \frac{1}{2\pi}\frac{(-i)^{\Delta_p+1}}{\beta_+^{\D_P-1}} e^{2\pi \beta_+ (\D_A-\frac{c_L}{2}+\frac{1}{12})}e^{\frac{2\pi }{\beta_+}(\frac{c_L}{2}-\Delta_\chi-\frac{1}{12})} \frac{1}{\frac{c_M-2\xi_\chi}{\beta_+}-\frac{\xi_p}{2\pi}}. \cr
 &\approx& \frac{1}{2\pi}\frac{(-i)^{\D_p+1}}{2\xi_A}\left(\frac{\xi_A}{\frac{c_M}{2}-\xi_{\chi}}\right)^{\frac{\D_p}{2}}e^{\frac{\xi_p}{2}\left(-\frac{\D_A-\frac{c_L}{2}+\frac{1}{12}}{\xi_A}-\frac{\frac{c_L}{2}-\Delta_\chi-\frac{1}{12}}{\frac{c_M}{2}-\xi_{\chi}}\right)}\cr
 &&\times e^{2\pi\left(\sqrt{\frac{\frac{c_M}{2}-\xi_{\chi}}{\xi_A}}(\D_A-\frac{c_L}{2}+\frac{1}{12})+\sqrt{\frac{\xi_A}{\frac{c_M}{2}-\xi_{\chi}}})(\frac{c_L}{2}-\Delta_{\chi}-\frac{1}{12})\right)}.
 \label{onepoint_non_zero_xi}
\eea
To find the asymptotic formula for density of states in the large $\xi_A$ limit, we put $\Delta_p=0=\xi_p$ directly in \eqref{onepoint_non_zero_xi} and also replace the primary field $\chi$ by the lightest primary field $\O$ present in the theory
\bea
\frac{D(\Delta_A,\xi_A)}{D(\Delta_\O,\xi_\O)}
 &\approx& - \frac{i}{4\pi\xi_A}
 e^{2\pi\left[(\D_A-\frac{c_L}{2}+\frac{1}{12}) \sqrt{\frac{\frac{c_M}{2}-\xi_\O}{\xi_A}}+(\frac{c_L}{2}-\Delta_\O-\frac{1}{12})\sqrt{\frac{\xi_A}{\frac{c_M}{2}-\xi_\O}}\right]}. \label{dos}
\eea
It is satisfying to note that in the limit $\D_A\gg c_L$ and with $\O$ as the vacuum (so $\D_\O= \xi_\O = 0$), taking the logarithm of the above equation \refb{dos}, we reproduce the Cardy formula for BMS primaries, which was worked out in the saddle point method in \cite{Bagchi:2019unf}. 
Substituting \refb{dos} back in \eqref{onepoint_non_zero_xi} we obtain the asymptotic formula of three-point coefficient for primary fields for large $\xi_A$

\bigskip

\noindent {\fbox{
\addtolength{\linewidth}{+2\fboxsep}%
 \addtolength{\linewidth}{+2\fboxrule}%
 \begin{minipage}{\linewidth}
 \bea \label{q}
C_{ApA}&\approx&(-i)^{\D_p}C_{\chi A \chi}\frac{D(\D_{\chi},\xi_{\chi})}{D(\D_\O,\xi_\O)}\left(\frac{\xi_A}{\frac{c_M}{2}-\xi_{\chi}}\right)^{\frac{\D_p}{2}}e^{\frac{\xi_p}{2}\left(-\frac{\D_A-\frac{c_L}{2}+\frac{1}{12}}{\xi_A}-\frac{\frac{c_L}{2}-\Delta_\chi-\frac{1}{12}}{\frac{c_M}{2}-\xi_{\chi}}\right)} \times \\
 &&e^{2\pi\left(\sqrt{\frac{\xi_A}{\frac{c_M}{2}-\xi_\chi}}(\frac{c_L}{2}-\frac{1}{12}-\Delta_\chi)-\sqrt{\frac{\xi_A}{\frac{c_M}{2}-\xi_\O}}(\frac{c_L}{2}-\frac{1}{12}-\Delta_\O)+(\sqrt{\frac{\frac{c_M}{2}-\xi_\chi}{\xi_A}}-\sqrt{\frac{\frac{c_M}{2}-\xi_\O}{\xi_A}})(\Delta_A+\frac{1}{12}-\frac{c_L}{2})\right)}. \non\\\non
\eea
\end{minipage}
}}

\bigskip

\noindent If we consider the limit where $c_M\gg \xi_\chi, \xi_\O$, then we see that the asymptotic structure constants for the BMS primaries matches the general analysis we have considered in earlier sections (e.g. comparing \refb{q} and \refb{impr}), with the identification 
\be \label{shift}
c_L \to c_L -\frac{1}{6}.
\ee
The shift in central charge due to descendants is similar to that which has been observed in 2d CFTs. This shift can be viewed as a one-loop renormalisation of the bulk central charge due to the presence of BMS descendants. It is interesting to see that the central term $c_M$ does not get shifted. This is reminiscent of the recent analysis of \cite{Merbis:2019wgk}, where the authors also found a one-loop shift of only $c_L$ and not $c_M$. In fact the shift of the central charge $c_L$ found in \cite{Merbis:2019wgk} exactly matches with our result above \refb{shift} (after accounting for the difference in normalisations between our work and \cite{Merbis:2019wgk}).{\footnote{The authors in \cite{Merbis:2019wgk} speak about two different shifts in the central extension $c_L$ corresponding to two different classes of co-adjoint orbits of the BMS group, viz. the vacuum orbit and a generic orbit. The case we are interested in is the generic orbit since on the boundary side we are interested in thermal states and in the bulk dual we deal with the FSC solution, and not the vacuum Minkowski solution.}}

\bigskip

\newpage

\section{A Quick Bulk analysis}\label{bulk}
\subsection{Flatspace Cosmologies}
In a holographic duality, the thermal states on the field theory are equivalent to geometries with horizons in the bulk dual. For a 2d CFT, thermal states are dual to BTZ black holes in AdS$_3$. For non-extremal BTZ solutions, the metric is given by
\be \label{btz}
ds^2_{\text{\tiny BTZ}} = - \frac{(r^2 - r_+^2)(r^2 - r_-^2)}{r^2 \ell^2}dt^2 + \frac{r^2 \ell^2}{(r^2 - r_+^2)(r^2 - r_-^2)}dr^2 + r^2\left(d\phi - \frac{r_+ r_-}{\ell r^2} dt\right)^2. 
\ee
Here $r_\pm$ are the outer and inner horizons, which are given in terms of the mass ($M$) and angular momentum ($J$) of the BTZ black hole by 
\be
r_\pm = \sqrt{2G\ell(\ell M +J)} \pm  \sqrt{2G\ell(\ell M -J)}.
\ee
In the above, $\ell$ is the radius of AdS$_3$. BTZ black hole solutions are locally AdS$_3$ and can be viewed as orbifolds of AdS$_3$. 

\begin{figure}[t]
\begin{center}
\includegraphics[scale=.022]{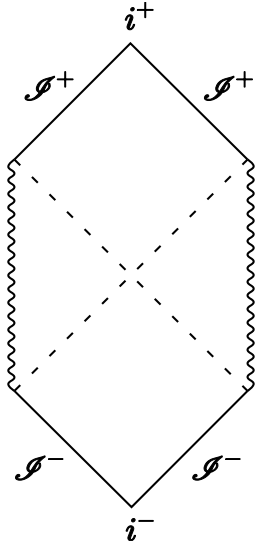}
\end{center}
\caption{Penrose diagram for Flat Space Cosmologies.}
\label{fig2} 
\end{figure}

The holographic duals of thermal states in 2d BMSFTs are the so-called Flat Space Cosmologies (FSC). The metric for these cosmological solutions are given by 
\be\label{fsc1}
ds^2_{\text{\tiny FSC$_1$}} = M du^2 - 2du dr + J du d\psi +r^2 d\psi^2
\ee 
Here $u$ is the retarded time and $M$ and $J$ label the mass and angular momentum of these solutions. These are the zero modes of the most generic metric 
\be
ds^2_{\text{\tiny FSC$_1$}} = \Theta(\phi) du^2 - 2du dr + \left[\Xi(\phi) + \p_\phi \Theta(\phi)\right] du d\psi +r^2 d\psi^2
\ee
for asymptotically flat boundary conditions for which one gets the BMS$_3$ algebra as the asymptotic symmetry algebra on the null boundary. 
$\Theta(\phi)$ and $\Xi(\phi)$ are functions called the mass and angular momentum aspects which reduce to $M$ and $J/2$ for the FSC solution. FSC solutions are locally flat and can also be obtained by quotienting 3d Minkowski spacetime by a boost and a translation. For this reason, they are also known as shifted-boost orbifolds \cite{Cornalba:2003kd}. The Penrose diagram of FSC spacetimes is given in Figure \ref{fig2}. 

\medskip

\noindent Minkowski spacetimes can be viewed as an infinite radius of AdS.  One can similarly obtain the FSC solutions by starting with \refb{btz} and taking $\ell \to \infty$. The metric here, written in $(t,r, \phi)$ coordinates, takes the form
\be\label{fsc2}
ds^2_{\text{\tiny FSC$_2$}}=\hat{r}^2_{+}dt^2-\frac{r^2}{\hat{r}^2_{+}(r^2-r_0^2)}dr^2 + 2\hat{r}_+ r_0 d\phi dt + r^2 d\phi^2, 
\ee
where
\be
\hat{r}_+ = \sqrt{8GM},\,\,\,r_0 = \sqrt{\frac{2G}{M}}J. 
\ee
Going from \refb{fsc1} to \refb{fsc2} is done by the following change of variables:
\be
d\psi = d\phi + \frac{r_0 dr}{\hat{r}_+(r^2-r_0^2)}, \quad du = dt + \frac{r^2 dr}{\hat{r}_+^2(r^2-r_0^2)}
\ee
The cosmological nature of the metric \refb{fsc2} manifests itself in the limit. When $\ell \to \infty$, the outer horizon of the original BTZ black hole goes out to infinity leaving behind just the interior of the black hole as the whole solution. Here thus the roles of the radial and temporal directions flip. The solution \refb{fsc2} is dependent on $r$, which in this case is the temporal direction.  $r=r_0$ is the remnant of the inner BTZ horizon that turns into a cosmological horizon in this singular limit. 

To understand the cosmological nature of the FSC intrinsically without the aid of a limiting procedure, it is best to look at another coordinate system where the metric can be rewritten as
\be\label{fsc3}
ds^2_{\text{\tiny FSC$_3$}}=-d\t^2 + \frac{(E \t)^2}{1+ (E \t)^2} dx^2 + \left(1+ (E \t)^2\right)\left( dy + \frac{(E \t)^2}{1+ (E \t)^2} dx\right)^2
\ee
In the above, $y$ is identified as $y \sim y + 2\pi r_0$. It can be checked that the above is a solution to vacuum Einstein's equation in three dimensions with a vanishing cosmological constant. The obvious temporal dependence makes this a cosmological solution. For positive $\t$, \refb{fsc3} describes an expanding universe from a cosmological horizon at $\t=0$. The map between \refb{fsc3} and \refb{fsc2} is the following
\be
\hat{r}_+ t = x, \quad r_0 \phi = y+x, \quad \left(\frac{r}{r_0}\right)^2 = 1+ (E \t)^2, \, \text{with} \, E= \frac{\hat{r}_+}{r_0}.
\ee
For our calculations in the next subsection, we will focus on using \refb{fsc2} as our metric of choice. 

\subsection{One point functions in FSC background}
We now wish to calculate $\<E|O|E\>$, for high energy $E$, from the bulk side. The dual to the finite temperature BMSFT thermal state is a FSC solution. The BMS weights of the FSC are 
\be
\xi_{\text{\tiny FSC}}=M+\frac{c_M}{2},\,\,\,\D_{\text{\tiny FSC}}=J+\frac{c_L}{2}. 
\ee
We also know that for 3d Einstein gravity $c_L=0,\,\,c_M=\frac{1}{4G}$. For large $M$ and $J$, we have
\be
\xi_{\text{\tiny FSC}} \approx M,\,\,\, r_0 \approx \frac{\D_{\text{\tiny FSC}}}{\sqrt{2c_M \xi_{\text{\tiny FSC}}}},\,\,\, \hat{r}_+ \approx \sqrt{\frac{2\xi_{\text{\tiny FSC}}}{c_M}}.
\ee
It is of interest to note here that with these identifications of the BMS weights of the FSC, the Bekenstein-Hawking entropy associated with the cosmological horizon matches precisely with the BMS-Cardy formula \refb{bcar} presented in Sec.~\refb{bmsft} \cite{Bagchi:2012xr}: 
\be
S_{\text{\tiny{Bekenstein-Hawking}}} = \frac{\text{Area of horizon}}{4G} = \frac{2 \pi r_0}{4G} = S_{\mbox{\tiny{BMS-Cardy}}}. 
\ee
We now wish to take this correspondence one step further and match the asymptotic formula for the structure constants that we have obtained to a bulk analysis. 

\bigskip

\noindent Our object of interest, $\<E|O|E\>$, in the bulk side denotes the calculation of a one-point function of a light operator $O$, or a probe, in the background of a heavy state $|E\>$, which is given by a FSC solution. Of course, we don't expect a single microstate to have a geometric description, so in a sense the FSC geometry arises on course graining over a family of such microstates. We are working in the probe limit and hence we will discount the backreaction of $O$ on the FSC geometry. 

\begin{figure}[t]
\begin{center}
\includegraphics[scale=1]{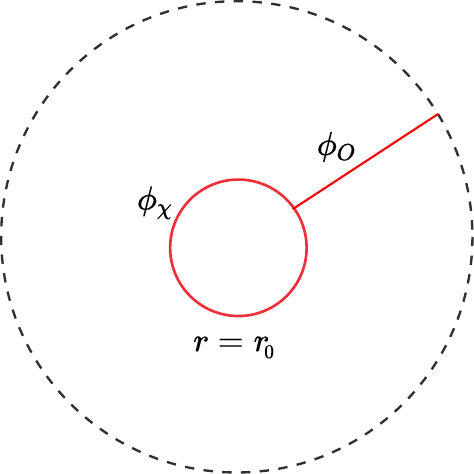}
\end{center}
\caption{One-loop contribution to $\<E|O|E\>$}
\label{oneloop}
\end{figure}

\medskip

\noindent We are interested in the contribution to $\<E|O|E\>$ that comes from the diagram shown in Figure \ref{oneloop}. The field $\phi_O$ comes from boundary at infinity and splits into a pair of $\phi_{\chi}$ which wraps around the cosmological horizon $r_0$. We are working in the probe limit where $\xi_O,\xi_{\chi} \gg 1$ but these fields are light in the sense that $\xi_O,\xi_{\chi} \ll c_M$. We also have $\xi_{\text{\tiny FSC}} \gg c_M$. Even in the case of asymptotically flat spacetimes, for bulk scalar field with mass $m$, the two point function is given by $e^{-mL}$, where $L$ is the length of the geodesic connecting the two points \cite{Hijano:2017eii}. So, for $\phi_{\chi}$, the contribution to the one loop diagram is given by
\be
\exp({-\xi_{\chi}2\pi r_0}) \approx \exp\left({-\frac{2\pi\xi_{\chi}\D_{\text{\tiny FSC}}}{\sqrt{2\xi_{\text{\tiny FSC}} c_M}}}\right). 
\ee
In the above equation we make use of the fact that for a primary field $(\D_b,\xi_b)$ on the boundary, the corresponding field in the bulk has a mass $\xi_b$. This follows from a matching of Casimirs from the point of the bulk and the boundary.

\medskip

\noindent Now let us calculate the contribution from the geodesic along the $r$ direction $r = r_0$ to $r = \infty$. In general, the length of this geodesic from $r = 0$ to $r=\Lambda$ is given by
\bea
L = \int_{r_0}^{\Lambda} \frac{r dr}{\hat{r}_+\sqrt{r^2-r_0^2}} = \frac{\sqrt{\Lambda^2-r_0^2}}{\hat{r}_+^2} \, \Rightarrow \log{L} = \frac{1}{2}\log \left(\frac{\Lambda^2}{r_0^2}-1\right) - \log\left(\frac{\hat{r}_+}{r_0}\right). 
\eea
In the above expression, we can see that $\log L$ diverges when $\Lambda$ and hence $\log\left(\frac{\Lambda}{r_0}-1\right)$ goes to infinity. We regulate this it by removing the first term. The renormalized geodesic length is thus given by 
\be{} 
\log L = -\log\left(\frac{\hat{r}_+}{r_0}\right) \implies L = \frac{r_0}{\hat{r}_+} \approx \frac{\D_{\text{\tiny FSC}}}{2\xi_{\text{\tiny FSC}}}.
\ee
So, the contribution of this geodesic to to the one-point function is given by
\be
\exp(-\xi_O L) \approx \exp\left(-{\xi_{O}\frac{r_0}{\hat{r}_+}}\right) \approx \exp\left(-\frac{\xi_{O}  \D_{\text{\tiny FSC}}}{2\xi_{\text{\tiny FSC}}}\right).  
\ee
Combining the above calculations, we have
\be
\boxed{\< E|O|E \> \approx \langle \chi|O|\chi \rangle \exp \left(-{\frac{\xi_{O}  \D_{\text{\tiny FSC}}}{2\xi_{\text{\tiny FSC}}}} -  {\frac{2\pi\xi_{\chi}\D_{\text{\tiny FSC}}}{\sqrt{2\xi_{\text{\tiny FSC}} c_M}}}\right).}
\label{1p_bulk}
\ee
This matches with the field theory calculation for asymptotic formula for three-point coefficient given in \eqref{three_point_ceoff_limit}. 

\medskip
\noindent From the other field theory calculation in \eqref{impr}, we have 
\bea
{\langle E|O|E} \rangle &\approx & \frac{\rho(\chi)}{\rho(O)}C_{\chi O \chi}\left(\frac{\xi}{\frac{c_M}{2}-\xi_{\chi}}\right)^{\frac{\D_O-1}{2}}\sqrt{\frac{\xi}{\frac{c_M}{2}-\xi_l}}\,\,e^{\frac{\xi_O}{2}\left(-\frac{\Delta-\frac{c_L}{2}}{\xi}+\frac{\frac{c_L}{2}-\Delta_\chi}{\frac{c_M}{2}-\xi_\chi}\right)} \\
&& \times e^{2\pi\left(\sqrt{\frac{\xi}{\frac{c_M}{2}-\xi_\chi}}\left(\frac{c_L}{2}-\Delta_\chi\right)-\sqrt{\frac{\xi}{\frac{c_M}{2}-\xi_l}}\left(\frac{c_L}{2}-\Delta_l\right)+\left(\sqrt{\frac{\frac{c_M}{2}-\xi_\chi}{\xi}}-\sqrt{\frac{\frac{c_M}{2}-\xi_l}{\xi}}\right)\left(\Delta-\frac{c_L}{2}\right)\right)}.\non
\label{1p_field_theory}
\eea
Here $l$ is the lightest field in the theory with weights $(\xi_l,\D_l)$. We will take this field to be the vacuum, so we have $\xi_l=\D_l=0$. Since we consider the case where the corresponding bulk fields $\phi_O$ and $\phi_{\chi}$ are scalars, we have $\D_O=\D_{\chi}=0$. For 3d Einstein gravity, we also have $c_L=0$. So, for these cases, \eqref{1p_field_theory} reduces to
\bea
\langle E|O|E \rangle &\approx & \frac{\rho(\chi)}{\rho(O)}C_{\chi O \chi}\,\exp\left(-{\frac{\xi_O\D_{\text{\tiny FSC}}}{2\xi_{\text{\tiny FSC}}}} {-\frac{2\pi\xi_{\chi}\D_{\text{\tiny FSC}}}{\sqrt{2\xi_{\text{\tiny FSC}} c_M}}}\right),
\eea
matching with the bulk analysis answer \eqref{1p_bulk}.

\newpage

\section{Conclusions}\label{conc}
\subsection{Summary}
Let us summarise for the reader what we have achieved in the current paper. We have been looking 2d field theories invariant under the BMS$_3$ algebra. These are theories which are putative duals to 3d asymptotically flat spacetimes. 

\medskip

\noindent We concentrated on modular properties of these 2d BMSFTs and specifically those of the torus one-point function. Using the BMS modular transformations of the torus one-point function, we computed an asymptotic formula for the three-point structure constants.  To begin with, we used two distinct methods, one a saddle-point analysis and the other which hinged on inverting integral transformations, to obtain expressions for the structure constants. We found that the second method worked for a wider range of parameters than the saddle point method. Both methods were however blind to the exact states that the trace was performed on. 

\medskip

\noindent 
We then computed the asymptotic form of structure constants when the states were in the BMS highest weight representation. In order to do this, we needed to develop quite a lot of new machinery for these BMSFTs. A particularly important development was the derivation of the BMS torus blocks. We found expressions for the leading and subleading terms in the BMS torus blocks in a particular limit of large weights. These developments were put in use when we recomputed the torus one-point function for the BMS primaries.   

\medskip

\noindent 
Finally, we used a bulk analysis that depended on a geodesic approximation in terms of a probe scalar in the background of a flat space cosmological solution to reproduce the field theory results. Additionally, we have a large number of appendices with a lot of more technical details. 

\subsection{Discussions and Future Directions}

We now turn to discuss various aspects and future directions of our findings in this paper. 

\medskip

\noindent {\em{More flat space holography}} 

\smallskip

\noindent One of our principal objectives in this work has been to further the advancement of holography for 3d asymptotically flat spacetimes. We have added the asymptotic three-point structure constants to the list of quantities that now match between the bulk and the putative boundary theory. Our bulk analysis can be improved by looking at a Witten diagram like procedure instead of the geodesic approximation \cite{Kraus:2016nwo, Kraus:2017ezw}. One can also look at the Chern-Simons formulation to understand this in a different way, following e.g. \cite{Alkalaev:2020yvq}. 

\smallskip

\noindent There are obvious generalisations to our work in this paper to theories with higher spin \cite{Afshar:2013vka, Gonzalez:2013oaa}, and supersymmetric BMS theories (see e.g. \cite{Barnich:2014cwa, Lodato:2016alv, Banerjee:2018hbl}), and other extensions of BMS e.g. with extra currents \cite{Basu:2017aqn}. 

\smallskip

\noindent An important direction is the generalisation to higher point functions on the torus. For the torus two point function, there are important applications. These would be connected to quasinormal modes in the dual bulk spacetime and to the eigenstate thermalisation hypothesis \cite{Brehm:2018ipf, Romero-Bermudez:2018dim, Hikida:2018khg} in the context of BMSFTs {\footnote{For further discussions of higher point torus functions for 2d CFTs and their holographic interpretation, the reader is pointed to \cite{Alkalaev:2017bzx}.}}.  

\smallskip

\noindent The modular bootstrap programme in 2d CFTs initiated in \cite{Hellerman:2009bu} helps constrain theories further. The S-modular transformation is particularly useful in this regard. A review of current developments is summarised in \cite{Brehm:2019pcx}. A similar programme is currently being attempted currently for 2d BMSFTs. This is something we hope to report on in the near future. 

\smallskip

\noindent It is of interest to see how the one-point torus amplitude calculation generalises in the case of Flatspace Chiral gravity (F$\chi$G) \cite{Bagchi:2012yk} (and its supersymmetric cousin \cite{Bagchi:2018ryy}). F$\chi$G relates a theory of Chern-Simons gravity with asymptotically flat boundary conditions to a chiral 2d conformal field theory. This theory admits FSC solutions. So it is natural to ponder what sort of bulk construction would reproduce what should be a chiral half of the Kraus-Maloney CFT answer.  

\bigskip

\noindent {\em{Black holes in general dimensions}} 

\smallskip

\noindent In \cite{Carlip:2017xne, Carlip:2019dbu}, Carlip has argued that the algebra of horizon preserving diffeomorphisms are enhanced to form BMS$_3$ for generic black holes and has used the BMS-Cardy formula \cite{Bagchi:2012xr} to derive the Bekenstein-Hawking entropy for these black holes. Along the same lines, it could thus be argued that the considerations of this paper would also be applicable to black holes in general dimensions. In particular, higher dimensional black holes will also get their ``spots" \cite{Kraus:2016nwo} through the mechanism elucidated in this paper.  

\bigskip

\noindent {\em{Tensionless string theory and BMS modular invariance}} 

\smallskip

\noindent BMS$_3$ algebra arises as residual gauge symmetries on the worldsheet of the tensionless bosonic closed string, in the equivalent of the conformal gauge \cite{Isberg:1993av, Bagchi:2013bga}. There has been recent efforts in trying to formulate the theory of tensionless strings from the point of BMS symmetries, mirroring the development of usual string theory by methods of conformal symmetry. Our modular explorations are hence of importance when one attempts to understand tensionless string theory on higher genus surfaces. One of the principal aims of the programme initiated in \cite{Bagchi:2013bga, Bagchi:2015nca} is the understanding of string amplitudes in this very high energy limit by purely worldsheet methods and comparing it to the seminal analysis of Gross and Mende \cite{Gross:1987kza, Gross:1987ar, Gross:1988ue}. Modular properties of 2d BMSFTs would be central to understanding one loop amplitudes in this tensionless regime. 

\smallskip

\noindent Relatedly, ambitwistor strings have been shown to be tensionless strings in disguise \cite{Casali:2016atr}. Efforts in attempting to construct partition functions for these ambitwistor strings have been made \cite{Casali:2017zkz} using the modular transformations we have worked extensively with in this work. It is conceivable that more can be understood for ambitwistor string theory using techniques developed in this paper. 

\bigskip

\newpage

\noindent {\em{Non-relativistic conformal symmetry}}

\smallskip

\noindent As remarked earlier, BMS$_3$ is isomorphic to the 2d Galilean conformal algebra (GCA) \cite{Bagchi:2010zz}, which can be obtained by a non-relativistic limit on the parent relativistic conformal theory. This remarkable isomorphism means that all our computations in this paper are equally valid for non-relativistic conformal systems in 2d. 

\smallskip

\noindent 
There has been recent interesting progress in the understanding of non-relativistic strong gravity \cite{Hansen:2020pqs}, where non-relativistic black hole solutions have been discussed. Building on the initial proposal of a non-relativistic limit of AdS/CFT, one could put our results in context. Starting with AdS$_3$/CFT$_2$, following \cite{Bagchi:2009my}, one would end up with a non-relativistic Newton-Cartan like AdS$_2 \times {\rm I\!R}$ which would be dual to a 2d Galilean CFT. There would presumably be solutions in the bulk corresponding to the BTZ black hole, which like the Schwarzschild solutions in \cite{Hansen:2020pqs} would have horizons. The modular properties described in this paper would help reproduce the entropy and probe reactions to these solutions. It would be interesting to investigate this in further detail. 

\smallskip

\noindent 
The 2d GCA also appears on non-relativistic string worldsheets corresponding to strings moving on target spaces with Newton-Cartan geometries \cite{Harmark:2018cdl}. Again, properties of 2d field theories with these symmetries, particularly modular properties discussed in this paper, would be important there. Construction of partition functions and modular invariance in the context of these string theories would rely heavily on the things we have developed here. 

\bigskip

\noindent {\em{Higher dimensional explorations}}

\smallskip

\noindent 
It would be very useful to try and generalise these considerations to higher dimensions. The general idea of holography stays the same, viz. we propose that the $d$-dimensional field theory would inherit the asymptotic symmetries of the $(d+1)$ dimensional asymptotically flat bulk, which is the BMS$_{d+1}$ group \cite{Bagchi:2016bcd, Bagchi:2019xfx} {\footnote{For other recent proposals, see e.g. \cite{Ball:2019atb, Laddha:2020kvp}. See \cite{Banerjee:2020kaa, Banerjee:2018gce} for interesting work in trying to understand scattering in four dimensional asymptotically flat spacetimes with a potential three dimensional dual building on the work relating scattering with a celestial 2d CFT in \cite{Pasterski:2016qvg, Pasterski:2017kqt}.}}. In particular, if some of these methods can be applied to 3d field theories with BMS$_4$ symmetries, we would be able to connect with quantities of very significant physical interest, S-matrix elements.  
BMS$_4$ symmetries however are more involved and the structure of $\mathscr{I}^\pm$ which is given by ${\rm I\!R}_u \times S^2$, now admits two copies of the (non-centrally extended) Virasoro algebra along with supertranslations {\footnote{This is one of the proposed infinite dimensional extensions of BMS$_4$ \cite{Barnich:2010eb}. For another proposal, involving the group Diff$(S^2)$, see \cite{Campiglia:2014yka}.}}. It is not entirely obvious how the methods described in this paper would generalise. 

\smallskip

\noindent As we mentioned, the isomorphism between the BMS$_3$ and the GCA$_2$ means that the results we derived here are equally applicable to non-relativistic conformal systems. In higher dimensions, the structure of GCA$_d$ remains very similar to the $d=2$ case, with only additional vectorial labels on the $M$ generators and the existence of rotations. These symmetries have been shown to exist in non-relativistic versions of electrodynamics \cite{Bagchi:2014ysa, Festuccia:2016caf} and Yang-Mills \cite{Bagchi:2015qcw}, with arbitrary massless matter couplings as well \cite{Bagchi:2017yvj}. Our methods developed in this paper may be more amenable to generalisation to higher dimensions in this context given the similar structures of the $d=2$ case with the general case. 
 
\bigskip 

\bigskip \bigskip

\subsection*{Acknowledgements}
It is a pleasure to thank Rudranil Basu, Daniel Grumiller, Max Reigler, Joan Simon for interesting conversations and comments on a draft of the paper. We also thank Konstantin Alkalaev for helpful correspondence and an anonymous referee for pointing out some inconsistencies in an earlier version of the paper. 

\medskip

\noindent AB's research is supported by a Swarnajayanti fellowship of the Department of Science and Technology and the Science and Engineering Research Board (SERB), India. AB is further supported by the following grants from SERB: EMR/2016/008037, ERC/2017/000873, MTR/2017/000740. PN is supported by a Fulbright-Nehru Doctoral Research fellowship (Grantee ID E0609781). PN would also like to thank Mukund Rangamani and the hospitality of UC Davis during the course of this work.  She would further like to thank Sankha Bhattacharya and Nivedita Bhattacharya for their unconditional support and hospitality at San Jose during a difficult time of emergency in middle of the worldwide pandemic.

\newpage

\section*{APPENDICES}

\bigskip

\appendix

\section{Review of torus one-point function in  2d CFT}\label{ApA}
In this appendix, we are giving a review of large $h_A$ limit of torus one point function for relativistic CFT$_2$ using mode expansion. Only for this section, we use the notation $L_n, \bar{L}_n$ to define the Virasoro generators giving the corresponding algebra.
\be{}
\begin{split}
[L_n,L_m]=&(n-m)L_{n+m}+\frac{c}{12}n(n^2-1)\delta_{m+n,0}\\
[\bar{L}_n,\bar{L}_m]=&(n-m)\bar{L}_{n+m}+\frac{\bar c}{12}n(n^2-1)\delta_{m+n,0}
\end{split}\ee
\subsection*{Mode expansion}
It is well known that a conformal field $\phi_p(z, \bar{z})$ of dimension $(h, \bar{h})$ on the 2-d Euclidean plane may be mode-expanded as below:
\begin{align}\label{2.1}
\phi_p(z,\bar{z})=\sum_{m\in\mathbb{Z}} \sum_{n\in\mathbb{Z}} z^{-m-h}\bar{z}^{-n-\bar{h}}\phi_{m,n}.
\end{align}
The quasi-primary modes satisfy the usual Hermitian-conjugation relation (on the real surface):
\begin{align}\label{2.2}
\phi^\dagger _{m,n}=\phi_{-m,-n}.
\end{align}
As the holomorphic and anti-holomorphic degrees of freedom of a CFT on a 2-d Euclidean plane decouples from each other, we define the mode-expansion for the holomorphic part (and exactly similarly, for the anti-holomorphic part as well) as:
\begin{align}
\label{2.3}
\phi_p(z)=\sum_{m\in\mathbb{Z}} z^{-m-h}\phi_{m},
\end{align}
with the Hermitian-conjugation: $\phi^\dagger _{m}=\phi_{-m}$, for quasi-primary modes. From the `plane'-representation theory of $CFT_2$, we know that for a local (holomorphic) primary field $\phi_p(z)$, we have (for $n\geqslant -1$):
\begin{align}\label{2.4}
[L_n, \phi_p(z)]=[h(n+1)z^n+z^{n+1}\partial_z]\phi_p(z).
\end{align}
Now, using the mode-expansion \eqref{2.3} into \eqref{2.4} and comparing the coefficients of $z^{-m-h}$ on both sides of the equation, we conclude that (for $n\geqslant -1$):
\begin{align}\label{2.5}
[L_n, \phi_m]=[n(h-1)-m]\phi_{m+n}.
\end{align}
A noteworthy point is that while the mode-expansion \eqref{2.1} or \eqref{2.3} is valid for any local fields, relation \eqref{2.5} is only applicable to the primary fields. 

\subsection*{Torus one-point function}
Exploiting the translation invariance of the torus one-point function of a local primary field, i.e. using the fact that it is independent of the `elliptic coordinate' $(w, \bar{w})$ on the torus, the torus one-point function is a further generalization of the partition function in the way it is defined (for a primary field $\phi_p$ on the plane, with dimensions $(h_p,\bar{h}_p)$):
\begin{align}\label{2.6}
\langle \phi\rangle_{(\tau, \bar{\tau})}&= \rm{Tr}\left[\phi_p(1,1)e^{2\pi i\tau(L_0-\frac{c}{24})}e^{-2\pi i\bar{\tau}(\bar{L}_0-\frac{\bar{c}}{24})}\right]\nonumber\\
&=\sum_{A} D(h_A,\bar{h}_A)q^{h_A-\frac{c}{24}}\bar{q}^{\bar{h}_A-\frac{\bar{c}}{24}}\langle h_A,\bar{h}_A| \phi_p(1,1)|h_A,\bar{h}_A\rangle F^{h_p,\bar{h}_p}_{h_A,\bar{h}_A,c,\bar{c}}(q,\bar{q}),
\end{align}
where $\{A\}$ denotes the collection of all the primary fields in the theory and $q=e^{2\pi i\tau}$. $D(h_A,\bar{h}_A)$ is the number of the primary fields with dimension $(h_A,\bar{h}_A)$, in the theory. The `F' functions, called `torus conformal blocks', which are completely determined by the symmetry algebra alone, are clearly defined (in the highest-weight representation of the Virasoro algebra) to be holomorphically factorizable as $F^{h_p,\bar{h}_p}_{h_A,\bar{h}_A,c,\bar{c}}(q,\bar{q})=F^{h_p}_{h_A,c}(q)F^{\bar{h}_p}_{\bar{h}_A,\bar{c}}(\bar{q})$, where:
\begin{align}\label{2.7}
F^{h_p}_{h_A,c}(q)&=q^{-h_A+\frac{c}{24}}\frac{Tr_{\mathcal{M}(h_A,c)}(\phi_p(1)q^{L_0-\frac{c}{24}})}{\langle h_A| \phi_p(1)|h_A\rangle}\nonumber\\
&=\sum_{N=0}^\infty \sum_{i,j\in \{N\}}q^N G^{ij}_{(N)}\frac{\langle h_A;N,i| \phi_p(1)|h_A;N,j\rangle}{\langle h_A| \phi_p(1)|h_A\rangle}:=\sum_{N=0}^\infty q^N F_N(h_p,h_A,c),
\end{align} 
along with a similar definition for the anti-holomorphic part. In the above, $\mathcal{M}(h_A,c)$ denotes the holomorphic sector of the Verma module of the primary field with dimensions $(h_A,\bar{h}_A)$, while $[G^{ij}_{(N)}]$ is the inverse Gram-matrix at level N of this holomorphic module, with $|h_A;N,i\rangle$ being a state at level N in the highest-weight representation. $\{N\}$ is the collection of all such states in the module.\\
To explicitly calculate the matrix elements as in \eqref{2.7}, let us consider the mode-expansion (2.3) at $z=1$:
\begin{align}\label{2.8}
\phi_p(z=1)=\sum_{m\in\mathbb{Z}}\phi_{m}.
\end{align}
Now, let's consider the following stream of calculations that will lead to a useful identity (for $N\geqslant0$ and $M\geqslant0$ ):
\begin{align}
& \langle h_A;N,i|L_0 \phi_p(1)|h_A;M,j\rangle\nonumber\\
=&\langle h_A;N,i|[L_0,\phi_p(1)]|h_A;M,j\rangle+\langle h_A;N,i| \phi_p(1)L_0|h_A;M,j\rangle\nonumber\\
\Rightarrow &\sum_{m\in\mathbb{Z}}(h_A+N)\langle h_A;N,i|\phi_m|h_A;M,j\rangle\nonumber\\
=&\sum_{m\in\mathbb{Z}}\langle h_A;N,i|(-m)\phi_m|h_A;M,j\rangle+\sum_{m\in\mathbb{Z}}(h_A+M)\langle h_A;N,i|\phi_m|h_A;M,j\rangle\nonumber\\
\Rightarrow &\sum_{m\in\mathbb{Z}}(m+N-M)\langle h_A;N,i|\phi_m|h_A;M,j\rangle=0\nonumber\\
\Rightarrow &\sum_{m\in\mathbb{Z}}m\langle h_A;N,i|\phi_{m-N+M}|h_A;M,j\rangle=0.
\label{2.9}
\end{align}
Using \eqref{2.8} and \eqref{2.9}, we may now proceed to calculate the relevant matrix elements in \eqref{2.7}. Firstly, we look at level 1 matrix element:
\be{}
\begin{split}
&\langle h_A|L_1 \phi_p(1)L_{-1}|h_A\rangle\nonumber\\
&=\langle h_A|[L_1,\phi_p(1)]L_{-1}|h_A\rangle+\langle h_A|\phi_p(1)L_1L_{-1}|h_A\rangle\nonumber\\
&=\sum_{m\in\mathbb{Z}}\langle h_A|(h_p-1-m)\phi_{m+1}L_{-1}|h_A\rangle+2h_A\langle h_A|\phi_p(1)|h_A\rangle\nonumber\\
&=\sum_{m\in\mathbb{Z}}(h_p-1){\langle h_A|L_1\phi_{-m-1}|h_A\rangle}^*+2h_A\langle h_A|\phi_p(1)|h_A\rangle\nonumber\\
\end{split}
\ee
\be{}
\begin{split}
&=\sum_{m\in\mathbb{Z}}(h_p-1){\langle h_A|(h_p+m)\phi_{-m}|h_A\rangle}^*+2h_A\langle h_A|\phi_p(1)|h_A\rangle\\
&=\sum_{m\in\mathbb{Z}}(h_p-1)(h_p+m){\langle h_A|\phi_{m}|h_A\rangle}+2h_A\langle h_A|\phi_p(1)|h_A\rangle\\
&=[h_p(h_p-1)+2h_A]\langle h_A|\phi_p(1)|h_A\rangle.
\end{split}
\ee
where, to go from 2nd to 3rd step and from 5th to 6th step, \eqref{2.9} was in use. As level 1 has only one state, we get $G^{11}_{(1)}=\frac{1}{2h_A}$ and hence
\begin{align}
F_1(h_p,h_A,c)=1+\frac{h_p(h_p-1)}{2h_A}.
\end{align}
Of course we may just blindly copy the above calculation and conclude that for $n\geqslant1$:
\begin{align}\label{2.10}
\langle h_A|L_n \phi_p(1)L_{-n}|h_A\rangle=[n^2 h_p(h_p-1)+2nh_A+(n^3-n)\frac{c}{12}]\langle h_A|\phi_p(1)|h_A\rangle.
\end{align}
A class of repeatedly occurring matrix elements that can be easily calculated recursively and have a neat `series-summation' form is of the general form $\langle h_A|L_n ^k \phi_p(1)L_{-n} ^k|h_A\rangle$ for $k>0$ and $n\geqslant1$. To evaluate this, we need the following commutation relation:
\begin{align}
[L_n,L_{-n} ^k]=\sum_{l=0}^{k-1} L_{-n} ^{k-1-l}(2nL_0+(n^3-n)\frac{c}{12})L_{-n} ^l.
\end{align}
The calculation is as below (with $h_{A}^\prime=h_A+(n^2-1)\frac{c}{24}$):
\be{}
\begin{split}
&\langle h_A|L_n ^k \phi_p(1)L_{-n} ^k|h_A\rangle\nonumber\\
& \non\\
&~=\langle h_A|L_n ^{k-1}[L_n,\phi_p(1)]L_{-n}^k|h_A\rangle+\langle h_A|L_n ^{k-1}\phi_p(1)L_nL_{-n}^k|h_A\rangle\nonumber\\
&\non\\
&~=\sum_{m\in\mathbb{Z}}\langle h_A|L_n ^{k-1}[n(h_p-1)-m]\phi_{m+n}L_{-n}^k|h_A\rangle +2n\sum_{l=0}^{k-1}(h_{A}^\prime+nl)\langle h_A|L_n ^{k-1}\phi_p(1)L_{-n}^{k-1}|h_A\rangle\nonumber\\
&\non\\
&~=[n(h_p-1)]\langle h_A|L_n ^{k-1}\phi_p(1)L_{-n}^k|h_A\rangle\ +nk[2h_{A}^\prime+n(k-1)]\langle h_A|L_n ^{k-1}\phi_p(1)L_{-n}^{k-1}|h_A\rangle\non\\
&\non\\
&~=n\sum_{l=0} ^1 \frac{1}{(1-l)!}\binom{k}{l} n^{l}(\frac{2h_{A}^\prime}{n}+k-l)_{l}(h_p-1+l)_{1-l}\langle h_A|L_n ^{k-1}\phi_p(1)L_{-n}^{k-l}|h_A\rangle\nonumber\\
\end{split}
\ee
\be{}
\begin{split}
& \non\\&~=[n(h_p-1)]\sum_{m\in\mathbb{Z}}\langle h_A|L_n ^{k-2}[n(h_p-1)-m]\phi_{m+n}L_{-n}^k|h_A\rangle +2nk[2h_{A}^\prime+n(k-1)][n(h_p-1)]\non\\
&~~~\quad \langle h_A|L_n ^{k-2}\phi_p(1)L_{-n}^{k-1}|h_A\rangle+n^2k(k-1)[2h_{A}^\prime+n(k-1)][2h_{A}^\prime+n(k-2)]\langle h_A|L_n ^{k-2}\phi_p(1)L_{-n}^{k-2}|h_A\rangle\nonumber\\
\end{split}
\ee
\be{}
\begin{split}
&\non\\
&~=[n(h_p-1)][n(h_p-2)]\langle h_A|L_n ^{k-2}\phi_p(1)L_{-n}^k|h_A\rangle +2nk[2h_{A}^\prime+n(k-1)][n(h_p-1)]\\
&~~~~~\langle h_A|L_n ^{k-2}\phi_p(1)L_{-n}^{k-1}|h_A\rangle +n^2k(k-1)[2h_{A}^\prime+n(k-1)][2h_{A}^\prime+n(k-2)]\langle h_A|L_n ^{k-2}\phi_p(1)L_{-n}^{k-2}|h_A\rangle\\
&\\
&~=n^2\sum_{l=0} ^2 \frac{2}{(2-l)!}\binom{k}{l} n^{l}(\frac{2h_{A}^\prime}{n}+k-l)_{l}(h_p-2+l)_{2-l}\langle h_A|L_n ^{k-2}\phi_p(1)L_{-n}^{k-l}|h_A\rangle\\
&\\
&\text{=~~~~~~~~~~~$\hdots\hdots\hdots\hdots$ recursive reduction of the power of $L_n$ $\hdots\hdots\hdots\hdots\hdots$} \label{2.11}\\ 
& \\
&~=n^k\sum_{l=0} ^k (k-l)!{\binom{k}{l}}^2 n^{k-l}(\frac{2h_{A}^\prime}{n}+l)_{k-l}(h_p-l)_l\langle h_A|\phi_p(1)L_{-n}^{l}|h_A\rangle.
\end{split}
\ee
Now, we are left with the undetermined $\langle h_A|\phi_p(1)L_{-n}^{l}|h_A\rangle$ which can be evaluated recursively (we note that $\phi_p ^\dagger(1)=\phi_p(1)$):
\begin{align}
&\hspace{-1cm}\langle h_A|\phi_p(1)L_{-n}^{l}|h_A\rangle={\langle h_A|L_{n}^{l}\phi_p(1)|h_A\rangle}^*\non\\
&\quad\quad = \sum_{m\in\mathbb{Z}}{\langle h_A|L_{n}^{l-1}[n(h_p-1)-m]\phi_{m+n}|h_A\rangle}^*\nonumber\\
& \quad\quad =\sum_{m\in\mathbb{Z}}{\langle h_A|L_{n}^{l-1}[nh_p-m]\phi_{m}|h_A\rangle}^* 
={\langle h_A|L_{n}^{l-1}[nh_p+n(l-1)]\phi_p(1)|h_A\rangle}^*\nonumber\\ 
&\quad \quad = n(h_p+l-1){\langle h_A|\phi_p(1)L_{-n}^{l-1}|h_A\rangle}=n^l(h_p)_l\langle h_A|\phi_p(1)|h_A\rangle. \label{2.12}
\end{align}
Thus finally we get:
\begin{align}
\frac{\langle h_A|L_n ^k \phi_p(1)L_{-n} ^k|h_A\rangle}{\langle h_A|\phi_p(1)|h_A\rangle}=n^{2k}\sum_{l=0} ^k (k-l)!{\binom{k}{l}}^2 (\frac{2h_{A}^\prime}{n}+l)_{k-l}(h_p-l)_{2l}\label{2.13}
\end{align}
We see that \eqref{2.10} is a special case of \eqref{2.12}. Now, for the purpose of book-keeping, let's find the relevant matrix elements at level 2. Using the above general formulae, we instantly get:
\begin{align}
&\frac{\langle h_A|L_1 ^2 \phi_p(1)L_{-1} ^2|h_A\rangle}{\langle h_A|\phi_p(1)|h_A\rangle}=4h_A(2h_A+1)+4h_p(h_p-1)(2h_A+1)+(h_p-2)_4\nonumber\\
&\frac{\langle h_A|L_2 \phi_p(1)L_{-2}|h_A\rangle}{\langle h_A|\phi_p(1)|h_A\rangle}=4h_p(h_p-1)+4h_A+\frac{c}{2}.
\end{align}
while a little calculation reveals that:

\begin{align}
{\langle h_A|L_2 \phi_p(1)L_{-1} ^2|h_A\rangle}& ={\langle h_A|L_1 ^2\phi_p(1)L_{-2}|h_A\rangle}\nonumber\\
&=\sum_{m\in\mathbb{Z}}[2(h_p-1)-m]\langle h_A|\phi_{m+2}L_{-1} ^2|h_A\rangle+6h_A\langle h_A| \phi_p(1)|h_A\rangle\nonumber\\
&=2(h_p-1)\langle h_A|\phi_p(1)L_{-1} ^2|h_A\rangle+6h_A\langle h_A| \phi_p(1)|h_A\rangle\nonumber\\
&=[2(h_p-1)h_p(h_p+1)+6h_A]\langle h_A| \phi_p(1)|h_A\rangle.
\end{align}
where, to go to the last step from the penultimate one, we used \eqref{2.11} directly. Now, we are ready to note down the level 2 contribution to the torus block:
\be{}
F_2(h_p,h_A,c) =2+\frac{4[(8h_A+\frac{c}{2})(2h_A+1)-6h_A(h_p+1)](h_p-1)_2+(4h_A+\frac{c}{2})(h_p-2)_4}{4h_A[(2h_A+1)(4h_A+\frac{c}{2})-9h_A]}
\ee
As an aside, noticing the utility of \eqref{2.11}, we may further generalize it into the following:
\be{}
\frac{\langle h_A|\phi_p(1)\prod_{i=1}^n L_{-i} ^{l_i}|h_A\rangle}{\langle h_A|\phi_p(1)|h_A\rangle}=\frac{\langle h_A|(\prod_{i=n}^1 L_{i} ^{l_i})\phi_p(1)|h_A\rangle}{\langle h_A|\phi_p(1)|h_A\rangle}^\star =\prod_{k=1}^n k^{l_k}(h_p+\frac{1}{k}\sum_{i=k+1}^n il_i)_{l_k}.\label{2.14}
\ee
In principle, following our derivation of \eqref{2.12}, one should be able to evaluate the most general matrix element $\langle h_A|(\prod_{j=m}^1 L_{j} ^{l_j})\phi_p(1)\prod_{i=1}^n L_{-i} ^{k_i}|h_A\rangle$, by recursively reducing the powers of all the $\{L_{j}\}$ to 0 and then applying Virasoro commutation relation and finally \eqref{2.14}. 

\subsection*{Global torus block}
In the highest-weight representation of the global Virasoro algebra, the descendant state at level N in the holomorphic module of the primary field with dimension $h_A$ is given by $L_{-1} ^N|h_A\rangle$. So, the global torus block can be computed very easily using \eqref{2.12} with $n=1$. Also, we have:
\begin{align}
\langle h_A|L_1 ^N L_{-1} ^N|h_A\rangle=N!(2h_A)_N.
\end{align}
These lead to (for $n=1$, $h^\prime _A=h_A$):
\begin{align}
F_N ^{global}(h_p,h_A)=\sum_{l=0} ^N \frac{1}{l!}\binom{N}{l} \frac{(h_p-l)_{2l}}{(2h_A)_l}.
\end{align}
And, finally, we evaluate the holomorphic global torus block, by evaluating the global version of \eqref{2.7} to obtain, for $h_A>0$:
\begin{align}
F^{global}_{h_p,h_A}(q)&=\sum_{N=0} ^{\infty}q^N\sum_{l=0} ^N \frac{1}{l!}\binom{N}{l} \frac{(h_p-l)_{2l}}{(2h_A)_l}=\sum_{l=0} ^{\infty}\sum_{N=0} ^{\infty}q^{N+l} \frac{1}{l!}\binom{N+l}{l} \frac{(h_p-l)_{2l}}{(2h_A)_l}\nonumber\\
&=\sum_{l=0} ^{\infty}\frac{q^l}{l!}[\sum_{N=0} ^{\infty}q^{N} \binom{N+l}{l}] \frac{(h_p-l)_{2l}}{(2h_A)_l}=\sum_{l=0} ^{\infty}q^l\frac{{(1-q)}^{-l-1}}{l!} \frac{(h_p-l)_{2l}}{(2h_A)_l}\nonumber\\
&=\frac{1}{1-q}\sum_{l=0} ^{\infty} \frac{(1-h_p)_l(h_p)_l}{l!(2h_A)_l}\Big(\frac{q}{q-1}\Big)^l\nonumber\\
&={(1-q)^{-1}}{}_2F_1\Big(h_p,1-h_p;2h_A;\frac{q}{q-1}\Big).\label{2.15}
\end{align}
We note that as $|q|<1$, the convergence issue is automatically taken care of.

\subsection*{Large $h_A$ limit of $F_N(h_p,h_A,c)$}
While deriving an asymptotic, Cardy-like formula for structure constant, we need to consider the torus block of a heavy exchanged operator with dimension $h_A\gg c,h_p$. We would like to find the leading and the sub-leading terms in the asymptotic expansion in $\frac{1}{h_A}$ of the torus block. The general matrix element at level $N$, $\langle h_A|(\prod_{j=N}^1 L_{j} ^{l_j})\phi_p(1)\prod_{i=1}^N L_{-i} ^{k_i}|h_A\rangle$, with $\sum_{j=1}^N jl_j=N=\sum_{i=1}^N ik_i$, is evaluated algorithmically as below that in essence is the recursive reduction of the Virasoro operator string at the left of $\phi_p(1)$. Being real quantities, these matrix elements is indifferent to the swapping of the left and right string according to the Hermittian conjugation. Among these two strings, the one which has less length than the other, is reduced recursively (for faster computation). In the following, without the loss of generality, we assume that $\sum_{j=1}^N l_j\leqslant\sum_{i=1}^N k_i$. 

\begin{align}
&\langle h_A|(\prod_{j=N}^1 L_{j} ^{l_j})\phi_p(1)\prod_{i=1}^N L_{-i} ^{k_i}|h_A\rangle=\langle h_A|(\prod_{i=N}^1 L_{i} ^{k_i})\phi_p(1)\prod_{j=1}^N L_{-j} ^{l_j}|h_A\rangle^*\nonumber\\
&\non\\
&=(h_p-1)\langle h_A|(\prod_{j=N}^2 L_{j} ^{l_j})L_{1} ^{l_1-1}\phi_p(1)\prod_{i=1}^N L_{-i} ^{k_i}|h_A\rangle+\langle h_A|(\prod_{j=N}^2 L_{j} ^{l_j})L_{1} ^{l_1-1}\phi_p(1)L_1\prod_{i=1}^N L_{-i} ^{k_i}|h_A\rangle\nonumber\\
&\non\\
&=(h_p-1)(h_p-2)\langle h_A|(\prod_{j=N}^2 L_{j} ^{l_j})L_{1} ^{l_1-2}\phi_p(1)\prod_{i=1}^N L_{-i} ^{k_i}|h_A\rangle\nonumber\\
&+2(h_p-1)\langle h_A|(\prod_{j=N}^2 L_{j} ^{l_j})L_{1} ^{l_1-2}\phi_p(1)L_1\prod_{i=1}^N L_{-i} ^{k_i}|h_A\rangle+\langle h_A|(\prod_{j=N}^2 L_{j} ^{l_j})L_{1} ^{l_1-2}\phi_p(1)L_1^2\prod_{i=1}^N L_{-i} ^{k_i}|h_A\rangle\nonumber\\
&\non\\
&=\sum_{a_1=0}^{l_1}\binom{l_1}{a_1}(h_p-a_1)_{a_1}\langle h_A|(\prod_{j=N}^2 L_{j} ^{l_j})\phi_p(1)L_1^{l_1-a_1}\prod_{i=1}^N L_{-i} ^{k_i}|h_A\rangle\nonumber\\
&\non\\
&=\sum_{a_2=0}^{l_2}\binom{l_2}{a_2}2^{a_2}(h_p-\frac{a_1}{2}-a_2)_{a_2}\sum_{a_1=0}^{l_1}\binom{l_1}{a_1}(h_p-a_1)_{a_1}\nonumber\\
&\qquad\times\langle h_A|(\prod_{j=N}^3 L_{j} ^{l_j})\phi_p(1)L_2^{l_2-a_2}L_1^{l_1-a_1}\prod_{i=1}^N L_{-i} ^{k_i}|h_A\rangle\nonumber\\
&\non\\
&=...\text{recursive reduction of the left string of $\{L_j\}$}...\nonumber
\end{align}

\begin{align}
&\non\\
&=\underset{(\sum_{m=1}^N a_m)>0}{\sum_{a_1=0}^{l_1}\sum_{a_2=0}^{l_2}...\sum_{a_N=0}^{l_N}}\langle h_A|\phi_p(1)[\prod_{j=N}^1\binom{l_j}{a_j}j^{a_j}(h_p-\frac{1}{j}\sum_{p=1}^j pa_p)_{a_j}L_j^{l_j-a_j}]\prod_{i=1}^N L_{-i} ^{k_i}|h_A\rangle\nonumber\\
&+\langle h_A|\phi_p(1)|h_A\rangle\langle h_A|(\prod_{j=N}^1L_j^{l_j})\prod_{i=1}^N L_{-i} ^{k_i}|h_A\rangle. \label{2.16}
\end{align}
From the first term of \eqref{2.15}, to make the $[\prod_{j=N}^1L_j^{l_j-a_j}]\prod_{i=1}^N L_{-i} ^{k_i}|h_A\rangle$ part a proper Virasoro highest-weight descendant state at level $\sum_{j=1}^N ja_j$, we use the Virasoro commutation relation and then using \eqref{2.13}, we obtain a final form as $g(h_A,h_p)\langle h_A|\phi_p(1)|h_A\rangle$, where $g(h_A,h_p)$ is a polynomial in both $h_A$ and $h_p$.
Clearly, the second term in the above identity is an element of the Gram-matrix $[G_{(N)ij}]$, upto the $\langle h_A|\phi_p(1)|h_A\rangle$ factor, while the $h_p$-dependence comes solely from the first term. Hence, the total contribution of this generic second term to $F_N(h_p,h_A,c)$ in \eqref{2.7} is simply ${\rm dim}_N=p(N)$.

As we are interested in the large $h_A$ limit, we would like to find the leading order term in $h_A$ from the first term of \eqref{2.15}. We note that the generic level $N$ Gram-matrix element, $\langle h_A|(\prod_{j=N}^1 L_{j} ^{l_j})\prod_{i=1}^N L_{-i} ^{k_i}|h_A\rangle$, is of the order $\mathcal{O}(h_A^\alpha)$, with $1\leqslant\alpha\leqslant min\{\sum_{j=1}^N l_j,\sum_{i=1}^N k_i\}$, while the corresponding element of the inverse Gram-matrix, ($\frac{\text{the co-factor of the mentioned element}}{\text{Kac determinant at level $N$}}$) is of the order $\mathcal{O}(h_A^{-\beta})$, with $\beta\geqslant max\{\sum_{j=1}^N l_j,\sum_{i=1}^N k_i\}$. Hence, the generic summation term in \eqref{2.15} is at most of the order $\mathcal{O}(h_A^{\sum_{j=1}^N (l_j-a_j)})$ in $h_A$, while the remaining descendant state at level $\sum_{j=1}^N ja_j$ gives a factor dependent on $h_p$ according to \eqref{2.13}. Clearly, the maximum possible order in $h_A$ for such a term is $\mathcal{O}(h_A^{(\sum_{j=1}^N l_j)-1})$ and we simplify this term as below:
\begin{align}
&\underset{(\sum_{m=1}^N a_m)=1}{\sum_{a_1=0}^{l_1}\sum_{a_2=0}^{l_2}...\sum_{a_N=0}^{l_N}}\langle h_A|\phi_p(1)[\prod_{j=N}^1\binom{l_j}{a_j}j^{a_j}(h_p-\frac{1}{j}\sum_{p=1}^j pa_p)_{a_j}L_j^{l_j-a_j}]\prod_{i=1}^N L_{-i} ^{k_i}|h_A\rangle\nonumber\\
&=(h_p-1){\sum_{j=1}^N}jl_j\langle h_A|\phi_p(1)(\prod_{p=N}^{j+1}L_p^{l_p})L_j^{l_j-1}(\prod_{q=j-1}^{1}L_q^{l_q})\prod_{i=1}^N L_{-i} ^{k_i}|h_A\rangle.
\end{align}    
Now, in any row or column of both the $[G_{(N)ij}]$ and its inverse, the diagonal element is the highest order term in $h_A$, though other terms may be of equal or lower order. Clearly, the diagonal element of the Gram matrix is of the order $\mathcal{O}(h_A^{\sum_{i=1}^N k_i})$, while the corresponding diagonal element of the inverse Gram matrix is of the order $\mathcal{O}(h_A^{-\sum_{i=1}^N k_i})$, i.e. only a diagonal element of $[G_{(N)ij}]$ and the corresponding one of $[G_{(N)}^{ij}]$ are of the same order in $h_A$ and $\frac{1}{h_A}$ respectively. By definition $G_{(N)}^{ij}G_{(N)jk}=\delta^i_k$ and if we naively evaluate the large $h_A$ limit of $[G_{(N)}^{ij}]$ and $[G_{(N)ij}]$ by keeping only the leading term in $h_A$ of each element and then do matrix multiplication, we see that the non-diagonal elements of the resulting matrix is at most $\mathcal{O}(h_A^{-1})$ while diagonal elements are $1+\mathcal{O}(h_A^{-1})$.

Thus if we multiply $[G_{(N)}^{ij}]$ and $[\langle h_A;N,i|\phi_p(1)|h_A;N,j\rangle]-\langle h_A|\phi_p(1)|h_A\rangle[G_{(N)ij}]$, by keeping only the highest order term in $h_A$ for each element, it is easy to deduce from the above arguments that the non-diagonal terms in the resultant matrix are at most $\mathcal{O}(h_A^{-2})$ while the diagonal terms are $\mathcal{O}(h_A^{-1})$. So, in the large $h_A$ limit, to find the $h_p$-dependent coefficient of the $\mathcal{O}(h_A^{-1})$ term in the $\frac{1}{h_A}$-expansion of the torus block, it suffices to consider only the diagonal elements: $\langle h_A|(\prod_{j=N}^1 L_{j} ^{k_j})\phi_p(1)\prod_{i=1}^N L_{-i} ^{k_i}|h_A\rangle$. Even for each of these elements also, we only need the leading order term in $h_A$, that also depends on $h_p$ (l.o.t.$\Rightarrow$ lower order terms in $h_A$):\\
\be{}
\begin{split}
&\text{$h_p$-dependent term in }\langle h_A|(\prod_{j=N}^1 L_{j} ^{k_j})\phi_p(1)\prod_{i=1}^N L_{-i} ^{k_i}|h_A\rangle\\
&=(h_p-1){\sum_{j=1}^N}jk_j\langle h_A|\phi_p(1)(\prod_{p=N}^{j+1}L_p^{k_p})L_j^{k_j-1}(\prod_{q=j-1}^{1}L_q^{k_q})\prod_{i=1}^N L_{-i} ^{k_i}|h_A\rangle+{l.o.t.}^{\prime\prime}\\
&=(h_p-1){\sum_{j=1}^N}[jk_j\langle h_A|\phi_p(1)k_jL_{-j}|h_A\rangle{\langle h_A|L_jL_{-j}|h_A\rangle}^{k_j-1}\\
&\qquad\times\prod_{i=1,i\neq j}^N{\langle h_A|L_iL_{-i}|h_A\rangle}^{k_i}]+{l.o.t.}^{\prime}\\
&=(h_p-1)h_p{\sum_{j=1}^N}[{(jk_j)}^2{(2jh_A)}^{k_j-1}{(k_j-1)}!\prod_{i=1,i\neq j}^N{(2ih_A)}^{k_i}{k_i}!]\langle h_A|\phi_p(1)|h_A\rangle+{l.o.t.}
\end{split}
\ee
Clearly, for our purpose, we only need to explicitly know only the leading order term in $h_A$ of only the diagonal elements of the inverse Gram-matrix. In the large $h_A$ limit, it is approximately given by:
\begin{align}
\langle h_A|(\prod_{j=N}^1 L_{j} ^{k_j})\prod_{i=1}^N L_{-i} ^{k_i}|h_A\rangle^{-1}=[\prod_{i=1}^N{(2ih_A)}^{k_i}{k_i}!]^{-1}+{l.o.t.}\nonumber
\end{align}
Thus, each diagonal element of the $\langle h_A;N,i|\phi_p(1)|h_A;N,i\rangle$ matrix gives the following $h_p$-dependent contribution to $F_N(h_p,h_A,c)$ in the large $h_A$ limit:
\begin{align}
&\text{$h_p$-dependent term in }\langle h_A|(\prod_{j=N}^1 L_{j} ^{k_j})\phi_p(1)\prod_{i=1}^N L_{-i} ^{k_i}|h_A\rangle\times\langle h_A|(\prod_{j=N}^1 L_{j} ^{k_j})\prod_{i=1}^N L_{-i} ^{k_i}|h_A\rangle^{-1}\nonumber\\
&=\frac{h_p(h_p-1)}{2h_A}\sum_{j=1}^N jk_j+\mathcal{O}(h_A^{-2})=N\frac{h_p(h_p-1)}{2h_A}+\mathcal{O}(h_A^{-2}).
\end{align}
So, each diagonal element gives equal contribution to the $\frac{1}{h_A}$-term and there are $p(N)$ such elements. Hence, collecting all the contributions in large $h_A$ limit:
\begin{align}
F_N(h_p,h_A,c)=p(N)\left(1+N\frac{h_p(h_p-1)}{2h_A}\right)+\mathcal{O}(h_A^{-2}).\label{2.17}
\end{align}
\subsection*{Special cases}
\textbf{1. $h_p=1$}: From \eqref{2.15}, it is evident that the first term which involves at least one commutation of the $\phi_p(1)$ with the Virasoro generators is proportional to $(h_p-1)$, so the first term is 0. So, the generic matrix element satisfies $\frac{\langle h_A|(\prod_{j=N}^1 L_{j} ^{l_j})\phi_p(1)\prod_{i=1}^N L_{-i} ^{k_i}|h_A\rangle}{\langle h_A|\phi_p(1)|h_A\rangle}=\langle h_A|(\prod_{j=N}^1L_j^{l_j})\prod_{i=1}^N L_{-i} ^{k_i}|h_A\rangle$, i.e. it is proportional to the Gram-matrix elements. Thus, clearly, we have an exact result: $F_N(h_p=1,h_A,c)=p(N)$.\\
\textbf{2. $h_p=0$}: To finally evaluate the first term of \eqref{2.15}, we must finally use \eqref{2.13} as described in the previous section. If $h_p=0$, the R.H.S. of \eqref{2.13} is 0. So, the generic matrix element is again proportional to the corresponding Gram-matrix element. Thus, here also, we have the same result: $F_N(h_p=0,h_A,c)=p(N)$.

\subsection*{Asymptotic structure constant for torus one point function:}
After having the expression for CFT torus conformal block we will now briefly summarise the analysis of \cite{Kraus:2016nwo} to calculate torus one point function, more specifically the asymptotic form of the three point co-efficient. We will be exploiting the modular transformation property of CFT torus one point function. Under S modular transformation, the one point function transforms as,
\be{}
\langle \phi\rangle_{(\tau, \bar{\tau})}= \tau^{-h} \bar{\tau}^{-\bar h}\langle \phi\rangle_{(-\frac{1}{\tau},-\frac{1}{\bar{\tau}})}.
\ee
 Following \eqref{2.6} we can write,
 \be{}\label{A.30}
\begin{split}
\langle \phi\rangle_{(\tau, \bar{\tau})}&=\sum_{A} D(h_A,\bar{h}_A) e^{2\pi i \tau(h_A-\frac{c}{24})} e^{-2\pi i \bar{\tau}(\bar{h}_A-\frac{\bar{c}}{24})} C_{ApA} F^{h_p,\bar{h}_p}_{h_A,\bar{h}_A,c,\bar{c}}(\tau,\bar{\tau})\\
&= \tau^{-h_p} \bar{\tau}^{-\bar{h_p}}\sum_{A} D(h_i,\bar{h}_i) C_{ipi} \: e^{\big(-\frac{2\pi i}{\tau}(h_i-\frac{c}{24})\big)} \: e^{\big(\frac{2 \pi i}{\bar \tau}(\bar{h}_i-\frac{\bar c}{24})\big)}.
\end{split}
\ee
Here we have used $q= e^{2\pi i \tau},\: \bar{q}=e^{-2\pi i \bar{\tau}}$. The LHS of the expression is written in terms of the torus block of the primaries, where as the RHS is expanded in arbitrary eigenstates of virasoro zero mode $L_0, \bar{L}_0$. We have also used the notation for the three-point coefficients $C_{ApA}=\langle h_A,\bar{h}_A| \phi_p(1,1)|h_A,\bar{h}_A\rangle, \; C_{ipi}=\langle h_i,\bar{h}_i| \phi_p(1,1)|h_i,\bar{h}_i\rangle$. \eqref{A.30} relates the high temperature limit of the theory to the low temperature one. \\
Next we take the limit $\tau=\frac{i\beta}{2\pi}\to i0^{+} ,\; \bar{\tau}=\frac{-i\beta}{2\pi}\to-i 0^{+}$ or $\beta \to 0^{+}$, and write the RHS in terms of the lightest state $\chi$ for which $\langle \chi | \phi | \chi \rangle\neq 0$. Then the RHS is given by,
\be{}
\text{RHS}\approx i^{h_p-\bar{h}_p}\left( \frac{2\pi}{\beta}\right)^{h_p+\bar{h}_p}D(h_\chi, \bar{h}_{\chi}) C_{\chi p \chi}e^{-\frac{4\pi^2}{\beta}(h_\chi+\bar{h}_\chi-\frac{c}{12})}.
\ee
We will rewrite the sum of the scaling dimension of the state $\chi$ as $E_\chi=h_\chi+\bar{h}_\chi$ and also use $c=\bar{c}$. Accordingly, we also write the contributions from the scaling dimension of the operator $\phi$ as $E_p=h_p+\bar{h}_p$ and $s_p=h_p-\bar{h}_p$. We also write $T_p(h_A,\bar{h}_{A})\equiv D(h_A,\bar{h}_A)C_{ApA}$ and $T_p(h_\chi,\bar{h}_\chi)\equiv D(h_A,\bar{h}_A)C_{\chi p \chi}$.\\
Now,  the small $\beta$ limit implies the torus blocks in the LHS to be approximated in the large $h_A,\bar{h}_A$ limit. Following \cite{Kraus:2016nwo}, the torus blocks can be written as,
\be{}
F^{h_p,\bar{h}_p}_{h_A,\bar{h}_A,c}(\beta) \approx \beta e^{\frac{1}{12}\left( -\beta+\frac{4\pi^2}{\beta}\right)}
\ee
This expression is approximated in small $\beta$ limit only. Finally, we can perform an inverse Laplace transform on \eqref{A.30} to find the form of $T_p(h_A,\bar{h}_A)$ as,
\be{}\label{A.33}
T_p(h_A,\bar{h}_A)=\frac{i^{s_p}}{2\pi} \int dh_A d{\bar{h}_A}T_p({h_{\chi},\bar{h}_\chi}) \left(\frac{2\pi}{\beta}\right)^{E_p+1} \text{Exp}\left[(h_A+\bar{h}_A-\frac{c-1}{12})\beta-(E_\chi-\frac{c-1}{12})\frac{4\pi^2}{\beta}\right]
\ee
At large $h_A$ limit, this integral can be approximated by the contribution from the saddle point at
\be{}
\beta \approx 2\pi \sqrt{\frac{\frac{c-1}{12}-E_{\chi}}{E-\frac{c-1}{12}}}+\frac{E_p+1}{2(E-\frac{c-1}{12})}+\hdots .
\ee
Here, we have used $E=h_A+\bar{h}_A$. The leading terms dominate the suppressed terms in the large $h_A,\bar{h}_A$ limit or $E\to \infty$ limit. The above equation is also valid as long as the operator is light that is $E_\chi< < \frac{c}{12}$. Putting this value back into \eqref{A.33},  we obtain
\be{}\label{A.35}
\begin{split}
\frac{T_p(h_A,\bar{h}_A)}{T_p(h_\chi,\bar{h}_\chi)}\approx& \frac{i^{s_p}}{\sqrt{2\pi}}\left( \frac{c-1}{12}-E_\chi\right)^{-\frac{E_p}{2}-\frac{1}{4}} \left( E-\frac{c-1}{12}\right)^{\frac{E_p}{2}-\frac{1}{4}}\\
&~~~~~~ \text{Exp}\left(4\pi\sqrt{\left( \frac{c-1}{12}-E_\chi\right) \left( E-\frac{c-1}{12}\right)} +\hdots \right)
\end{split}
\ee
At this point, to find the asymptotic formula for density of states in the large $h_A$ limit, we put $E_p=0=s_p$ directly in the above equation and also replace the primary field $\chi$ by the lightest primary field $\O$ present in the theory.
\be{}
\begin{split}
\frac{D_p(h_A,\bar{h}_A)}{D_p(h_\mathcal{O},\bar{h}_\mathcal{O})}\approx& \frac{1}{\sqrt{2\pi}}\left( \frac{c-1}{12}-E_\mathcal{O}\right)^{-\frac{1}{4}} \left( E-\frac{c-1}{12}\right)^{-\frac{1}{4}}\\
&~~~~~~ \text{Exp}\left(4\pi\sqrt{\left( \frac{c-1}{12}-E_\mathcal{O}\right) \left( E-\frac{c-1}{12}\right)} +\hdots \right)
\end{split}
\ee
We can see that in the large $E\to \infty$ limit and taking $\mathcal{O}$ to be the vacuum $E_\mathcal{O}=0$, we match this expression with that of  \cite{Kraus:2016nwo}. Substituting this back to \eqref{A.35}, we finally obtain the asymptotic formula of three-point coefficient for primary fields for large $h_A,\bar{h}_A$ or $E\to \infty$, 
\be{}
\begin{split}
\text{C}_{ApA}\approx& i^{s_p}\frac{D{(h_\chi,\bar{h}_\chi)}}{D{(h_\mathcal{O},\bar{h}_{\mathcal O)}}}C_{\chi p \chi}\left(\frac{c-1}{12}-E_\mathcal{O}\right)^{\frac{1}{4}}\left(E-\frac{c-1}{12}\right)^{\frac{E_p}{2}}\left(\frac{c-1}{12}-E_\chi\right)^{-\frac{E_p}{2}-\frac{1}{4}}\\
&~~~~~\text{Exp}\left[-\frac{\pi (c-1) }{12}\left(\sqrt{1-\frac{12E_\mathcal{O}}{c-1}}-\sqrt{1-\frac{12E_\chi}{c-1}} \right)\sqrt{\frac{12E}{c-1}-1} \right]    
\end{split}
\ee
If we assume there is no degeneracy in the system along with $\mathcal O$ to be the vacuum state, we get back the same result as \cite{Kraus:2016nwo}.

\newpage

\section{Finite BMS transformations}\label{Finite ${BMS}_3$ transformations}
In this section, our main goal is to derive the transformation rule for the $BMS_3$ primary fields under finite $BMS_3$ transformation. We will do this by deriving the finite $GCA_2$ transformation of primary fields by taking non-relativistic contraction of the corresponding rule in $CFT_2$. The result for $BMS$ is obtained form that of $GCA$ by flipping the coordinates. 

Under a local conformal transformation $z\rightarrow w(z)$ and $\bar{z}\rightarrow \bar{w}(\bar{z})$, the $CFT_2$ primary field with conformal weights ($h$, $\bar{h}$) transforms like:
\begin{align}\label{eq1}
\tilde{\phi}(w,\bar{w})=(\frac{dw}{dz})^{-h}(\frac{d\bar{w}}{d\bar{z}})^{-\bar{h}}{\phi}(z,\bar{z})
\end{align}
In order to look at non-relativsitic contraction, we write the complex co-ordinates as $z=u+v$ and $\bar{z}=u-v$, where $u$ is the Euclidean time and $v$ is the space coordinate. Now, we have defined our conformal transformation in such a way that $w(z)$ is a holomorphic function of $z$ and $\bar{w}(\bar{z})$ is an anti-holomorphic function of $\bar{z}$. Let 
\begin{align}
&w(z)= {u}^\prime(u, v)+{v}^\prime(u, v) \nonumber\\
&\bar{w}(\bar{z})= {u}^\prime(u, v)-{u}^\prime(u, v)
\end{align}
By using the fact that $\frac{\partial w(z)}{\partial\bar{z}}=\frac{\partial \bar{w}(\bar{z})}{\partial z}=0$, we get the Cauchy-Riemann equation 
\be 
\partial_u u^\prime=\partial_v v^\prime,\quad\quad \partial_v u^\prime=\partial_u v^\prime.
\label{cauchy_riemann}
\ee
Using this, we can deduce that 
\begin{align}
&\frac{dw}{dz}=\partial_u u^\prime+\partial_u v^\prime, \quad\quad \frac{d\bar{w}}{d\bar{z}}=\partial_u u^\prime-\partial_u v^\prime.
\end{align}

Now, let us consider the non-relativistic contraction of the Virasoro symmetry algebra of the relativistic CFT, by the scaling $u \rightarrow  u$ and $v \rightarrow \epsilon v$ with $\epsilon\rightarrow0$ and similar scaling is applied to the primed coordinates. Also, conformal weights are scaled as $h\rightarrow\frac{1}{2}(\Delta-\frac{\xi}{\epsilon})$ and $\bar{h}\rightarrow\frac{1}{2}(\Delta+\frac{\xi}{\epsilon})$. Thus, we proceed to apply the contraction to \eqref{eq1}:
\begin{align}
(\frac{dw}{dz})^{-h}(\frac{d\bar{w}}{d\bar{z}})^{-\bar{h}}&=
\lim_{\epsilon\rightarrow0}(\partial_u u^\prime+\epsilon\partial_u v^\prime)^{-\frac{1}{2}(\Delta-\frac{\xi}{\epsilon})}(\partial_u u^\prime-\epsilon\partial_u v^\prime)^{-\frac{1}{2}(\Delta+\frac{\xi}{\epsilon})}\nonumber\\
&=(\partial_u u^\prime)^{-\Delta}e^{\xi\frac{\partial_u v^\prime}{\partial_u u^\prime}}
\end{align}

If we apply the non-relativistic contraction to the Cauchy-Riemann equations \eqref{cauchy_riemann}, we get 
\be 
\partial_u u^\prime=\partial_v v^\prime,\quad\quad \lim_{\epsilon\rightarrow0}\frac{1}{\epsilon}\partial_v u^\prime=-\lim_{\epsilon\rightarrow0}\epsilon\partial_u v^\prime.
\ee 
Clearly, the second condition tells us that $\partial_v u^\prime$ must be 0, which in turn implies that $u^\prime=u^\prime(u)$. By solving the remaining equation, we get $v^\prime(u,v)=v\partial_u u^\prime+f(u)$, where $f(u)$ is any arbitrary well-behaved function.
Thus, under the finite $GCA_2$ transformation $u\rightarrow u^\prime = u^\prime(u)$ and $x\rightarrow v^\prime=v\partial_u u^\prime+f(u)$, the  primary field with $GCA_2$ weights ($\Delta$, $\xi$) transforms as:
\begin{align}
\tilde{\phi}(u^\prime, v^\prime)=(\partial_u u^\prime)^{-\Delta}e^{\xi\frac{\partial_u v^\prime}{\partial_u u^\prime}}\phi(u, v).
\label{finite_gca2}
\end{align}

The way to obtain the finite $BMS_3$ transformation would be to use the commutation of the $BMS_3$ generators with the primary field. We know that the difference in the commutation of the primary field with the generators in $GCA_2$ and $BMS_3$ is that $u$ and $v$ are exchanged. 
Thus for finite $BMS_3$ transformation, we just have to exchange $u$ and $v$ in the previous equation \eqref{finite_gca2} giving us
\begin{align}
\tilde{\phi}(u^\prime, v^\prime)=(\partial_v v^\prime)^{-\Delta}e^{\xi\frac{\partial_v u^\prime}{\partial_v v^\prime}}\phi(u, v).
\end{align}
As for the coordinate, finite $BMS_3$ transformation are given by
\be
u \rightarrow u^{\prime} = u\partial_v f(v) + g(v),\quad \quad v \rightarrow v^{\prime} = f(v).  
\ee
Let us consider the mapping from 2-d Euclidean plane (spanned by $(t,x)$) to cylinder (spanned by $(u,\phi)$) in $BMS_3$, given by: $t=-iue^{i\phi}$ and $x=e^{i\phi}$. The primary field in the cylinder $\phi_{\rm cyl}(u,\phi)$ with $(\Delta, \xi)$ are related to the primary field on the plane $\phi(t,x)$ as
\be
 \phi_{\rm cyl}(u,\phi) = A e^{i\phi\D}e^{-iu\xi}\phi_p(t,x).
\ee where p in subscript denotes `plane' and $A$ is constant which can be absorbed to $\phi_p(t,x)$. This is the same transformation rule as in \eqref{ptceq}.

\newpage

\section{Some useful identities and their derivations}
\label{appendix:identities}

\subsection{Some useful identities}
\label{sec:useful_identities}
In this section we will list out useful identities for the computation of the matrix elements $\<A;N,i|\phi_p(0,1)e^{2\pi i \rho M_0}|A;N,j\>$ in \eqref{F_N}. These identities are derived using the commutation relation \eqref{commutation_mode} and their derivations are given in the next subsection. The primary field $\phi_p$ is at $t=0,x=1$ and the mode expansion \eqref{mode_exp} at this point is given by
\begin{align}
\phi_p(0,1)=\sum_{m\in \mathbb{Z}}\phi_{m,0}.
\label{phi10_mode_expansion}
\end{align}
The first identity is given by 
\begin{align}
\sum_{m\in\mathbb{Z}}(m+N-M)\langle A;N,i|\phi_{m,n}|A;M,j\rangle &=0\nonumber\\
\implies \sum_{m\in\mathbb{Z}}m\langle A;N,i|\phi_{m-N+M,n}|A;M,j\rangle &=0 \text{    , (for any $n\in \mathbb{Z}-\mathbb{N}$)}.
\label{first_identity}
\end{align} 
This is for $N,M\geqslant0$. In the following, $M_{-\vec{i}}$ is any string of $M$-operators only (0-length string is also allowed), that acts on the highest-weight state $|\Delta_A,\xi_A\rangle\equiv|A\rangle$ to produce a descendant state. 
The second identity is given by

\begin{align}
\sum_{m\in\mathbb{Z}}\langle A|M_{\vec{j}}\phi_{m,n}M_{-\vec{i}}|A\rangle &=0   \text{    , (for any $n\in \mathbb{Z^-}$)},
\label{second_identity}
\end{align}
or, as a consequence
\begin{align}
&\sum_{m\in\mathbb{Z}}\langle A|\phi_{m,n}|A\rangle=0   \text{    , (for any $n\in \mathbb{Z^-}$)}.
\end{align}
Next, we note down a useful commutation relation
\begin{align}
[M_0,L_{-n}^{l_n}]=nl_nM_{-n}L_{-n}^{l_n-1}.
\end{align}
Using the above commutation relation and the BCH-lemma, we get the third useful identity given below
\begin{align}
&e^{2\pi i\rho M_0}\prod_{n=1}^N L_{-n}^{l_n}=(\prod_{n=1}^Ne^{2\pi i\rho M_0}L_{-n}^{l_n}e^{-2\pi i\rho M_0})e^{2\pi i\rho M_0}\nonumber\\
&=\{\prod_{n=1}^N(\sum_{k=0}^{\infty}\frac{{(2\pi i\rho)}^k}{k!}[\underbrace{M_0,[M_0,[M_0,...[M_0}_\text{$k$ times},L_{-n}^{l_n}]]]...])\}e^{2\pi i\rho M_0}\nonumber\\
&=\{\prod_{n=1}^N(\sum_{k=0}^{l_n}\frac{{(2\pi in\rho)}^k}{k!}\frac{{l_n}!}{(l_n-k)!}L_{-n}^{l_n-k}M_{-n}^{k})\}e^{2\pi i\rho M_0}\nonumber\\
&=\left(\prod_{n=1}^N{(L_{-n}+2\pi in\rho M_{-n})}^{l_n}\right)e^{2\pi i\rho M_0}.
\label{bch}
\end{align}
These are all the necessary identities for calculating the matrix elements $\<A,N,i|\phi_p(0,1)|A,N,j\>$.
We also note that $\phi_p^{\dagger}(0,1)=\phi_p(0,1)$. Let us we now show below how to use these identities to calculate  the three matrix elements in level 1 using \eqref{phi10_mode_expansion}-\eqref{bch}:
\be{}
\begin{split}
&\langle A|M_1\phi_p(0,1)y^{M_0}M_{-1}|A\rangle\cr
&~=y^{\xi_A}\sum_{m\in \mathbb{Z}}\langle A|M_1 \phi_{m,0}M_{-1}|A\rangle\\
&~=y^{\xi_A}\sum_{m\in \mathbb{Z}}\langle A|[M_1,\phi_{m,0}]M_{-1}|A\rangle y^{\xi_A}+\sum_{m\in \mathbb{Z}}\langle A|\phi_{m,0}M_{1}M_{-1}|A\rangle\cr
&~=y^{\xi_A}\sum_{m\in \mathbb{Z}}\langle A|(\xi_p\phi_{m+1,0}+\phi_{m+1,-1})M_{-1}|A\rangle + 0\cr
&~ = y^{\xi_A}\sum_{m\in \mathbb{Z}}\langle A|\xi_p\phi_{m+1,0}M_{-1}|A\rangle +\langle A|\phi_{m+1,-1}M_{-1}|A\rangle \cr
& ~=  y^{\xi_A}\sum_{m\in \mathbb{Z}}\langle A|\xi_p [\phi_{m+1,0},M_{-1}]|A\rangle +  y^{\xi_A}\sum_{m\in \mathbb{Z}}\langle A|M_{-1}\xi_p\phi_{m+1,0}|A\rangle + 0 \cr
&~ = y^{\xi_A}\xi_p^2\sum_{m\in \mathbb{Z}}\langle A|\phi_{m,0}|A\rangle + y^{\xi_A}\xi_p\sum_{m\in \mathbb{Z}}\langle A|\phi_{m,-1}|A\rangle + 0\cr
&~=y^{\xi_A}{\xi_p}^2\langle A|\phi_p(0,1)|A\rangle,
\end{split}
\ee

\begin{align}
&\langle A|L_1\phi_p(1,0)y^{M_0}M_{-1}|A\rangle\cr
&=y^{\xi_A}\sum_{m\in \mathbb{Z}}\langle A|[L_1, \phi_{m,0}]M_{-1}|A\rangle+y^{\xi_A}(2\xi_A)\langle A|\phi_p(1,0)|A\rangle\non
\end{align}
\begin{align}
&=y^{\xi_A}\sum_{m\in \mathbb{Z}}(\Delta_p-1-m)\langle A|\phi_{m+1,0}M_{-1}|A\rangle+y^{\xi_A}(2\xi_A)\langle A|\phi_p(1,0)|A\rangle\cr
&=y^{\xi_A}[\xi_p(\Delta_p-1)+2\xi_A]\langle A|\phi_p(1,0)|A\rangle\\
&\text{                             }\cr
&\langle A|M_1\phi_p(1,0)y^{M_0}L_{-1}|A\rangle\cr
&=y^{\xi_A}\langle A|M_1\phi_p(1,0)(L_{-1}+2\pi i\rho M_{-1})|A\rangle\cr
&=y^{\xi_A}[{\langle A|L_1\phi_p(1,0)M_{-1}|A\rangle}^*+\langle A|M_1 2\pi i\rho\phi_p(1,0) M_{-1}|A\rangle]\cr
&=y^{\xi_A}[(2\pi i\rho){\xi_p}^2+\xi_p(\Delta_p-1)+2\xi_A]\langle A|\phi_p(1,0)|A\rangle
\end{align}
These results are summarised in \eqref{eq5}.\\

Using the commutation relation \eqref{commutation_mode} and the identities given above, we could deduce a very useful identity

\be{}
\begin{split}\label{matrix_alma}
&\hspace{-1.2cm}\langle A|M_n^j L_n^{N-j} \phi_p(0,1)L_{-n}^{N-k}M_{-n}^{k}|A\rangle\\
&=n^{N-j}\sum_{m=0}^{N-j} \binom{N-j}{m}m!(\Delta_p+m-N+j)_{N-j-m}\\
&~~~~\times \sum_{l=0}^m\binom{k}{l}\binom{N-k}{m-l}{(2\xi_{A}^\prime)}^ln^{N-k}(\frac{2\Delta_{A}^\prime}{n}+N+k-m)_{m-l}\\
&~~~~\times \sum_{r=0}^{N-k-m+l}{(2\xi_{A}^\prime)}^rr!\binom{j}{r}\binom{N-k-m+l}{r}(\Delta_p+k-l-j+r)_{N-k-m+l-r}\\
&~~~~~~{(n\xi_p)}^{k-l+j-r}\langle A|\phi_p(0,1)|A\rangle,
\end{split}
\ee
where $\Delta_{A}^\prime=\Delta_A+(n^2-1)\frac{c_L}{2}$ and $\xi_{A}^\prime=\xi_A+(n^2-1)\frac{c_M}{2}$. Here, we have also used the Pochhammer symbol 
\be
(A)_i=\Pi_{m=0}^{i-1}(A+m).
\ee
Using this, we could derive the following repeatedly occurring matrix element
\begin{align}
&\langle A|M_n^j L_n^{N-j} \phi_p(0,1)e^{2\pi i \rho M_0}L_{-n}^{N-k}M_{-n}^{k}|A\rangle\nonumber\\
&\non\\
&= e^{2\pi i \rho\xi_A}\langle A|M_n^j L_n^{N-j} \phi_p(0,1)(L_{-n}+2\pi i \rho n M_{-n})^{N-k}M_{-n}^{k}|A\rangle\nonumber\\
&\non\\
&=e^{2\pi i \rho\xi_A}\sum_{s=0}^{N-k}\binom{N-k}{s}(2\pi i \rho n)^{N-k-s}\langle A|M_n^j L_n^{N-j} \phi_p(0,1)L_{-n}^{s} M_{-n}^{N-s}|A\rangle.
\end{align}
and from there, we only require to substitute $k$ by $N-s$ in \eqref{matrix_alma}. 
Hence,

\begin{equation}\label{matrix_fi}
\begin{split}
&\langle  A|M_n^j L_n^{N-j} \phi_p(0,1)e^{2\pi i \rho M_0}L_{-n}^{N-k}M_{-n}^{k}|A\rangle\\
&\\
&~~=e^{2\pi i \rho\xi_A}\sum_{m=0}^{N-j} \binom{N-k}{s} (2\pi i \rho n)^{N-k-s} \big[ n^{N-j} \sum_{m=0}^{N-j}\binom{N-j}{m}m! (\Delta_P+m-N+j)_{N-j-m}\\
&\\
&~~~~\times\sum_{l=0}^m \binom{N-s}{l} \binom{s}{m-l}(2\xi_A^\prime)^l n^s \left(\frac{2\Delta_A^\prime}{n}+2n-s-m\right)_{m-l}\\
&\\
&~~~~\times\sum_{r=0}^{s-m-l}(2\xi_A^\prime)^r r! \binom{j}{r} \binom{s-m+l}{r}(\Delta_P+N-s-l-j+r)_{s-m+l-r}(n\xi_p)^{N-s-l+j-r}\langle A| \phi_p(0,1)|A \rangle\big].
\end{split}
\end{equation}Detail derivations of all these identities are given in below.

\subsection{Derivations}
In this appendix we will give a derivation of the identities listed above.
First of all, for $N,M\geqslant0$, we have
\bea 
&&\langle A;N,i|L_0 \phi_p(0,1)|A;M,j\rangle
=\langle A;N,i|[L_0,\phi_p(0,1)]|A;M,j\rangle +\langle A;N,i| \phi_p(0,1)L_0|A;M,j\rangle \cr
&&\non\\
&&\implies \sum_{m\in\mathbb{Z}}(\Delta_A+N)\langle A;N,i|\phi_{m,0}|A;M,j\rangle
=\sum_{m\in\mathbb{Z}}\langle A;N,i|(-m)\phi_{m,0}|A;M,j\rangle\non\\
 &&\hspace{7cm}~ +\sum_{m\in\mathbb{Z}}(\Delta_A+M)\langle A;N,i|\phi_{m,0}|A;M,j\rangle \cr
&&\implies  \sum_{m\in\mathbb{Z}}(m+N-M)\langle A;N,i|\phi_{m,0}|A;M,j\rangle = 0\\
&& \implies \sum_{m\in\mathbb{Z}}m\langle A;N,i|\phi_{m-N+M,0}|A;M,j\rangle = 0. \non
\eea
In the above, had we considered $\sum_{m\in\mathbb{Z}} \langle A;N,i|L_0 \phi_{m,n}|A;M,j\rangle$ as the starting point (for any $n\leqslant0$), we would have been led to \eqref{first_identity} (which is more general than the above)
\begin{align}
&\sum_{m\in\mathbb{Z}}(m+N-M)\langle A;N,i|\phi_{m,n}|A;M,j\rangle=0\nonumber\\
\Rightarrow&\sum_{m\in\mathbb{Z}}m\langle A;N,i|\phi_{m-N+M,n}|A;M,j\rangle=0 \text{    , (for any $n\in \mathbb{Z}-\mathbb{N}$)}.
\end{align} 
For any $n\in \mathbb{Z}-\mathbb{N}$, we have:

\begin{eqnarray}
&&\sum_{m\in\mathbb{Z}}\langle A|M_{\vec{j}}M_0 \phi_{m,n}M_{-\vec{i}}|A\rangle
 =\sum_{m\in\mathbb{Z}}[\langle A|M_{\vec{j}}[M_0,\phi_{m,n}]M_{-\vec{i}}|A\rangle+\langle A|M_{\vec{j}} \phi_{m,n}M_0M_{-\vec{i}}|A\rangle]\cr
&&\implies  \sum_{m\in\mathbb{Z}}(1-n)\langle A|M_{\vec{j}}\phi_{m,n-1}M_{-\vec{i}}|A\rangle =0.
\end{eqnarray}
This give us the second identity \eqref{second_identity}
\begin{align}
\sum_{m\in\mathbb{Z}}\langle A|M_{\vec{j}}\phi_{m,n}M_{-\vec{i}}|A\rangle &=0   \text{    , (for any $n\in \mathbb{Z^-}$)}.
\end{align}

Now, let us describe how to obtain \eqref{matrix_alma}. 
In the following, we use the notation $\Delta_{A}^\prime=\Delta_A+(n^2-1)\frac{c_L}{2}$ and $\xi_{A}^\prime=\xi_A+(n^2-1)\frac{c_M}{2}$ (without the loss of generality we take $j\geqslant k$, for faster evaluation and $n\geqslant1$).
We will also use the identities below for the next calculation.
\begin{equation}
\begin{split}
&[L_0,M^k_{-n}]=nkM^k_{-n},\\
&[L_0,L^k_{-n}]=nkL^k_{-n}\\
&[L_n,L^{N-k}_{-n}]M^k_{-n}=n(N-k)\big[n(N+k-1)L^{N-k-1}_{-n}M^{k}_{-n}\\
&~~~~~~~~~~~~~~~~~~~~~+2L_{-n}^{N-k-1}M^k_{-n}L_0+(n^2-1)c_L L^{N-k-1}_{-n}M^{k}_{-n}\big]
\end{split}
\end{equation}

\be{}
\begin{split}
&\langle A|M_n^j L_n^{N-j} \phi_p(0,1)L_{-n}^{N-k}M_{-n}^{k}|A\rangle\\
&\\
&={\langle A|M_n^k L_n^{N-k} \phi_p(0,1)L_{-n}^{N-j}M_{-n}^{j}|A\rangle}^*\cr
&\\
&=n(N-k)[2\Delta_{A}^\prime+n(N+k-1)]\langle A|M_n^j L_n^{N-j-1} \phi_p(0,1)L_{-n}^{N-k-1}M_{-n}^{k}|A\rangle+nk(2\xi_{A}^\prime)\cr
&\\
&\quad \langle A|M_n^j L_n^{N-j-1} \phi_p(0,1)L_{-n}^{N-k}M_{-n}^{k-1}|A\rangle+n(\Delta_p-1)\langle A|M_n^j L_n^{N-j-1} \phi_p(0,1)L_{-n}^{N-k}M_{-n}^{k}|A\rangle\non
\end{split}
\ee
\be{}
\begin{split}
&=n\sum_{m=0}^1 \binom{1}{m}m!(\Delta_p+m-1)_{1-m}\sum_{l=0}^m{(2\xi_{A}^\prime)}^l\binom{k}{l}\binom{N-k}{m-l}n^{m-l}\cr
&\qquad\qquad\times(\frac{2\Delta_{A}^\prime}{n}+N+k-m)_{m-l}\langle A|M_n^j L_n^{N-j-1} \phi_p(0,1)L_{-n}^{N-k-m+l}M_{-n}^{k-l}|A\rangle\cr
&\cr
&=\hdots\hdots\hdots\text{recursive reduction of the power of $L_n$}\hdots\hdots\hdots\cr
&\cr
&=n^{N-j}\sum_{m=0}^{N-j} \binom{N-j}{m}m!(\Delta_p+m-N+j)_{N-j-m}\sum_{l=0}^m\binom{k}{l}\binom{N-k}{m-l}\cr
&\qquad\qquad\qquad\times {(2\xi_{A}^\prime)}^l n^{m-l}(\frac{2\Delta_{A}^\prime}{n}+N+k-m)_{m-l}\langle A|M_n^j\phi_p(0,1)L_{-n}^{N-k-m+l}M_{-n}^{k-l}|A\rangle.
\end{split}
\ee
Here, we have used the Pochhammer symbol 
\be
(A)_i=\prod_{m=0}^{i-1}(A+m).
\ee
By similar kind of calculations, for $p,q,r,\{s_i\},\{t_j\} \in \mathbb{N}\cup\{0\}$, we are led to the following two identities:

\begin{align}
&\langle A|M_n^pL_{n}^{q}\phi_p(0,1)M_{-n}^{r}|A\rangle\\
&\non\\
&~={\langle A|M_{n}^r\phi_p(0,1)L_{-n}^{q}M_{-n}^{p}|A\rangle}^*\nonumber\\
&\non\\
&~=n(2\xi_{A}^\prime)r\langle A|M_n^pL_{n}^{q-1}\phi_p(0,1)M_{-n}^{r-1}|A\rangle
+n(\Delta_p+p+q-r-1)\langle A|M_n^pL_{n}^{q-1}\phi_p(0,1)M_{-n}^{r}|A\rangle\nonumber\\
&\non
\end{align}
\begin{align}
&~=n\sum_{l=0}^1{(2\xi_{A}^\prime)}^ll!\binom{r}{l}\binom{1}{l}(\Delta_p+p+q-r-1+l)_{1-l}\langle A|M_n^pL_{n}^{q-1}\phi_p(0,1)M_{-n}^{r-l}|A\rangle\nonumber\\
&\non\\
&~=\hdots\hdots\hdots\text{recursive reduction of the power of $L_n$}\hdots\hdots\hdots\nonumber\\
&\non\\
&~=n^q\sum_{l=0}^q{(2\xi_{A}^\prime)}^ll!\binom{r}{l}\binom{q}{l}(\Delta_p+p-r+l)_{q-l}\langle A|M_n^p\phi_p(0,1)M_{-n}^{r-l}|A\rangle.
\end{align}
Also,
\be{}
\begin{split}
&\hspace{-0.5cm}\langle A|(\prod_{i=1}^P M_{n_i}^{s_i})\phi_p(0,1)\prod_{j=1}^Q M_{-m_j}^{t_j}|A\rangle=\langle A|(\prod_{j=1}^Q M_{m_j}^{t_j})\phi_p(0,1)\prod_{i=1}^P M_{-n_i}^{s_i}|A\rangle^*\\
&=\sum_{m\in \mathbb{Z}}\langle A|(\prod_{i=2}^P M_{n_i}^{s_i})M_{n_1}^{s_1-1}(n_1\xi_p\phi_{m+n_1,0}+\phi_{m+n_1,-1})\prod_{j=1}^Q M_{-m_j}^{t_j}|A\rangle\\
&=n_1\xi_p\langle A|(\prod_{i=2}^P M_{n_i}^{s_i})M_{n_1}^{s_1-1}\phi_p(0,1)\prod_{j=1}^Q M_{-m_j}^{t_j}|A\rangle+0\\
&={(n_1\xi_p)}^{s_1}{\langle A|(\prod_{i=2}^P M_{n_i}^{s_i})\phi_p(0,1)\prod_{j=1}^Q M_{-m_j}^{t_j}|A\rangle}=(\prod_{i=1}^P{(n_i\xi_p)}^{s_i})\langle A|(\prod_{j=1}^Q M_{m_j}^{t_j})\phi_p(0,1)|A\rangle^*\\
&=(\prod_{i=1}^P{(n_i\xi_p)}^{s_i})(\prod_{j=1}^Q{(m_j\xi_p)}^{t_j}){\langle A|\phi_p(0,1)|A\rangle},
\label{eq4}
\end{split}
\ee
which imply that
\begin{equation}
 \langle A|M_n^pL_{n}^{q}\phi_p(0,1)M_{-n}^{r}|A\rangle = n^q\sum_{l=0}^q{(2\xi_{A}^\prime)}^ll!\binom{r}{l}\binom{q}{l}(\Delta_p+p-r+l)_{q-l}{(n\xi_p)}^{p+r-l}\langle A|\phi_p(0,1)|A\rangle.
\end{equation}

\noindent Thus finally we obtain \eqref{matrix_alma}
\begin{align}
&\langle A|M_n^j L_n^{N-j} \phi_p(0,1)L_{-n}^{N-k}M_{-n}^{k}|A\rangle\nonumber\\
&\cr
&~~=n^{N-j}\sum_{m=0}^{N-j} \binom{N-j}{m}m!(\Delta_p+m-N+j)_{N-j-m}\sum_{l=0}^m\binom{k}{l}\binom{N-k}{m-l}{(2\xi_{A}^\prime)}^l\nonumber\\
&\qquad\qquad\qquad\times  n^{m-l}(\frac{2\Delta_{A}^\prime}{n}+N+k-m)_{m-l}{\langle A|M_n^{k-l}L_{n}^{N-k-m+l}\phi_p(0,1)M_{-n}^{j}|A\rangle}^*\nonumber
\end{align}
\begin{align}
&~~=n^{N-j}\sum_{m=0}^{N-j} \binom{N-j}{m}m!(\Delta_p+m-N+j)_{N-j-m}\sum_{l=0}^m\binom{k}{l}\binom{N-k}{m-l}{(2\xi_{A}^\prime)}^l\nonumber\\
&\qquad\qquad\qquad\times  n^{N-k}(\frac{2\Delta_{A}^\prime}{n}+N+k-m)_{m-l}\sum_{r=0}^{N-k-m+l}{(2\xi_{A}^\prime)}^rr!\binom{j}{r}\binom{N-k-m+l}{r}\nonumber\\
&\qquad\qquad\qquad\times(\Delta_p+k-l-j+r)_{N-k-m+l-r}{(n\xi_p)}^{k-l+j-r}\langle A|\phi_p(0,1)|A\rangle.
\end{align}

Using this, we arrive at our desired, repeatedly occurring matrix element:
\begin{align}
&\langle A|M_n^j L_n^{N-j} \phi_p(1,0)e^{2\pi i \rho M_0}L_{-n}^{N-k}M_{-n}^{k}|A\rangle\nonumber\\
&\cr
&~~~=e^{2\pi i \rho\xi_A}\langle A|M_n^j L_n^{N-j} \phi_p(1,0)(L_{-n}+2\pi i \rho n M_{-n})^{N-k}M_{-n}^{k}|A\rangle\nonumber\\
&\cr
&~~~=e^{2\pi i \rho\xi_A}\sum_{s=0}^{N-k}\binom{N-k}{s}(2\pi i \rho n)^{N-k-s}\langle A|M_n^j L_n^{N-j} \phi_p(1,0)L_{-n}^{s} M_{-n}^{N-s}|A\rangle
\end{align}

and from there, we only require to substitute $k$ by $N-s$ in the last line.
Finally we obtain \eqref{matrix_fi},
\begin{equation}
\boxed{
\begin{split}
&\langle  A|M_n^j L_n^{N-j} \phi_p(0,1)e^{2\pi i \rho M_0}L_{-n}^{N-k}M_{-n}^{k}|A\rangle\\
&=e^{2\pi i \rho\xi_A}\sum_{m=0}^{N-j} \binom{N-k}{s} (2\pi i \rho n)^{N-k-s} \big[ n^{N-j} \sum_{m=0}^{N-j}\binom{N-j}{m}m! (\Delta_P+m-N+j)_{N-j-m}\\
&\sum_{l=0}^m \binom{N-s}{l} \binom{s}{m-l}(2\xi_A^\prime)^l n^s \left(\frac{2\Delta_A^\prime}{n}+2n-s-m\right)_{m-l}\\
&\sum_{r=0}^{s-m-l}(2\xi_A^\prime)^r r! \binom{j}{r} \binom{s-m+l}{r}(\Delta_P+N-s-l-j+r)_{s-m+l-r}\\
& (n\xi_p)^{N-s-l+j-r}\langle A| \phi_p(0,1)|A \rangle\big]
\end{split}
}
\end{equation}
Using the identities given above we calculate for level 2 the matrix elements that contribute to $\mathcal{F}_{2}(\Delta_p,\xi_p;\Delta_A,\xi_A|c_L,c_M|\rho)$. These are listed below

\begin{flalign}
&\frac{\langle A|L_1^{2} \phi_p(0,1)e^{2\pi i \rho M_0}M_{-1}^{2}|A\rangle}{\langle A|\phi_p(0,1)e^{2\pi i \rho M_0}|A\rangle}=8{\xi_A}^2+8\xi_A\xi_p(\Delta_p-1)+{\xi_p}^2(\Delta_p-2)_2,
\end{flalign}
\begin{flalign}
&\frac{\langle A|M_1^{2} \phi_p(0,1)e^{2\pi i \rho M_0}L_{-1}^{2}|A\rangle}{\langle A|\phi_p(0,1)e^{2\pi i \rho M_0}|A\rangle}=8{\xi_A}^2+8\xi_A\xi_p(\Delta_p-1)+{\xi_p}^2(\Delta_p-2)_2\nonumber\\
&\qquad\qquad\qquad\qquad\qquad\qquad\qquad+2(2\pi i\rho){\xi_p}^2[4\xi_A+\xi_p(\Delta_p-1)]+(2\pi i\rho)^2{\xi_p}^4,
\end{flalign}
\begin{flalign}
&\frac{\langle A|L_2 \phi_p(0,1)e^{2\pi i \rho M_0}M_{-2}|A\rangle}{\langle A|\phi_p(0,1)e^{2\pi i \rho M_0}|A\rangle}=4\xi_A+6c_M+4\xi_p(\Delta_p-1),
\end{flalign}
\begin{flalign}
&\frac{\langle A|M_2 \phi_p(0,1)e^{2\pi i \rho M_0}L_{-2}|A\rangle}{\langle A|\phi_p(0,1)e^{2\pi i \rho M_0}|A\rangle}=4\xi_A+6c_M+4\xi_p(\Delta_p-1)+(2\pi i\rho)8{\xi_p}^2,
\end{flalign}
\begin{flalign}
&\frac{\langle A|M_1L_1 \phi_p(0,1)e^{2\pi i \rho M_0}L_{-1}M_{-1}|A\rangle}{\langle A|\phi_p(0,1)e^{2\pi i \rho M_0}|A\rangle}=4\xi_A^2+4\xi_A\xi_p(\Delta_p-1)+\xi_p^2[2(\Delta_A+1)+\Delta_p(\Delta_p-1)]\nonumber\\
&\qquad\qquad\qquad\qquad\qquad\qquad\qquad\qquad\quad+(2\pi i\rho){\xi_p}^2[4\xi_A+\xi_p(\Delta_p-1)],
\end{flalign}
\begin{flalign}
&\frac{\langle A|M_2 \phi_p(0,1)e^{2\pi i \rho M_0}M_{-2}|A\rangle}{\langle A|\phi_p(0,1)e^{2\pi i \rho M_0}|A\rangle}=4\xi_p^2,
\end{flalign}
\begin{flalign}
&\frac{\langle A|M_2 \phi_p(0,1)e^{2\pi i \rho M_0}M_{-1}^2|A\rangle}{\langle A|\phi_p(0,1)e^{2\pi i \rho M_0}|A\rangle}=\frac{\langle A|M_1^2 \phi_p(0,1)e^{2\pi i \rho M_0}M_{-2}|A\rangle}{\langle A|\phi_p(0,1)e^{2\pi i \rho M_0}|A\rangle}=2\xi_p^3,\text{	(using \eqref{eq4})},
\end{flalign}
\begin{flalign}
&\frac{\langle A|M_1^2 \phi_p(0,1)e^{2\pi i \rho M_0}M_{-1}^2|A\rangle}{\langle A|\phi_p(0,1)e^{2\pi i \rho M_0}|A\rangle}=\xi_p^4,
\end{flalign}
\begin{flalign}
&\frac{\langle A|L_2 \phi_p(0,1)e^{2\pi i \rho M_0}M_{-1}^2|A\rangle}{\langle A|\phi_p(0,1)e^{2\pi i \rho M_0}|A\rangle}=2(\Delta_p-1)\xi_p^2,
\end{flalign}
\begin{flalign}
&\frac{\langle A|M_1^2 \phi_p(0,1)e^{2\pi i \rho M_0}L_{-2}|A\rangle}{\langle A|\phi_p(0,1)e^{2\pi i \rho M_0}|A\rangle}=2(\Delta_p-1)\xi_p^2+4(2\pi i\rho)\xi_p^3,
\end{flalign}
\begin{flalign}
&\frac{\langle A|M_1L_1 \phi_p(0,1)e^{2\pi i \rho M_0}M_{-1}^2|A\rangle}{\langle A|\phi_p(0,1)e^{2\pi i \rho M_0}|A\rangle}=\xi_p^2[4\xi_A+\xi_p(\Delta_p-1)],
\end{flalign}
\begin{flalign}
&\frac{\langle A|M_1^2 \phi_p(0,1)e^{2\pi i \rho M_0}L_{-1}M_{-1}|A\rangle}{\langle A|\phi_p(0,1)e^{2\pi i \rho M_0}|A\rangle}=\xi_p^2[4\xi_A+\xi_p(\Delta_p-1)]+(2\pi i\rho)\xi_p^4,
\end{flalign}
\begin{flalign}
&\frac{\langle A|M_1L_1 \phi_p(0,1)e^{2\pi i \rho M_0}M_{-2}|A\rangle}{\langle A|\phi_p(0,1)e^{2\pi i \rho M_0}|A\rangle}=2(\Delta_p-1)\xi_p^2+3\xi_p^2=(2\Delta_p+1)\xi_p^2,
\end{flalign}
\begin{flalign}
&\frac{\langle A|M_2 \phi_p(0,1)e^{2\pi i \rho M_0}L_{-1}M_{-1}|A\rangle}{\langle A|\phi_p(0,1)e^{2\pi i \rho M_0}|A\rangle}=(2\Delta_p+1)\xi_p^2+2(2\pi i\rho)\xi_p^3.
\end{flalign}

\bigskip

\section{Matrix elements for level 1}
\label{sec:matrxi_ele_level1}
In this appendix we give another way of calculating $\<\Psi_i|\phi_p(0,1)e^{2\pi i \rho M_0}|\Psi_j\>$ for level 1. For the first matrix elements we have
\bea
\<\Psi_1|\phi_p(0,1)e^{2\pi i M_0}|\Psi_2\> &=&  \<\D_A,\xi_A|L_1\phi_p(0,1)e^{2\pi i M_0}M_{-1}|\D_A,\xi_A\>\cr
&=& \<\D_A,\xi_A|L_1\phi_p(0,1)M_{-1}e^{2\pi i M_0}|\D_A,\xi_A\>\cr
&=& e^{2\pi i \xi_A} \<\D_A,\xi_A|L_1\phi_p(0,1)M_{-1}|\D_A,\xi_A\>.
\eea
The three point function
\be
 \<\D_A,\xi_A|L_1\phi_p(0,1)M_{-1}|\D_A,\xi_A\>,
\ee
can be calculated using
 \be
\<\D_A,\xi_A|\phi_p(t,x)|\D_A,\xi_A\> = C_{ApA}\,x^{-\D_p}e^{\xi_p\frac{t}{x}},
\ee
and 
\bea
[L_n,\phi_{\D,\xi}(t,x)]&=&[x^{n+1}\partial_x + (n+1)x^n t\partial_t +(n+1)(\D x^n -n \xi x^{n-1} t )]\phi_{\D,\xi}(t,x) \cr
[M_n,\phi_{\D,\xi}(t,x)] &=& [-x^{n+1}\partial_t + (n+1)\xi x^n]\phi_{\D,\xi}(t,x).
\eea
More precisely, we have
\bea
&&\hspace{-1.5cm} \<\D_A,\xi_A|L_1\phi_p(t,x)M_{-1}|\D_A,\xi_A\> \cr
&=& \<\D_A,\xi_A|L_1[\phi_p(t,x),M_{-1}]|\D,\xi\> + \<\D_A,\xi_A|L_1M_{-1}\phi_p(t,x)|\D_A,\xi_A\> \cr
&=&\partial_t \<\D_A,\xi_A|[L_1,\phi_p(t,x)]|\D_A,\xi_A\> + \<\D_A,\xi_A|2M_0\phi_p(t,x)|\D_A,\xi_A\> \cr
&=&\partial_t(t^2\partial_x+2xt\partial_t+2(\D_p x -\xi_p t ))x^{-\D_p}e^{\xi_p\frac{t}{x}} +
2\xi x^{-\D_p}e^{\xi_p\frac{t}{x}} \cr
&=& -x^{-1}\xi _p \left(-\Delta _p x + x+\xi _p t\right) \<\D,\xi|\phi_p(t,x)|\D,\xi\> + 2\xi_A\<\D_A,\xi_A|\phi_p(t,x)|\D_A,\xi_A\>.~
\eea
Thus
\be
 \<\D_A,\xi_A|L_1\phi_p(0,1)M_{-1}|\D_A,\xi_A\> = C_{ApA}(\xi_p\D_p - \xi_p + 2\xi_A).
\ee
So, finally
\be
 \<\Psi_1|\phi_p(0,1)e^{2\pi i M_0}|\Psi_2\> = C_{ApA}\,y^{\xi_A} (\xi_p\D_p - \xi_p + 2\xi_A).
\ee
Using \eqref{bch} and calculating the three-point functions like what we have just done, we get
\bea
&& \hspace{-0.6cm}\<\D_A,\xi_A|M_1\phi_p(0,1)e^{2\pi i M_0}L_{-1}|\D_A,\xi_A\> \cr
 &&=\<\D_A,\xi_A|M_1\phi_p(0,1)L_{-1}e^{2\pi i M_0}|\D_A,\xi_A\> + 2\pi i \rho \<\D_A,\xi_A|M_1\phi_p(0,1)M_{-1}e^{2\pi i M_0}|\D_A,\xi_A\>\cr
&&= e^{2\pi i \xi_A} \<\D_A,\xi_A|M_1\phi_p(0,1)L_{-1}|\D_A,\xi_A\> + 2\pi i \rho e^{2\pi i \xi_A} \<\D_A,\xi_A|M_1\phi_p(0,1)M_{-1}|\D_A,\xi_A\>\cr
&&= y^{\xi_A} C_{ApA}(\xi_p\D_p-\xi_p + 2\xi_A + 2\pi i \rho \xi_p^2).
\eea
For the fourth term in \eqref{block_level1_exp}, we have
\bea
&& \hspace{-1.6cm} \<\D_A,\xi_A|M_1\phi_p(0,1)e^{2\pi i M_0}M_{-1}|\D_A,\xi_A\>  \cr
&&= \<\D_A,\xi_A|M_1\phi_p(0,1)M_{-1}e^{2\pi i M_0}|\D_A,\xi_A\>  \cr
&&= y^{\xi_A} \<\D_A,\xi_A|M_1\phi_p(0,1)M_{-1}|\D_A,\xi_A\> \cr
&&=  y^{\xi_A} C_{ApA} \, \xi_p^2. 
\eea

\section{Proof for minimum number of uncontracted $M$-operators}
\label{sec:minimum_uncontracted}
In this appendix we will proof that for the generic $BMS_3$ Gram-matrix element with all $N_j\geqslant k_j\geqslant 0$, $N_j^\prime\geqslant k_j^\prime\geqslant 0$ and $\sum_{j=1}^N jN_j=\sum_{j=1}^N jN_j^\prime=N$:
\begin{align}
\langle A|(\prod_{j=N}^1 M_j^{k_j^\prime})(\prod_{j=N}^1 L_j^{N_j^\prime-k_j^\prime})(\prod_{j=1}^N L_{-j}^{N_j-k_j})(\prod_{j=1}^N M_{-j}^{k_j})|A\rangle,
\end{align}
the minimum number of `uncontracted' $M$-operators inside the bra-ket of inner-product is 2.

Let's assume that the total number of `uncontracted' $M$-operators inside the bra-ket of inner product is $m\geqslant2$ and that collection is denoted by $\{M_{i_r}\}$ with $1\leqslant r\leqslant m$ and $i_r\in \mathbb{Z}-\{0\}$. Now, define the sum of $L$-indexes of a collection $L$ of $n$-numbers of $L$-operators $\{L_j\}$ with $j\in \mathbb{Z}-\{0\}$ simply as $\sum_{L_j\in L} j$; the sum of $M$-indexes is similarly defined. To get an $M_j$ with $j\in \mathbb{Z}-\{0\}$ contracted, a collection of $L$-operators with sum of indexes equaling to $-j$ is needed. Thus,
\be  
\text{sum of}\,M\text{-indexes that have been contracted} =- \text{sum of}\,L\text{-indexes used in contraction}.
\ee
 Also, according to the starting assumption, along with the uncontracted $\{M_{i_r}\}$, there must also remain a collection (including the empty one) of $L$-operators, the elements of which can't be grouped into sub-collections with sum of indexes $-i_r$ for any $1\leqslant r\leqslant m$, to leave the whole $\{M_{i_r}\}$ uncontracted. But, as the two states giving rise to the inner product are at the same level of the $BMS_3$ module, we have:
\begin{align}
&\sum_{j=1}^N j(k_j^\prime-k_j)=-\sum_{j=1}^N j[(N_j^\prime-k_j^\prime)-(N_j-k_j)]\nonumber\\
\Rightarrow & \text{(sum of all $M$-indexes)}=-\text{(sum of all $L$-indexes)}\nonumber\\
\Rightarrow & \text{(sum of `contracted' $M$-indexes)}+\sum_{r=1}^m i_r
=-[\text{(sum of `contracted' $L$-indexes)}\cr
& \qquad\qquad\qquad\qquad\qquad \qquad\qquad\qquad\qquad\quad +\text{(sum of remaining $L$-indexes)}]\nonumber\\
\Rightarrow& \text{ }-\sum_{r=1}^m i_r=\text{(sum of remaining $L$-indexes)}
\end{align}
Even though no summand from the L.H.S. equals to any summand from the R.H.S., it is always the case that the two sums are equal to one another for $m\geqslant2$. But for $m=1$, i.e. when only one $M_i$ is uncontracted, we also have a collection (including the empty one) of $L$-operators, with sum of indexes $-k$, with $k\neq i$ to leave that $M_i$ uncontracted. But the condition of same level implies that $i=k$. Thus, we are led to a contradiction for $m=1$ and hence, we conclude that at least 2 among all the $M$-operators must be left uncontracted.

\section{Global torus blocks}

As an aside, to highlight an application of the rather cumbersome formula \eqref{matrix_alma}, we outline the steps to calculate the global $BMS_3$ torus block. In the highest-weight representation of the global $BMS_3$ algebra, the descendant states at level N in the $BMS_3$ module of the primary field with dimension $(\Delta_A,\xi_A)$ are given by $\{L_{-1}^{N-k}M_{-1}^k|A\rangle\}$, with $0\leqslant k\leqslant N$; so, there are $N+1$ descendants at level N. The generic matrix elements in the global version of \eqref{F_N} can be computed very easily using \eqref{matrix_alma} with $n=1$ (for convenience, we assume $j\leqslant k$):
\begin{align}
&\frac{\langle A|M_1^j L_1^{N-j} \phi_p(0,1)e^{2\pi i \rho M_0}L_{-1}^{N-k}M_{-1}^{k}|A\rangle}{\langle A|\phi_p(0,1)e^{2\pi i \rho M_0}|A\rangle}\nonumber\\
=&\sum_{s=0}^{N-k}\binom{N-k}{s}(2\pi i \rho)^{N-k-s}\sum_{m=0}^{N-j} \binom{N-j}{m}m!(\Delta_p+m-N+j)_{N-j-m}\nonumber\\
&\times\sum_{l=0}^m{(2\xi_{A})}^{m-l}\binom{N-s}{m-l}\binom{s}{l}(2\Delta_{A}+2N-s-m)_{l}\sum_{r=0}^{s-l}{(2\xi_{A})}^rr!\binom{j}{r}\nonumber\\
&\times\binom{s-l}{r}(\Delta_p+N-j-m-s+l+r)_{s-l-r}{\xi_p}^{N-s+l+j-m-r}.
\end{align}
While deriving the previous equation we have replaced $k$ by $N-s$, and $l$ by $m-l$ in \eqref{matrix_alma}.
Also, it can be easily shown that the generic element of the global $BMS_3$ Gram-matrix takes the following form:
\begin{align}
&\langle A|M_1^j L_1^{N-j}L_{-1}^{N-k}M_{-1}^{k}|A\rangle=\langle A|M_1^k L_1^{N-k}L_{-1}^{N-j}M_{-1}^{j}|A\rangle\nonumber\\
&={(2\xi_A)}^{j+k}\sum_{l=0}^j\binom{j}{l}\frac{(N-j)!(N-k)!}{(N-j-k-l)!}(2\Delta_A+k+l)_{N-j-k-l}.
\end{align}
From the above expression, it is evident that $\langle A|M_1^j L_1^{N-j}L_{-1}^{N-k}M_{-1}^{k}|A\rangle=0$, when $N<j+k$, i.e. for the global $BMS_3$ algebra also, the Gram-matrix can be brought into an upper anti-triangular form. Hence, the inverse Gram-matrix must be lower anti-triangular. Thus, for $N>j+k$, the $\langle A|M_1^j L_1^{N-j} \phi_p(0,1)e^{2\pi i \rho M_0}L_{-1}^{N-k}M_{-1}^{k}|A\rangle$ matrix elements don't contribute to the global torus block. Finally, putting all the components in right places as in the global version of \eqref{F_N}, we get the expression for the global torus block. This method thus offers a different route to calculate these global block as compared to the differential equation way of doing this outlined in the main text. 

\newpage

\section{Alternate derivation: Large $\D_A,\xi_A$ limit of one-point blocks}
In this appendix we gave an alternative derivation of the leading term in the large $\xi_A$ limit. We also explicitly find the subleading term for level 2. 
\subsection{Leading term}
We insert the identity operator for states at level $N$
\be
{\mathbb{1}}_N = \sum_{k,l} K^{kl}_{(N)}  |\D_A,\xi_A;N,k\>\<\D_A,\xi_A;N,l|
\ee
to the RHS of \eqref{F_N} giving us
\bea
&& \mathcal{F}_{N}(\Delta_p,\xi_p;\Delta_A,\xi_A|c_L,c_M|\rho)\cr
&=& \frac{1}{y^{\xi_A}C_{ApA}}\sum_{i,j,k,l} K^{ij}_{(N)} K^{kl}_{(N)}\<\D_A,\xi_A;N,i|\phi_p(0,1)|\D_A,\xi_A;N,k\>\cr
&&\quad\quad\quad\quad\quad\quad \quad\quad\quad\quad \quad \times \<\D_A,\xi_A;N,l|e^{2\pi i M_0}|\D_A,\xi_A;N,j\>.
\label{F_N_exp}
\eea
Let us schematically write the states as
\be
 |\D_A,\xi_A;N,k\> = L_{\vec{r}_k}M_{\vec{s}_k}|\D_A,\xi_A\>,
\ee
where $\vec{r}_k$ and $\vec{s}_k$ represents a concatenation of the the $L$'s and $M$'s for the $k$-th state. In this notation
\begin{align}
& \<\D_A,\xi_A;N,i|\phi_p(0,1)|\D_A,\xi_A;N,k\> \cr
= & \<\D_A,\xi_A|(M_{\vec{s}_i})^{\dagger}(L_{\vec{r}_i})^{\dagger}\phi_p(0,1) L_{\vec{r}_k}M_{\vec{s}_k}|\D_A,\xi_A\> \cr
=&  \<\D_A,\xi_A|(M_{\vec{s}_i})^{\dagger}(L_{\vec{s}_i})^{\dagger}L_{\vec{r}_k}M_{\vec{s}_k}\phi_p(0,1) |\D_A,\xi_A\> \cr
& + \<\D_A,\xi_A|(M_{\vec{s}_i})^{\dagger}(L_{\vec{r}_i})^{\dagger}[\phi_p(0,1),L_{\vec{r}_k}M_{\vec{s}_k}]|\D_A,\xi_A\> \cr
=&  \<\D_A,\xi_A|(M_{\vec{s}_i})^{\dagger}(L_{\vec{r}_i})^{\dagger}L_{\vec{r}_k}M_{\vec{s}_k}|\D_A,\xi_A\>\<\D_A,\xi_A|\phi_p(0,1) |\D_A,\xi_A\> \cr
& + \<\D_A,\xi_A|(M_{\vec{s}_i})^{\dagger}(L_{\vec{r}_i})^{\dagger}[\phi_p(0,1),L_{\vec{r}_k}M_{\vec{s}_k}]|\D_A,\xi_A\>\cr
=& K_{ik}^{(N)} C_{ApA} + \<\D_A,\xi_A|(M_{\vec{q_i}})^{\dagger}(L_{\vec{r}_i})^{\dagger}[\phi_p(0,1),L_{\vec{s}_k}M_{\vec{s}_k}]|\D_A,\xi_A\>.
\end{align}
Substituting this in \eqref{F_N_exp}, we have
\bea 
&& \mathcal{F}_{N}(\Delta_p,\xi_p;\Delta_A,\xi_A|c_L,c_M|\rho)\cr
&=& \frac{1}{y^{\xi_A}}\sum_{i,j,k,l} K^{ij}_{(N)} K^{kl}_{(N)} K_{ik}^{(N)} \<\D_A,\xi_A;N,l|e^{2\pi i M_0}|\D_A,\xi_A;N,j\>\cr
&& + \frac{1}{y^{\xi_A}C_{ApA}}\sum_{i,j,k,l} K^{ij}_{(N)} K^{kl}_{(N)}\<\D_A,\xi_A|(M_{\vec{q_i}})^{\dagger}(L_{\vec{q_i}})^{\dagger}[\phi_p(0,1),L_{\vec{p_k}}M_{\vec{q_k}}]|\D_A,\xi_A\>\cr
&&\quad\quad\quad\quad\quad\quad\quad\quad \times\<\D_A,\xi_A;N,l|e^{2\pi i M_0}|\D_A,\xi_A;N,j\>. 
\eea
The presence of a commutator inside a three-point function decrease the exponent of $\D_A$ or $\xi_A$. Thus the first term which does not have a commutator is the leading order in $\D_A$ and $\xi_A$. As we will see this is zeroth order in $\D_A$ and $\xi_A$. The other term contain subleading term (order $\mathcal{O}(\frac{1}{\D_A},\frac{1}{\xi_A})$)and higher order corrections.
We can simplify the first term
 using $\sum_{k} K^{kl}_{(N)} K_{ik}^{(N)} = \delta^l_i$ and using that fact that tracing the operator $e^{2\pi i \rho M_0}$ over the states in level $N$ gives $y^{\xi_A}$ times the number of states in level $N$
 \bea
 \hspace{-0.5cm}  \frac{1}{ y^{\xi_A}}\sum_{i,j,l} K^{ij}_{(N)} \delta^l_i \<\D_A,\xi_A;N,l|e^{2\pi i M_0}|\D_A,\xi_A;N,j\> 
  &=& \frac{1}{ y^{\xi_A}}\sum_{i,j} K^{ij}_{(N)}  \<\D_A,\xi_A;N,i|e^{2\pi i M_0}|\D_A,\xi_A;N,j\> \cr
  &=& \frac{1}{ y^{\xi_A}} {\rm Tr}_{N} e^{2\pi i M_0}  =  \frac{1}{ y^{\xi_A}} \widetilde{\rm \dim}_N \,y^{\xi_A} =  \widetilde{\rm \dim}_N.
 \eea
 Thus the leading term of $\mathcal{F}_{N}$ is just the number states at level $N$ 
 \be
  \mathcal{F}_{N}(\Delta_p,\xi_p;\Delta_A,\xi_A|c_L,c_M|\rho) = \widetilde{\rm \dim}_N + \mathcal{O}(\frac{1}{\xi_A},\frac{\D_A}{\xi_A^2})+...
 \ee
 
\subsection{Sub-leading term for level 2}
We can read off the subleading term in level 1 from \eqref{level1_bms}, it is given by
\be
\frac{1}{\xi_A}\left(\D_p\xi_p-\xi_p -\frac{\D_A}{2\xi^2_A}\xi_p^2 + \pi i \rho \xi_p^2\right).
\ee
In this subsection we will explicitly obtain the subleading term for level 2. For this level the basis states are 
\be
\begin{array}{l}
|\Psi_{1}\rangle=L_{-1}^{2}|\Delta_A,\xi_A\rangle,\,\,\,|\Psi_{2}\rangle=L_{-2}|\Delta_A,\xi_A\rangle,\,\,\,|\Psi_{3}\rangle=L_{-1}M_{-1}|\Delta_A,\xi_A\rangle,\,\,\, |\Psi_{4}\rangle = M_{-2}|\Delta_A,\xi_A\rangle\\
|\Psi_{5}\rangle=M_{-1}^{2}|\Delta_A,\xi_A\rangle.
\end{array}.
\ee
From \eqref{F_N_exp}, we have
\bea
&& \mathcal{F}_{N}(\Delta_p,\xi_p;\Delta_A,\xi_A|c_L,c_M|\rho)\cr
&=& \frac{1}{y^{\xi_A}C_{ApA}}\sum_{i,j,k,l} K^{ij}_{(N)} K^{kl}_{(N)}\mathcal{P}_{ik}\<\Psi_l|e^{2\pi i \rho M_0} |\Psi_j\>,
\eea
where $\mathcal{P}_{ik} = \<\Psi_i|\phi_p(0,1)|\Psi_k\>$. We rewrite the last factor as
\bea
\<\Psi_l|e^{2\pi i \rho M_0} |\Psi_j\> &=& \<\Psi_l|e^{2\pi i \rho M_0}L_{\vec{r}_j} M_{\vec{s}_j}|\D_A,\xi_A\> \cr
&=& \<\Psi_l|L_{\vec{r}_j} M_{\vec{s}_j} e^{2\pi i \rho M_0} |\D_A,\xi_A\> + \<\Psi_l|[e^{2\pi i \rho M_0},L_{\vec{r}_j} M_{\vec{s}_j}]|\D_A,\xi_A\> \cr
&=& e^{2\pi i \rho \xi_A} \<\Psi_l|L_{\vec{r}_j} M_{\vec{s}_j}|\D_A,\xi_A\> + \<\Psi_l|[e^{2\pi i \rho M_0},L_{\vec{r}_j} M_{\vec{s}_j}]|\D_A,\xi_A\> \cr
&=& y^{\xi_A} K_{lj}^{(N)}+ \<\Psi_l|[e^{2\pi i \rho M_0},L_{\vec{r}_j} M_{\vec{s}_j}]|\D_A,\xi_A\>.
\eea
Substituting back in the previous equation and using $\sum_{l} K^{kl}_{(N)}K_{lj}^{(N)}=\delta^k_j$, we have
\bea
&& \mathcal{F}_{N}(\Delta_p,\xi_p;\Delta_A,\xi_A|c_L,c_M|\rho) \cr
= && \frac{1}{C_{ApA}}\sum_{i,j,k} K^{ij}_{(N)}\mathcal{P}_{ik} \delta^k_j + \frac{1}{y^{\xi_A}C_{ApA}}\sum_{i,j,k,l} K^{ij}_{(N)} K^{kl}_{(N)}\mathcal{P}_{ik} \<\Psi_l|[e^{2\pi i \rho M_0},L_{\vec{r}_j} M_{\vec{s}_j}]|\D_A,\xi_A\>\cr
&=& \frac{1}{C_{ApA}}\sum_{i,j} K^{ij}_{(N)}\mathcal{P}_{ij} 
+ \frac{1}{y^{\xi_A}C_{ApA}}\sum_{i,j,k,l} K^{ij}_{(N)} K^{kl}_{(N)}\mathcal{P}_{ik} \<\Psi_l|[e^{2\pi i \rho M_0},L_{\vec{r}_j} M_{\vec{s}_j}]|\D_A,\xi_A\>.\cr
&&
\eea

The Gram Matrix for level 2 is
\bea
K^{(2)} = \left[  \begin{array}{ccccc}
4\Delta_A(2\Delta_A+1) & 6\Delta_A & 4\xi_A(2\Delta_A+1) & 6\xi_A & 8\xi^{2}_A\\
6\Delta_A & 4\Delta_A+6c_L & 6\xi_A & 4\xi_A+6c_M & 0\\
4\xi_A(2\Delta_A+1) & 6\xi_A & 4\xi^{2}_A & 0 & 0\\
6\xi_A & 4\xi_A+6c_M & 0 & 0 & 0\\
8\xi^{2}_A & 0 & 0 & 0 & 0
\end{array} \right].
\label{GM_level2} 
\eea 
For the inverse of the matrix, all the entries are of order $\frac{1}{\xi_A}$ or less. To order in $\frac{1}{\xi^2_A}$, we have
\be
K_{(2)}=
\left(
\begin{array}{ccccc}
 0 & 0 & 0 & 0 & \frac{1}{8 \xi^2_A} \\
 0 & 0 & 0 & \frac{1}{4 \xi_A } & -\frac{3}{16 \xi^2_A} \\
 0 & 0 & \frac{1}{4 \xi^2_A} & -\frac{3}{8 \xi^2_A} & -\frac{\Delta_A }{(4 \xi_A ) \xi^2_A} \\
 0 & \frac{1}{4 \xi_A } & -\frac{3}{8 \xi^2_A} & -\frac{\Delta_A }{(4 \xi_A ) \xi_A } & \frac{3 \Delta_A }{(8 \xi_A ) \xi^2_A} \\
 \frac{1}{8 \xi^2_A} & -\frac{3}{16 \xi^2_A} & -\frac{\Delta_A }{(4 \xi_A ) \xi^2_A} & \frac{3 \Delta_A }{(8 \xi_A ) \xi^2_A} & \frac{\Delta^2_A}{\left(8 \xi^2_A\right) \xi^2_A} \\
\end{array}
\right).
\ee
Next, for $\mathcal{P}_{ij} \equiv \<\Psi_i|\phi_p(0,1)|\Psi_i\>$ we have
\bea
\mathcal{P} 
&=& C_{ApA}\left(\begin{array}{ccccc}
- & - & - & - & K_{15}\\
- & - & - & K_{24} & K_{25}\\
- & - & K_{33} & K_{34} & K_{35}\\
- & K_{42} & K_{43} & K_{44} & K_{45}\\
K_{51} & K_{52} & K_{53} & K_{54} & K_{55}
\end{array}\right) \cr
&& +C_{ApA}\left(\begin{array}{ccccc}
- & - & - & - & 8\xi_A\left(\Delta_{p}-1\right)\xi_{p}\\
- & - & - & 4\left(\Delta_{p}-1\right)\xi_{p} & 0\\
- & - & 4\xi_A\left(\Delta_{p}-1\right)\xi_{p}+2\Delta\xi_{p}^{2} & 0 & 4\xi_A\xi_{p}^{2}\\
- & 4\left(\Delta_{p}-1\right)\xi_{p} & 0 & 4\xi_{p}^{2} & 0\\
8\xi\left(\Delta_{p}-1\right)\xi_{p} & 0 & 4\xi_A\xi_{p}^{2} & 0 & 0
\end{array}\right), \cr
&&
\eea
where we have shown only the terms needed to calculate to order $\frac{1}{\xi_A}$. We already know that the leading order is $5$, the number of sates at level $2$. Using the above matrices, we can find the sub-leading term in $\sum_{i,j} K^{ij}_{(2)}\mathcal{P}_{ij}$
\be
\frac{1}{C_{ApA}}\sum_{i,j} K^{ij}_{(2)}\mathcal{P}_{ij} = 5 + \frac{5}{\xi_A}\left( \D_p\xi_p-\xi_p-\frac{\D_A}{2\xi_A}\xi_p^2 \right).  
\ee

Now, let us calculate the sub-leading order in $\sum_{i,j,k,l} K^{ij}_{(2)} K^{kl}_{(2)}\mathcal{P}_{ik} \<\Psi_l|[e^{2\pi i \rho M_0},L_{\vec{r}_j} M_{\vec{s}_j}]|\D_A,\xi_A\>$. For this we have
\bea
\<\Psi_l|[ e^{2\pi i \rho M_0},L_{-1}L_{-1}]|\D,\xi\> &=& 4\pi i \rho\, y^{\xi_A} K_{l3} + (2\pi i \rho)^2 \,y^{\xi_A} K_{l5} \cr
\<\Psi_l|[ e^{2\pi i \rho M_0},L_{-2}]|\D,\xi\> &=& 4\pi i \rho \,y^{\xi_A} K_{l4}\cr
\<\Psi_l|[ e^{2\pi i \rho M_0},L_{-1}M_{-1}]|\D,\xi\> &=& 2\pi i \rho \,y^{\xi_A} K_{l5}\cr
\<\Psi_l|[ e^{2\pi i \rho M_0},M_{-2}]|\D,\xi\> &=& 0 \cr
\<\Psi_l|[ e^{2\pi i \rho M_0},M_{-1}M_{-1}]|\D,\xi\> &=& 0.
\eea
Using $\sum_{cd}K^{ad}K_{d\lambda} = \delta^c_{\lambda}$, we have
\bea
&& \sum_{i,j,k,l}K^{ij}K^{kl}\mathcal{P}_{ik}\<\Psi_l|[e^{2\pi i \rho M_0},L_{\vec{r}_j}M_{\vec{s}_j}]|\D,\xi\> \cr
&=&y^{\xi_A} \sum_{i}4\pi i \rho K^{i1}\mathcal{P}_{i3}  + (2\pi i \rho)^2 K^{i1}\mathcal{P}_{i5} + 4\pi i \rho K^{i2}\rho(\Psi_i,\Psi_4) + 2\pi i \rho K^{i3}\mathcal{P}_{i5} \cr
&=& y^{\xi_A} \left(4\pi i \rho K^{51}\mathcal{P}_{53} + 4\pi i \rho K^{42}\mathcal{P}_{44}  + 2\pi i \rho  K^{33}\mathcal{P}_{35}  \right)   \cr
&=& y^{\xi_A} 8\pi i \rho \frac{\xi_p^2}{\xi_A}.
\eea
So, the $\frac{1}{\xi_A}$ order contribution to level 2 is given by 
\be
\frac{5}{\xi_A}\left( \D_p\xi_p-\xi_p-\frac{\D_A}{2\xi_A}\xi_p^2 \right) + 8 \pi i \rho \frac{\xi_p^2}{\xi_A} .
\ee

\bigskip \bigskip

\newpage

\section{BMS torus blocks as a limit of CFT}\label{gensub}
The form of the BMS one-point torus blocks (in large $\xi_A$ limit) up to subleading term for general level was obtained through an intrinsics analysis  in Section \ref{sec:large_xi} (equation \eqref{onepoint5}). In this appendix we give another derivation of \eqref{onepoint5} by taking the non-relativistic limit of the known CFT result \eqref{2.17}
\be
F_{h_A}(q) = \sum_{n=0}^{\infty} F_n(h_p,h_A,c) q^n =\sum_{n=0}^{\infty}\left(1+\frac{h_p(h_p-1)}{2h_A}n+\mathcal{O}(h^{-2})\right)p(n)q^n,
\ee
where we use $F_{h_A}(q) \equiv F_{h_A,c}^{h_p}(q)$ to shorten the notation.
Combining this with the anti-holomorphic part, the full torus block up to subleading term is
\bea
 && F_{h_A}(q)F_{\bar{h}_A}(\bar{q}) \cr
 = &&  \sum_{n=0}^{\infty} \sum_{\bar{n}=0}^{\infty} \left(1+\frac{h_p(h_p-1)}{2h_A}n +\frac{\bar{h}_p(\bar{h}_p-1)}{2\bar{h}_A}\bar{n}+\mathcal{O}(h^{-2}_A,\bar{h}^{-2}_A,h^{-1}_A,\bar{h}^{-1}_A)\right)p(n)p(\bar{n})q^n\bar{q}^{\bar{n}}.\cr
 &&
\eea
Let us focus only on the subleading term 
\be
\sum_{n=0}^{\infty} \sum_{\bar{n}=0}^{\infty}\left(\frac{h_p(h_p-1)}{2h_A}n +\frac{\bar{h}_p(\bar{h}_p-1)}{2\bar{h}_A}\bar{n}\right) p(n)p(\bar{n})e^{2\pi i \tau n} e^{-2\pi i \bar{\tau}\bar{n}}. 
\ee
If we take the ultra-relativistic limit of this, it would give us the BMS block in the induced representation. In order to get the BMS block in the highest weight representation we have to take the non-relativistic limit. Thus, we use the following contraction  
\be 
(h,\bar{h})\rightarrow \frac{1}{2}\left(\D \mp \frac{\xi}{\epsilon}\right),\,\,\,\,(\tau,\bar{\tau})\rightarrow \pm \sigma - \epsilon \rho,
\ee 
which give us
\begin{align}
&\sum_{n=0}^{\infty} \sum_{\bar{n}=0}^{\infty}\left( \frac{(\D_p-\frac{\xi_p}{\e})(\frac{1}{2}(\D_p-\frac{\xi_p}{\e})-1)}{2(\D_A-\frac{\xi_p}{\e})}n  
+ \frac{(\D_p+\frac{\xi_p}{\e})(\frac{1}{2}(\D_p+\frac{\xi_p}{\e})-1)}{2(\D_A+\frac{\xi_A}{\e})} \bar{n} \right)\cr
& \quad \quad\quad \quad  \times p(n)p(\bar{n})e^{2\pi i \sigma (n+\bar n)} e^{2\pi i \epsilon \rho(\bar{n}-n)} \cr
=& \sum_{n=0}^{\infty} \sum_{\bar{n}=0}^{\infty}\left(\frac{(\epsilon\Delta_{p}-\xi_{p})(\Delta_{p}-\frac{\xi_{p}}{\epsilon}-2)}{4(\epsilon\Delta_A-\xi_A)}n+\frac{(\epsilon\Delta_{p}+\xi_{p})(\Delta_{p}+\frac{\xi_{p}}{\epsilon}-2)}{4(\epsilon\Delta_A+\xi_A)}\bar{n}\right) \cr
& \quad \quad\quad \quad  \times p(n)p(\bar{n})e^{2\pi i \sigma (n+\bar n)} e^{2\pi i \epsilon \rho(\bar{n}-n)} \cr
=& \sum_{n=0}^{\infty} \sum_{\bar{n}=0}^{\infty}\frac{1}{4\xi_A}\left(\frac{(-\epsilon\Delta_{p}+\xi_{p})(\Delta_{p}-\frac{\xi_{p}}{\epsilon}-2)}{(1-\epsilon\frac{\Delta_A}{\xi_A})}n+\frac{(\epsilon\Delta_{p}+\xi_{p})(\Delta_{p}+\frac{\xi_{p}}{\epsilon}-2)}{(1+\epsilon\frac{\Delta_A}{\xi_A})}\bar{n}\right)\cr
& \quad \quad\quad \quad  \times p(n)p(\bar{n})e^{2\pi i \sigma (n+\bar n)} (1+2\pi i \epsilon \rho(\bar{n}-n)+\mathcal{O}(\epsilon^2)) \cr
=& \sum_{n=0}^{\infty} \sum_{\bar{n}=0}^{\infty}\frac{e^{2\pi i \sigma (n+\bar n)}}{4\xi_A}\left((n+\bar{n})\left(2\D_p\xi_p-2\xi_p-\frac{\D_A}{\xi_A}\xi_p^2\right)+2\pi i \rho (\bar{n}-n)^2\xi_p^2+\mathcal{O}(\epsilon)\right)\cr
& \quad \quad\quad \quad  \times p(n)p(\bar{n}).
\end{align}
In the above equation we take the limit $\epsilon \rightarrow 0$. We also convert the summation: we sum over $n$ and $\bar{n}$ such that $n+\bar{n}=N$ and sum over $N$ from $0$ to infinity. Doing this we get
\begin{align}
& \sum_{N=0}^{\infty} \sum_{k=0}^{N}\frac{e^{2\pi i \sigma N}}{4\xi_A}\left(N\left(2\D_p\xi_p-2\xi_p-\frac{\D_A}{\xi_A}\xi_p^2\right)+2\pi i \rho (N-2k)^2 \xi_p^2+\mathcal{O}(\epsilon)\right) p(N-k)p(k) \cr
=&  \sum_{N=0}^{\infty} e^{2\pi i \sigma N}\left(\frac{N}{2\xi_A}\left(\D_p\xi_p-\xi_p-\frac{1}{2}\frac{\D_A}{\xi_A}\xi_p^2\right)\sum_{k=0}^{N}p(N-k)p(k) \right. \cr
& \left. \quad\quad\quad\quad\quad\quad\quad\quad\quad\quad+\frac{\pi i \rho }{2} \frac{\xi_p^2}{\xi_A} \sum_{k=0}^N p(N-k)p(k)(N-2k)^2 \right) \cr
=& \sum_{N=0}^{\infty} e^{2\pi i \sigma N}\left(\frac{N}{2\xi_A}\left(\xi_p(\D_p-1)-\frac{1}{2}\frac{\D_A}{\xi_A}\xi_p^2\right)\widetilde{\text{dim}}_N\ +\frac{\pi i \rho }{2} \frac{\xi_p^2}{\xi_A} \sum_{k=0}^N p(N-k)p(k)(N-2k)^2 \right). 
\end{align}
Thus the subleading correction to $\mathcal{F}_N$ is given by
\be
 \frac{N}{2\xi_A}\left(\xi_p(\D_p-1)-\frac{1}{2}\frac{\D_A}{\xi_A}\xi_p^2\right)\widetilde{\text{dim}}_N\ +\frac{\pi i \rho }{2} \frac{\xi_p^2}{\xi_A} \sum_{k=0}^N p(N-k)p(k)(N-2k)^2 .
\ee
If we take $\D_A << \xi_A$, then the subleading term is 
 \be
 \frac{N}{2\xi_A}\xi_p(\D_p-1)\widetilde{\text{dim}}_N\ +\frac{\pi i \rho }{2} \frac{\xi_p^2}{\xi_A} \sum_{k=0}^N p(N-k)p(k)(N-2k)^2 .
 \label{sublead_from_cft1}
\ee
We note that 
\be
\frac{1}{2}\sum_{k=0}^N  p(N-k)p(k)(N-2k)^2 = \sum_{k=0}^N  p(N-k)p(k)(N-k)(N-2k). 
\ee
Thus, we can also write the subleading term \eqref{sublead_from_cft1} as
\be
 \frac{N}{2\xi_A}\xi_p(\D_p-1)\widetilde{\text{dim}}_N\ +\pi i \rho \frac{\xi_p^2}{\xi_A} \sum_{k=0}^N  p(N-k)p(k)(N-k)(N-2k) .
 \label{sublead_from_cft2}
\ee

\bigskip \bigskip

\newpage

\section{Asymptotic form of partition of integers }
In this appendix we will use Cardy like analysis to find the asymptotic form of the partition of integers $p(N)$, the Hardy-Ramanujan partition formula.The generating function of partition of integers $p(N)$ is given by
\be
\prod_{N=1}^{\infty}\frac{1}{1-q^N}=\sum_{N=0}^{\infty}p(N)q^N. 
\label{gfpart}
\ee
In terms of the Dedekind eta function 
\be
\eta(q) = q^{\frac{1}{24}} \prod_{N=1}^{\infty}(1-q^N),
\ee
we have 
\be
\frac{q^{\frac{1}{24}}}{\eta(q)}=\sum_{n=0}^{\infty}p(n)q^n. 
\ee
Then we define $\mathcal{Z}(q)\equiv \frac{q^{\frac{1}{24}}}{\eta(q)}$. So we have
\be
\mathcal{Z}(q) = \sum_{N=0}^{\infty}p(N)q^N.
\ee
We could invert the above relation to express $p(N)$ in terms of $\mathcal{Z}(q)$ 
\be
p(N) = \frac{1}{2\pi i}\int \frac{dq}{q^{N+1}} \mathcal{Z}(q). 
\ee
In term of $q=e^{2\pi i \tau}$, the above equation is
\be
p(N) = \int d\tau e^{-2\pi i \tau (N+1)} \mathcal{Z}(\tau).
\label{partcon}
\ee
Under the S modular transformation $\eta(\tau)$ transforms as
\be
\eta\left(-\frac{1}{\tau}\right)=\sqrt{-i\tau}\eta(\tau). 
\ee
So we have 
\bea
 e^{\frac{2\pi i \tau}{24}}\frac{e^{-\frac{2\pi i}{24 \tau}}}{\eta\left(-\frac{1}{\tau}\right)}&=& e^{-\frac{2\pi i}{24 \tau}}\frac{e^{\frac{2\pi i \tau}{24}}}{\sqrt{-i\tau}\eta(\tau)} \cr
 \implies  e^{\frac{2\pi i \tau}{24}}\mathcal{Z}\left(-\frac{1}{\tau}\right) &=& \frac{e^{-\frac{2\pi i}{24 \tau}}}{\sqrt{-i\tau}}\mathcal{Z}(\tau) \cr
 \implies \mathcal{Z}(\tau) &=& e^{\frac{\pi i \tau}{12}}e^{\frac{\pi i}{12 \tau}}\mathcal{Z}\left(-\frac{1}{\tau}\right)\sqrt{-i\tau}.
 \label{modtranz}
\eea
Substituting this in \eqref{partcon}, we obtain
\bea
p(N) &=&  \int d\tau e^{-2\pi i \tau (N+1)} e^{\frac{\pi i \tau}{12}}e^{\frac{\pi i}{12 \tau}}\mathcal{Z}\left(-\frac{1}{\tau}\right)\sqrt{-i\tau}\cr
& \equiv & \int d\tau e^{f(\tau)}\mathcal{Z}\left(-\frac{1}{\tau}\right)\sqrt{-i\tau},
\eea
where 
\be
f(\tau) = -2\pi i \tau (N+1)+\frac{\pi i\tau}{12} + \frac{\pi i}{12\tau}. 
\ee
For large $N$ we can use saddle point method to find $p(N)$. It is given by 
\be
p(N) \approx \sqrt{\frac{2\pi}{-f^{\prime\prime}(\tau_0)}} e^{f(\tau_0)} \mathcal{Z}(-\frac{1}{\tau_o}) \sqrt{-i\tau_0},
\ee
where $\tau_0$ is the saddle point
\be
 \frac{df(\tau)}{d\tau}|_{\tau_0}=0.
\ee
The saddle point equation is given by
\be
24 N \tau_0 ^2+23 \tau^2_0+1 = 0, 
\ee
with solutions
\be
\tau_0 = i\frac{1}{\sqrt{23+24N}},\,\, -i\frac{1}{\sqrt{23+24N}}.
\ee
Since $N$ is large
\be 
\tau_0 \approx i\frac{1}{\sqrt{24N}},\,\, -i\frac{1}{\sqrt{24N}}.
\ee
We have to choose
\be\tau_0 \approx i\frac{1}{\sqrt{24N}}, 
\ee
since this give the maxima for $f(\tau)$. More precisely
\be
f(\tau_0) \approx e^{\sqrt{\frac{2}{3}} \pi  \sqrt{N}}. 
\ee
We also have 
\bea
\sqrt{\frac{2\pi}{-f^{\prime\prime}(\tau_0)}}&=&\frac{1}{6^{\frac{1}{4}}2N^{\frac{3}{4}}},\cr
\sqrt{-i\tau_0} &=&  \frac{1}{(24)^{\frac{1}{4}}N^{\frac{1}{4}}}.
\eea
Thus, we get
\be
 \sqrt{\frac{2\pi}{-f^{\prime\prime}(\tau_0)}} \sqrt{-i\tau_0} = \frac{1}{4\sqrt{3}N}.
\ee
Now 
\be
\mathcal{Z}(-\frac{1}{\tau_0}) = \sum_n p(n)e^{-2\pi i\frac{1}{\tau_0}n}= \sum_n p(n) e^{-4\pi \sqrt{6N}n}\approx P(0) = 1. 
\ee
So, for large $N$
\be
p(N)\approx  \sqrt{\frac{2\pi}{-f^{\prime\prime}(\tau_0)}} e^{f(\tau_0)} \mathcal{Z}(\tau_o) \sqrt{-i\tau_0} \approx 
\frac{1}{4\sqrt{3}N} e^{\sqrt{\frac{2N}{3}} \pi }.
\ee
This is the Hardy-Ramanujan formula.

\bigskip \bigskip
\bigskip \bigskip

\section{Asymptotic form of $\widetilde{\rm{dim}}_n$}

$\widetilde{\rm{dim}}_n$ is equal to the number of ways to partition an integer $N$ using two colors. For example, for the integer 3, the possibilities are
\be
\textcolor{cyan}{3},\,\textcolor{cyan}{12},\,\textcolor{cyan}{111},\,\textcolor{cyan}{2}\textcolor{red}{1},\,\textcolor{cyan}{11}\textcolor{red}{1},\,\textcolor{cyan}{1}\textcolor{red}{2},\,\textcolor{cyan}{1}\textcolor{red}{11},\,\textcolor{red}{111},\,\textcolor{red}{12},\,\textcolor{red}{3}, 
\ee 
giving us $\widetilde{\rm{dim}}_3=10$. In general, we have
\be
\widetilde{\rm{dim}}_n = \sum_{k=0}^{n}p(n-k)p(k). 
\ee
It is easy to see that
\bea
\sum_{n_1=0}^{\infty}p(n_1)q^{n_1} \sum_{n_2=0}^{\infty}p(n_2)q^{n_2} &=& \sum_{n=0}^{\infty} \sum_{k=0}^{n}p(n-k)p(k)q^n \cr 
&=&\sum_{n=0}^{\infty} \widetilde{\rm{dim}}_n q^n.
\eea
Substituting \eqref{gfpart} in the above equation
\be
 \left(\prod_{n=1}^{\infty}\frac{1}{1-q^n}\right)^2=\sum_{n=0}^{\infty} \widetilde{\rm{dim}}_n q^n.
\ee
Expressing this in terms of the Dedekind eta function
\bea
\frac{q^{\frac{1}{12}}}{\eta(q)^2} &=& \sum_{n=0}^{\infty} \widetilde{\rm{dim}}_n q^n \cr
\implies \frac{q^{\frac{1}{12}}}{\eta(q)^2}= \mathcal{Z}(q)^2 \equiv \widetilde{\mathcal{Z}}(q) &=& \sum_{n=0}^{\infty} \widetilde{\rm{dim}}_n q^n.
\eea
Inverting the above equation, we obtain
 \be
\widetilde{\rm{dim}}_N = \int d\tau e^{-2\pi i \tau (N+1)} \widetilde{\mathcal{Z}}(\tau).
\label{partbms}
\ee
We could see that
\be
\widetilde{\mathcal{Z}} = \mathcal{Z}^2. 
\ee
Using \eqref{modtranz} in the above equation, we get
\be{} \label{assympz}
\widetilde{\mathcal{Z}}(\tau) = e^{\frac{\pi i \tau}{6}}e^{\frac{\pi i}{6 \tau}}\widetilde{\mathcal{Z}}\left(-\frac{1}{\tau}\right)(-i\tau).
\ee
Substituting this in \eqref{partbms}, we have
\bea
\widetilde{\rm{dim}}_N &=& \int d\tau e^{-2\pi i \tau (N+1)} e^{\frac{\pi i \tau}{6}}e^{\frac{\pi i}{6 \tau}}\widetilde{\mathcal{Z}}\left(-\frac{1}{\tau}\right)(-i\tau) \cr
& \equiv & \int d\tau e^{\tilde{f}(\tau)}\widetilde{\mathcal{Z}}\left(-\frac{1}{\tau}\right)(-i\tau),
\eea
where 
\be
\tilde{f}(\tau) = -2\pi i \tau (N+1)+\frac{\pi i\tau}{6} + \frac{\pi i}{6\tau}. 
\ee 
For large $N$, using the saddle point analysis 
 \be
\widetilde{\rm{dim}}_N \approx \sqrt{\frac{2\pi}{-\tilde{f}^{\prime\prime}(\tau_0)}} e^{\tilde{f}(\tau_0)} \widetilde{\mathcal{Z}}(-\frac{1}{\tau_o}) (-i\tau_0),
\ee
where the saddle point equation $\frac{d\tilde{f}(\tau)}{d\tau}|_{\tau_0}=0$ is 
\be
1+11\tau_0^2 + 12N\tau_0^2=0, 
\ee
whose solutions are 
\be
\tau_0 = i\frac{1}{\sqrt{12N+11}},\, -i\frac{1}{\sqrt{12N+11}}.
\ee
For large $N$
\be
\tau_0 \approx i\frac{1}{\sqrt{12N}},\, -i\frac{1}{\sqrt{12N}}.
\ee
Once again, the maxima is given by the point $\tau_0 \approx i\frac{1}{\sqrt{12N}}$.
Thus 
\be
e^{\tilde{f}(\tau_0)} =  e^{\sqrt{\frac{N}{3}} 2\pi}.
\ee
We also have 
\bea
\sqrt{\frac{2\pi}{-\tilde{f}^{\prime\prime}(\tau_0)}}&\approx&\frac{1}{3^{\frac{1}{4}}2N^{\frac{3}{4}}},\cr
-i\tau_0 &\approx&  \frac{1}{2\sqrt{3N}}\cr
\implies \sqrt{\frac{2\pi}{-\tilde{f}^{\prime\prime}(\tau_0)}}(-i\tau_0) &\approx & \frac{1}{4\ 3^{\frac{3}{4}} N^{\frac{5}{4}}}.
\eea
Finally, the large $N$ asymptotic formula for  $ \widetilde{\rm{dim}}_N$ is given by
\be
 \widetilde{\rm{dim}}_N \approx  \frac{1}{4\ 3^{\frac{3}{4}} N^{\frac{5}{4}}} e^{\sqrt{\frac{N}{3}} 2\pi}.\label{dimn}
\ee

\newpage

\section{Condition for leading term to dominate asymptotics of torus blocks}\label{justif}
From \eqref{onepoint5}, we could see that the large $\xi_A$ expansion of the torus block upto first sub-leading term is given by
 \be
\sum_{N=0}^{\infty}\left[\left(1+\frac{\xi_p(\Delta_p-1)}{2\xi_A}N\right)\widetilde{\text{dim}}_N+\pi i\rho\frac{\xi_p^2}{\xi_A} \sum_{k=0}^Np(N-k)p(k)(N-k)(N-2k) \right] q^N.
\ee
We would like to find the condition which would allow us to keep only the leading term $\sum_{n=0}^{\infty}\widetilde{\text{dim}}_N q^N$ and neglect the higher order terms in $\frac{1}{\xi_{A}}$. 
If we look at the term inside the first bracket in the summand above, we can neglect $\frac{N}{\xi_A}$ in comparison to 1 if $N$ is finite and $\xi_A$ is large. However, $N$ is unbounded as we have to sum over $N$ from $0$ to infinity. So, however large $\xi_A$ is, we always have a $N$ comparable and larger than $\xi_A$ so that we cannot neglect the term $\frac{N}{\xi_A}$. So, we have to make $q=e^{-2\pi\beta}$ small i.e, make $\beta$ small enough such that the sum can be truncated at some value $N<<\xi_A$. 

Using the asymptotic form of $\widetilde{\text{dim}}_N$ \eqref{dimn}, we have  $ \widetilde{\text{dim}}_N \,q^N \sim N^{-\frac{5}{4}} e^{\sqrt{\frac{N}{3}} 2\pi} e^{-2\pi\beta N}$. This has a saddle point at $N=N_{\star}=\frac{-15 \beta+\sqrt{\pi } \sqrt{\pi -30 \beta}+\pi }{24 \pi  \beta^2}$. Since $\beta<<1$ we may approximate this as $N_{\star} \approx \frac{1}{24 \beta^2}$. As we can see, $N_{\star}\gg1$. Since $\widetilde{\text{dim}}_N q^N$ is maximum at $N_{\star}$, we at least have to sum upto $N_{\star}$. So, in order to truncate the sum at some value $N<<\xi_A$, we at least require $N_{\star}<< \xi_A$ which is same as the requirement that $\beta^2\xi_A >>1 $. If we consider the second term in $\frac{1}{\xi_A}$ and perform the similar analysis, we obtain the same condition. So, we have
\be{}
\mathcal{F}_{\D_A,\xi_A,c_L,c_M}^{\D_p,\xi_p}(\sigma,\rho) \approx \sum_{N=0}^{\infty}q^N\widetilde{\text{dim}}_N=\frac{q^\frac{1}{12}}{\eta(\sigma)^2}, ~~~~~~\text{provided~}\beta\ll1,\beta^2 \xi_A \gg 1.   
\ee

\bigskip \bigskip
\bigskip \bigskip

\newpage

\section{Details regarding number theory conjecture}\label{secconjecture}
In this section, we will provide some details regarding the validity of conjecture \eqref{conjecture}. This conjecture is important as it matches our intrinsic analysis of sub-leading term of BMS torus block with the limiting analysis from known CFT result. Though we have not been able to prove it, we have checked for its validity using two colours. We define $\widetilde{\rm{dim}}_N$ to be the  partition of an integer N using two colours . 
Thus,
\be{}
\begin{split}
&\widetilde{\rm{dim}}_1=2: ~~1,\txb{1}\\
&\widetilde{\rm{dim}}_2=5: ~~2,11,1\txb{1},\txb{11},\txb{2}\\
&\widetilde{\rm{dim}}_3=10:~~3,12,111,11\txb{1},1\txb{11},\txb{111},1\txb{2},\txb{12},\txb{3}\\
&\widetilde{\rm{dim}}_4=20:~~4,31,211,1111,111\txb{1},11\txb{11},1\txb{111},\txb{1111},21\txb{1},2\txb{11},\txb{2}11,\txb{21}1,\txb{211},3\txb{1},\txb{3}1,\txb{31},\txb{4},22,2\txb{2},\txb{22}\\
&\widetilde{\rm{dim}}_5=36:~~ 5,41,32,311,11111,1111\txb{1},111\txb{11},11\txb{111},1\txb{1111},\txb{11111},31\txb{1},3\txb{11},\txb{3}11,\txb{31}1,\txb{311},3\txb{2},\txb{3}2,\txb{32},\\
&~~~~~~~~~~~~~~~~~~4\txb{1},\txb{4}1,\txb{41},221,22\txb{1},2\txb{21},\txb{2}2\txb{1},\txb{221},\txb{22}1,1112,111\txb{2},11\txb{1}2,11\txb{12},1\txb{112},\txb{1112},\txb{11}12,\txb{111}2,\txb{5}.
\end{split}
\ee 
We also have the partition of integer $N$  defined as $p(N)$. Thus,
\be{}
\begin{split}
&\widetilde{\text{dim}}_N =\sum_{k=0}^n p(N-k)p(k).\\
&\text{where,~}p(0)=1,~p(1)=1,~p(2)=2,~p(3)=3,~p(4)=5,~p(5)=7.
\end{split}
\ee
The conjecture described in \eqref{conjecture} is
\begin{align}
\sum_{\{\{l_j\},\{m_j\}\}\in A}\,\,\sum_{j=1}^N j^2l_j(m_j+1)&=\frac{1}{2}\sum_{k=0}^Np(N-k)p(k)(N-2k)^2\cr
&= \sum_{k=0}^Np(N-k)p(k)(N-k)(N-2k).
\end{align}
We have checked for the validity of the conjecture upto $N=5$. We will describe here the calculation involved in level 5 as an example. For the RHS,
\be{}
\begin{split}\label{rcon}
&\frac{1}{2}\sum_{k=0}^5 p(N-k)p(k)(N-2k)^2\\
&=\frac{1}{2}\Big[2p(5)p(0)5^2+2p(2)p(3)1^2+2p(1)p(4)3^2\Big]\\
&= \sum_{k=0}^5p(N-k)p(k)(N-k)(N-2k)=226.\\
\end{split}
\ee
Now, we will concentrate on the LHS.
We have \be{}\label{cons}\sum_{\{\{l_j\},\{m_j\}\}\in A}\,\,\sum_{j=1}^5 j(l_j+ m_j)=5.\ee Our aim is to calculate the quantity $$\mathcal{I}= j^2l_j(m_j+1)$$ for the values of $\{j,l_j,m_j\}$ for $(j= 1,\hdots,5)$ satisfying \eqref{cons}. The table below shows the allowed values of $\{j, l_j, m_j\}$ and $\mathcal{I}$.
\begin{small}
\begin{center}
 \begin{tabular}{|c c || c c|| c c |} 
 \hline
$(j\;l_j\;m_j)$ & $\mathcal{I}$ & $(j\;l_j\;m_j)$ & $\mathcal{I}$ & $(j\;l_j\;m_j)$ & $\mathcal{I}$  \\ [0.5ex] 
 \hline\hline
 (150),(200),(300),(400),(500)& 5 &  (101),(220),(300),(400),(500) & 8 & (100),(210),(310),(400),(500) & 13\\ 
 (105),(200),(300),(400),(500) & 0 & (110),(220),(300),(400),(500) & 9 & (100),(210),(301),(400),(500) & 4\\
 (114),(200),(300),(400),(500) & 5 & (101),(202),(300),(400),(500) & 0 & (100),(201),(310),(400),(500) & 9\\
  (141),(200),(300),(400),(500)  & 8 & (110),(202),(300),(400),(500) & 1 & (100),(201),(301),(400),(500) & 0\\
  \cline{5-6}
   (123),(200),(300),(400),(500) & 8  & (101),(211),(300),(400),(500) & 8 & (111),(200),(310),(400),(500) &11\\
    (132),(200),(300),(400),(500)& 9 & (110),(211),(300),(400),(500) & 9 &(120),(200),(310),(400),(500) &11\\
    \cline{1-4}
 (110),(200),(300),(401),(500)&1& (103),(201),(300),(400),(500)& 0  &(102),(200),(310),(400),(500) &9\\
    (110),(200),(300),(410),(500)&17 & (103),(210),(300),(400),(500)& 4 &(111),(200),(301),(400),(500) &2\\
 (101),(200),(300),(401),(500)&0  & (130),(201),(300),(400),(500)& 3  &(120),(200),(301),(400),(500) &2\\
 (101),(200),(300),(410),(500)&16  & (130),(210),(300),(400),(500)& 7  &(102),(200),(301),(400),(500) &0\\
 \hline
 (100),(200),(300),(400),(501)& 0 & (112),(210),(300),(400),(500) & 7 & (112),(201),(300),(400),(500) & 3\\
  (100),(200),(300),(400),(510)& 25 & (121),(210),(300),(400),(500) & 8 &(121),(201),(300),(400),(500) & 4
   \\ [0.4ex] 
 \hline
\end{tabular}
\end{center}
\end{small}

\noindent Summing over the values of $\mathcal{I}= j^2l_j(m_j+1)$ we obtain 226. It matches with the RHS of \eqref{rcon}.

\end{document}